\documentclass[acmlarge]{acmart}
\AtBeginDocument{%
  }

\usepackage{multirow}
\usepackage{siunitx}
\usepackage{booktabs}
\usepackage{adjustbox}
\usepackage{tabularx}
\usepackage{colortbl}
\usepackage{xcolor}
\usepackage[normalem]{ulem}
\usepackage{makecell}
\usepackage{pifont}
\usepackage[linesnumbered,ruled,vlined]{algorithm2e}
\usepackage{ulem}
\usepackage{CJKulem}
\usepackage{soul}
\usepackage{paralist}
\usepackage{subcaption}
\usepackage{caption}
\usepackage{adjustbox}
\usepackage{graphicx}
\usepackage{mwe}
\usepackage{soul}
\usepackage{enumitem}
\usepackage{array}
\usepackage{tcolorbox}
\tcbuselibrary{listings, skins, breakable}
\usepackage{placeins}
\usepackage[left]{lineno}

\usepackage{color}
\usepackage{listings}
\usepackage{float}
\usepackage{xspace}
\newcommand{\workname}{PerCache\xspace}
\newcommand{\modify}[1]{{\color{black} #1}} 
\usepackage[subtle]{savetrees}
\newcommand{\final}[1]{{\color{black}#1}}

\newtcolorbox{promptbox}[1][]{
    breakable,
    enhanced,
    colback=gray!5,
    colframe=gray!70,
    fonttitle=\bfseries,
    fontupper=\small\ttfamily,
    before upper={\setlength{\parskip}{0.8em}},
    title={#1}
}

\setcopyright{cc}
\setcctype{by-nc-nd}
\acmJournal{IMWUT}
\acmYear{2026} \acmVolume{10} \acmNumber{4} \acmArticle{239}
\acmMonth{12} \acmDOI{10.1145/3857927}

\makeatletter
\renewcommand\@fnsymbol[1]{%
  \ifcase#1\or $\dagger$\or $*$\or $\ddagger$\or $\S$\or $\P$%
  \else\@ctrerr\fi}
\makeatother

\begin{document}
% \linenumbers

\title[PerCache: Predictive Hierarchical Cache for RAG Applications on Mobile Devices]
{\workname: Predictive Hierarchical Cache for RAG Applications on\\Mobile Devices}

\author{Kaiwei Liu}
\orcid{0009-0002-4108-0898}
\affiliation{
  \institution{The Chinese University of Hong Kong}
  \city{Hong Kong}
  \country{China}}
\email{kaiweiliu@link.cuhk.edu.hk}

\author{Liekang Zeng}
\orcid{0000-0003-4800-8768}
\affiliation{
  \institution{The Chinese University of Hong Kong}
  \city{Hong Kong}
  \country{China}}
\email{lkzeng@cuhk.edu.hk}

\author{Lilin Xu}
\orcid{0009-0007-5203-7496}
\affiliation{
  \institution{Columbia University}
  \city{New York}
  \country{United States}}
\email{lilin.xu@columbia.edu}

\author{Bufang Yang}
\orcid{0000-0003-0032-2539}
\affiliation{
  \institution{The Chinese University of Hong Kong}
  \city{Hong Kong}
  \country{China}}
\email{bfyang@link.cuhk.edu.hk}

\author{Zhenyu Yan}
\authornote{Corresponding author.}
\orcid{0000-0002-4433-5211}
\correspondingauthor
\affiliation{
  \institution{The Chinese University of Hong Kong}
  \city{Hong Kong}
  \country{China}}
\email{zyyan@ie.cuhk.edu.hk}

\begin{abstract}
Retrieval-augmented generation (RAG) has \modify{become} a \textit{de facto} paradigm in various large language model (LLM)-driven applications on mobile devices, such as mobile assistants leveraging personal emails or meeting records. However, due to lengthy prompts and resource constraints, mobile RAG systems exhibit significantly high response latency. \modify{To address this issue}, one promising approach is to reuse intermediate computational results across different queries to eliminate redundant computation. But most existing approaches, such as \modify{key--value (KV)} cache reuse and semantic cache reuse, are designed for cloud settings and \final{perform poorly on mobile devices as they overlook the distinctive characteristics of mobile RAG.}

We propose \workname, a novel hierarchical cache \modify{framework} designed for reducing end-to-end latency of personalized RAG applications on mobile platforms. \workname~adopts a hierarchical architecture that progressively matches similar queries and \modify{query--key--value (QKV)} cache to maximize the reuse of intermediate results at different computing stages. To improve cache hit rate, \workname~applies a predictive method to populate \modify{the} cache with queries that are likely to be raised in the future. In addition, \workname~can adapt its configurations to dynamic system loads, \modify{aiming to maximize caching utility} with constrained resource consumption. We implement \workname~on top of an existing mobile LLM inference engine \modify{and deploy it on} commodity mobile phones. 
\modify{
Extensive evaluations show that \workname~achieves the lowest average end-to-end latency across five datasets and two LLM architectures, reducing latency by up to 63.5\% over the strongest baseline while maintaining comparable answer quality.
}
\end{abstract}

\makeatletter
\def\@correspondingauthormark{}
\def\@mkauthorsaddresses{%
  \ifnum\num@authors>1\relax
  Authors' \else Author's \fi
  Contact Information:
  \bgroup
  \def\streetaddress##1{\unskip\ignorespaces}%
  \def\postcode##1{\unskip\ignorespaces}%
  \def\position##1{\unskip\ignorespaces}%
  \gdef\@ACM@institution@separator{, }%
  \def\institution##1{\unskip\@ACM@institution@separator
    ##1\gdef\@ACM@institution@separator{ and }}%
  \def\city##1{\unskip, ##1}%
  \def\state##1{\unskip, ##1}%
  \renewcommand\department[2][0]{\unskip\@addpunct, ##2}%
  \def\country##1{\unskip, ##1}%
  \def\and{\unskip; \gdef\@ACM@institution@separator{, }}%
  \def\@author##1{##1}%
  \def\@correspondingauthormark{\unskip\ (corresponding author)}%
  \def\email##1##2{\unskip, ##2}
  \addresses
  \egroup}
\makeatother




\maketitle

\renewcommand{\shortauthors}{Liu et al.}
\section{INTRODUCTION}

\begin{figure}[t]
\centering
\captionsetup{skip=0pt}
\resizebox{0.60\columnwidth}{!}{
\includegraphics{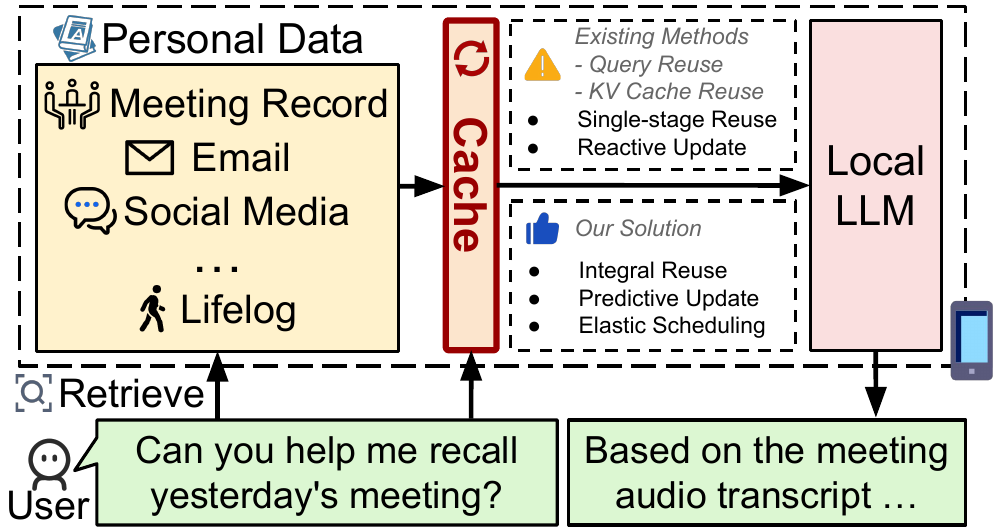}}
\vspace{1em}
\caption{Workflow of mobile RAG applications.}
\label{fig:application_of_speccache}
\vspace{-1em}
\end{figure}

Emerging deployments of large language models (LLMs)~\cite{yi2024phonelm,hu2024minicpm,yao2024minicpm,team2024gemma,bellagente2024stable,abdin2024phi, allal2025smollm2,lu2024small, qwen3-0.6b} \modify{have expanded to} various mobile platforms, driving numerous mobile applications such as personal assistants and mailing helpers~\cite{christmann2025recursive, christmann2025reqap, li2024personal, park2025mobilerag, seemakhupt2024edgerag}.
An example in Figure~\ref{fig:application_of_speccache} shows that such applications often use retrieval-augmented generation (RAG)~\cite{lewis2020retrieval} to generate responses based on personal data from the user's mobile devices.
By deploying both the LLM and personal database locally on mobile devices, these mobile RAG applications can prevent sensitive data leakage and reduce communication latency~\cite{li2024personal}.
However, the prompts containing the retrieved personal data are often extremely lengthy, which can be more than 10$\times$ \modify{as many tokens as} the original query~\cite{jin2024ragcache}, resulting in significant computational overhead and response latency, 
hindering their usability in applications requiring timely responses. For instance, a meeting assistant needs to quickly retrieve and summarize historical meeting information.

To accelerate RAG workflows, prior works have explored the following \modify{approaches}:
(1) Model optimization, such as knowledge distillation~\cite{chen2025drag, bezerra2025llmquoter}, reduces model size and computational cost in RAG workflows.
However, these methods introduce additional retraining overhead and suffer from degraded model performance.
(2) Hardware exploitation~\cite{quinn2025drex, liu2025heterrag} accelerates RAG workflows through specialized hardware architectures. However, relying on specific hardware design undermines their generalizability and requires complex scheduling mechanisms.
(3) RAG retrieval optimization~\cite{park2025mobilerag, seemakhupt2024edgerag, jiang2023chameleon, jiang2024piperag, zhang2024accelerating} focuses on reducing the latency of knowledge retrieval but cannot address the LLM inference latency caused by long prompts, which constitutes the primary bottleneck in single-user mobile RAG scenarios.

To address this gap, in this work, we focus on the primary latency bottleneck in \textbf{\textit{mobile}} RAG applications: LLM inference. 
Our observations show that it is possible to share intermediate computational results within the mobile RAG workflows across different user queries.
Thus, we consider accelerating mobile RAG workflows by caching reusable results to avoid repeated computation.
Some prior works, such as RAGCache~\cite{jin2024ragcache} and Prompt Cache~\cite{gim2024prompt}, reuse the key-value or attention state tensors across different queries to skip portions of attention computation during LLM prefilling.
Other works, such as GPTCache~\cite{bang2023gptcache} and MeanCache~\cite{gill2024privacy}, reuse the LLM outputs of previous \modify{queries that show} high semantic similarities with the new queries to skip inference.
However, these methods are limited to reusing results at a single computing stage. 
KV cache reuse can accelerate prefilling, but fails to avoid redundant computation caused by highly similar outputs during LLM decoding.
Semantic cache \modify{alone} cannot reduce the latency for cache-miss queries even if their prompts contain identical retrieved knowledge texts in the RAG workflow.
\modify{
This is because semantic caching reuses only a previously generated answer after a query-level match; once the match fails, it does not retain or reuse the QKV states of the otherwise identical retrieved knowledge, and the full prefilling and decoding process is still required.
}
For mobile RAG, the limitations of single-stage reuse are \modify{exacerbated} as mobile devices have limited parallel computing capabilities, causing both LLM prefilling and decoding to contribute significantly to overall latency.
Moreover, single-user mobile RAG applications typically receive much sparser queries than multi-user server applications, making it difficult to populate the cache adequately. This sparsity leads to low cache hit rates and limited efficiency gains.

In this paper, we explore a novel cache system to reduce end-to-end response latency for personalized mobile RAG applications. It effectively leverages reusable results at multiple stages within the RAG pipeline across different queries, achieving high cache hit rates even with sparse user queries.
Realizing this goal faces multiple challenges. 
\textit{First}, it requires fully utilizing the reusable results from each stage to maximize the end-to-end latency reduction. 
\textit{Second}, we need a new cache population mechanism to improve cache hit rates for single-user mobile RAG applications with sparse query patterns.
The reactive population mechanisms adopted by existing solutions make cache hit rates entirely dependent on the received queries.
The new mechanism needs to minimize this dependence while improving the hit rate. \textit{Finally}, since mobile devices have severely limited and dynamic computational and storage resources, \workname~must be able to adapt its cache storage and population strategy to effectively utilize system resources \final{while minimizing the impact on performance}.
 
\modify{To address these challenges, we design \workname, a caching framework tailored to fully on-device RAG applications for individual mobile users.}
\workname~adopts a hierarchical cache architecture that stores both \modify{query-answer (QA)} pairs and QKV caches of retrieved knowledge chunks. This design enables comprehensive reuse of intermediate results across mobile RAG workflow stages while providing enhanced flexibility through multiple cache layers.
To improve cache hit rates when sparse user queries fail to adequately populate the cache, \workname~employs a query prediction mechanism that predicts future user queries based on knowledge content and historical queries, enabling predictive cache population during device idle time.
\workname~also includes a cache scheduler to manage the cache population strategy and the conversion between QA bank and QKV cache tensors, enabling \workname~to effectively utilize dynamically available computational and storage resources while maintaining optimal latency performance.
\modify{
These mechanisms are tightly coordinated to maximize the reduction in RAG response latency under limited mobile resources.
}

\modify{We implement \workname on top of mllm~\cite{yi2023mllm} and evaluate it on five datasets comprising 608 evaluation units and 59,367 benchmark queries, including MISeD, EnronQA, PerLTQA, and two self-collected datasets.
Experiments using Llama-3.2-3B and Qwen-1.5-1.8B show that \workname achieves the lowest average end-to-end latency across all datasets while maintaining answer quality comparable to the Naive baseline.
On PerLTQA, \workname reduces average latency by 63.5\% (2.74$\times$) over the strongest baseline.
Three micro-benchmarks further show that \workname~can elastically bypass portions of the cache population, reducing computational overhead by 14.12\%, and can optimize latency through cross-layer cache conversion under varying resource constraints.}

In summary, this paper makes the following contributions:
\begin{itemize}[leftmargin=*]
\item
\final{We propose the first predictive hierarchical caching framework designed to reduce response latency in fully on-device, single-user mobile RAG applications.}

\item
We design a query prediction mechanism that leverages query history and personal data content to predict future user queries, enabling predictive cache population to improve hit rates despite sparse user queries.

\item 
We analyze latency-computation and latency-storage trade-offs in the cache system. Based on this analysis, we design a cache scheduler that manages the cache population strategy and controls cross-layer cache conversions to effectively utilize dynamically available system resources \final{while minimizing the impact on performance.}

\item
\modify{
An implementation and comprehensive evaluation of \workname show that it achieves the lowest average end-to-end latency across five datasets and two LLM architectures, reducing latency by up to 63.5\% over the strongest baseline while maintaining comparable answer quality.
}

\end{itemize}

\section{BACKGROUND AND MOTIVATION}

\subsection{Mobile LLM Inference with RAG}
LLMs have demonstrated remarkable capabilities across numerous tasks, such as question answering (QA)~\cite{yue2025survey}. 
Many researchers explore deploying small language models on mobile platforms~\cite{xu2025fast,yin2024elms,yin2024llm,yuan2024mobile,xue2024powerinfer,llamacpp2023,yi2023mllm}. 
Compared to cloud deployment, mobile deployment effectively reduces data transmission latency and protects personal data.
The development of retrieval-augmented generation (RAG)~\cite{lewis2020retrieval} further enhances LLM capabilities by enabling LLMs to generate high-quality responses using external knowledge~\cite{jiang2025rago}.

To improve RAG efficiency, researchers have explored the reuse of intermediate results across queries to avoid redundant computation. 
Some studies reuse KV cache or attention states of frequently retrieved documents \cite{jin2024ragcache, gim2024prompt, yao2025cacheblend}, while others cache past queries and answers~\cite{bang2023gptcache, couturier2025semantic, mohandoss2024context, li2024scalm, gill2024privacy} or compressed document summaries~\cite{couturier2025semantic}.
However, these approaches narrowly focus on reusing results at individual stages and \modify{overlook} query sparsity in single-user scenarios, leading to suboptimal performance in \textbf{\textit{mobile}} RAG applications.
Sections~\ref{sec:intermediate_result_reusing} and~\ref{sec:query_sparsity} present these issues in detail and clarify the goal of our system.

\subsection{Reusable Results in Mobile RAG}
\label{sec:intermediate_result_reusing}
We identify two types of reusable results in mobile RAG: similar user queries and repeatedly retrieved knowledge. 

\begin{figure}[t]
\centering
\captionsetup{skip=0pt}
\begin{minipage}[t]{0.42\textwidth}
    \centering
    \begin{subfigure}[t]{0.48\columnwidth}
        \centering
        \includegraphics[width=\textwidth]{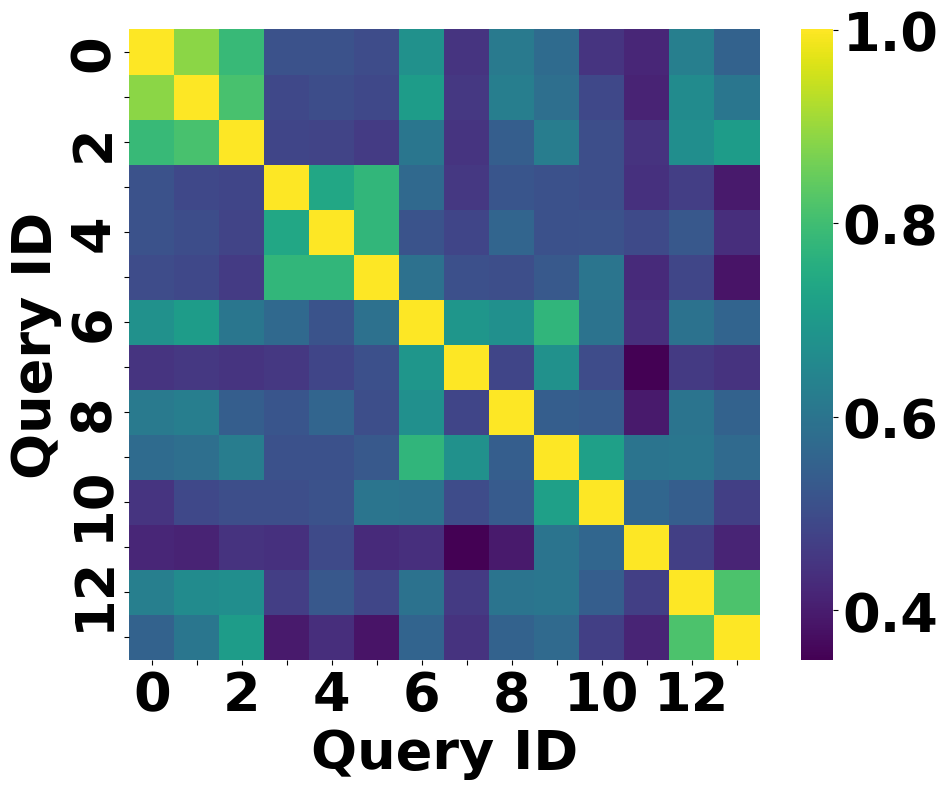}
        \caption{Email Dataset}
        \label{fig:question_similarities_email}
    \end{subfigure}
    \hfill
    \begin{subfigure}[t]{0.48\columnwidth}
        \centering
        \includegraphics[width=\textwidth]{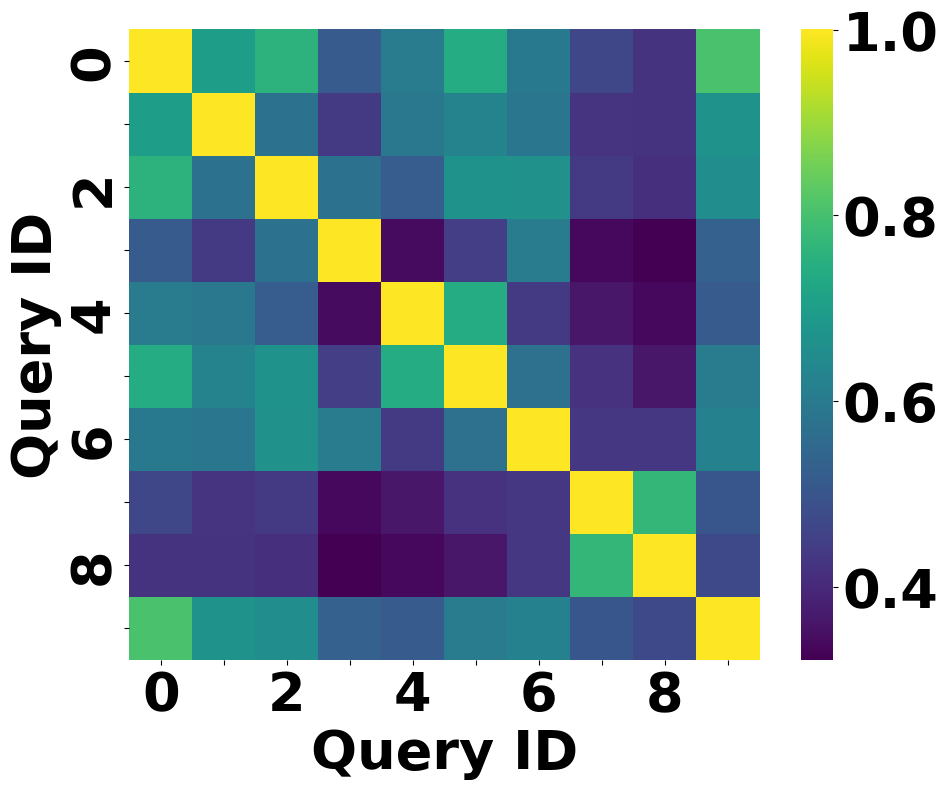}
        \caption{Dialog Dataset}
        \label{fig:question_similarities_dialog}
    \end{subfigure}
    \caption{Pairwise semantic similarity of queries from two distinct users in the Email and Dialog datasets.}
    \label{fig:question_similarities}
\end{minipage}
\hfill
\begin{minipage}[t]{0.57\textwidth}
    \centering
    \begin{subfigure}[t]{0.48\columnwidth}
        \centering
        \includegraphics[width=\textwidth]{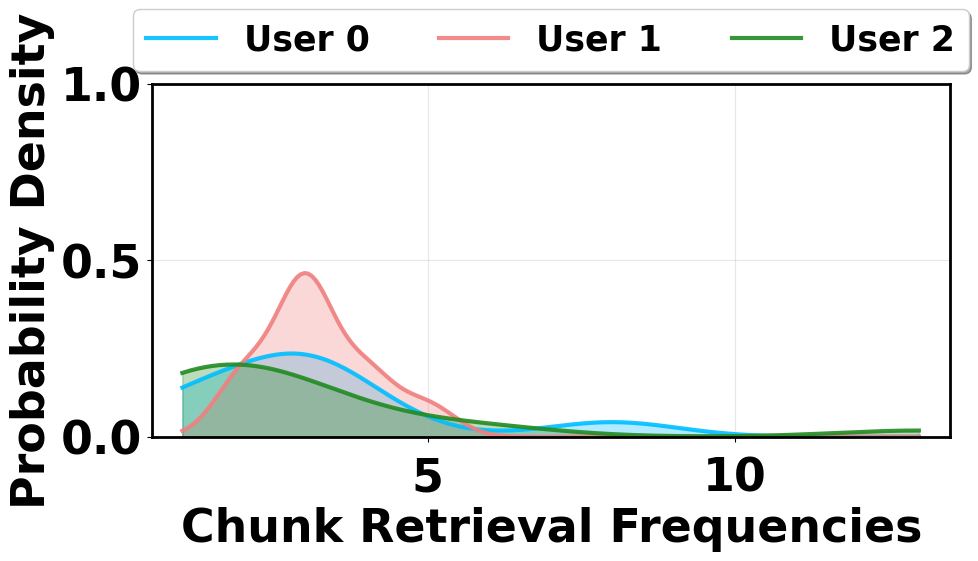}
        \caption{Email Dataset}
        \label{fig:email_retrieval_times}
    \end{subfigure}
    \hfill
    \begin{subfigure}[t]{0.48\columnwidth}
        \centering
        \includegraphics[width=\textwidth]{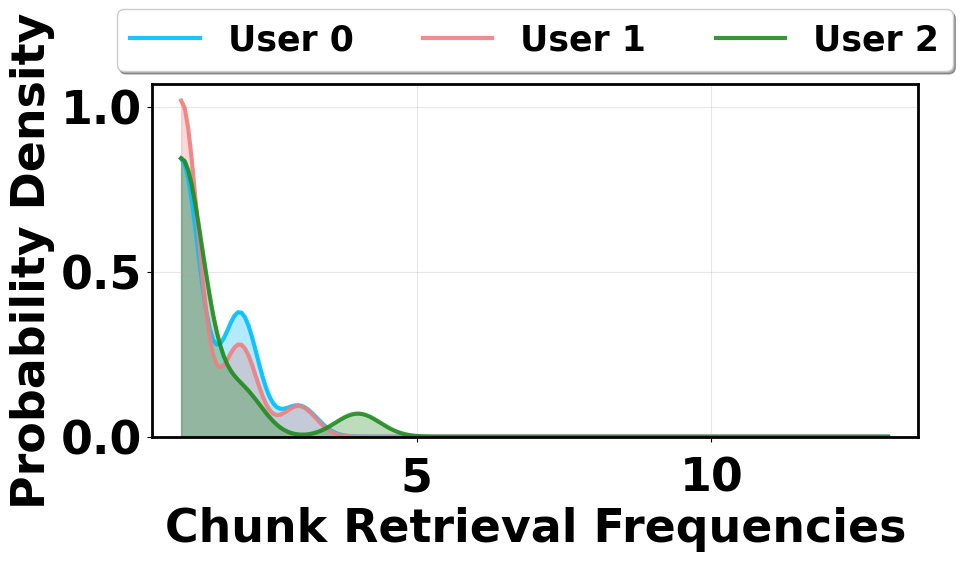}
        \caption{Dialog Dataset}
        \label{fig:dialog_retrieval_times}
    \end{subfigure}
    \caption{Probability density distribution of knowledge chunk retrieval frequencies.}
    \label{fig:retrieval_distribution}
\end{minipage}
\end{figure}

\noindent \textbf{Similar User Queries.} 
Some queries of individual users are highly semantically similar.
We conduct a motivation study using two applications where the LLM serves as a mobile assistant answering questions based on personal daily dialog and email messages, respectively (see Section~\ref{sec:experiment_setup} for dataset details). 
For these two datasets, we recruit separate groups of volunteers: one group uses portable microphones to record their daily dialogues, which are then transcribed into text, while the other group provides their personal emails.
For each dataset, volunteers log their queries and the corresponding answers when accessing their dialogue transcripts or emails.
We analyze two users, each randomly selected from one dataset, measuring the pairwise semantic similarities among each user's queries.
The results in Figure~\ref{fig:question_similarities} show that some queries exhibit high semantic similarities. 
For instance, for the user from the Email dataset, $\text{Query}_{12}$ (\textit{``When will the presentation rehearsal take place?''}) and $\text{Query}_{13}$ (\textit{``Is time of presentation rehearsal given?''}) show a similarity score of 0.815.
This indicates the potential to cache historical queries and skip LLM inference for highly similar future queries.

\noindent \textbf{Repeatedly Retrieved Knowledge.} 
In mobile RAG applications, LLMs answer queries based on personal data, where the same knowledge chunks in the personal database may be retrieved repeatedly by different queries.
Compared to multi-user RAG applications using public knowledge like Wikipedia, personal knowledge bases on mobile platforms are much smaller and more focused in content scope.
Moreover, individual users exhibit consistent language patterns, 
making repeated retrieval more likely.
As shown in Figure~\ref{fig:retrieval_distribution}, we measure the retrieval frequency of knowledge chunks for each user by retrieving the top-2 relevant knowledge chunks per query and recording how often each chunk is accessed.
\modify{
We set the number of retrieved knowledge chunks to \(k=2\) to provide the local LLM with as much relevant knowledge as possible while keeping the prompt within mobile input-length and memory constraints.
}
The results show that many chunks are retrieved multiple times, particularly in \modify{the} Email dataset, where every chunk retrieved by $\text{User}_{1}$ is retrieved more than once.
Such repeated retrieval leads to substantial computational redundancy, as identical intermediate results, such as the KV cache, are repeatedly generated during LLM inference.
A natural idea to mitigate this is to store the intermediate results of these repeatedly retrieved chunks for future reuse.

\begin{figure}[t]
\centering
\captionsetup{skip=0pt}
\resizebox{0.6\columnwidth}{!}{
\includegraphics{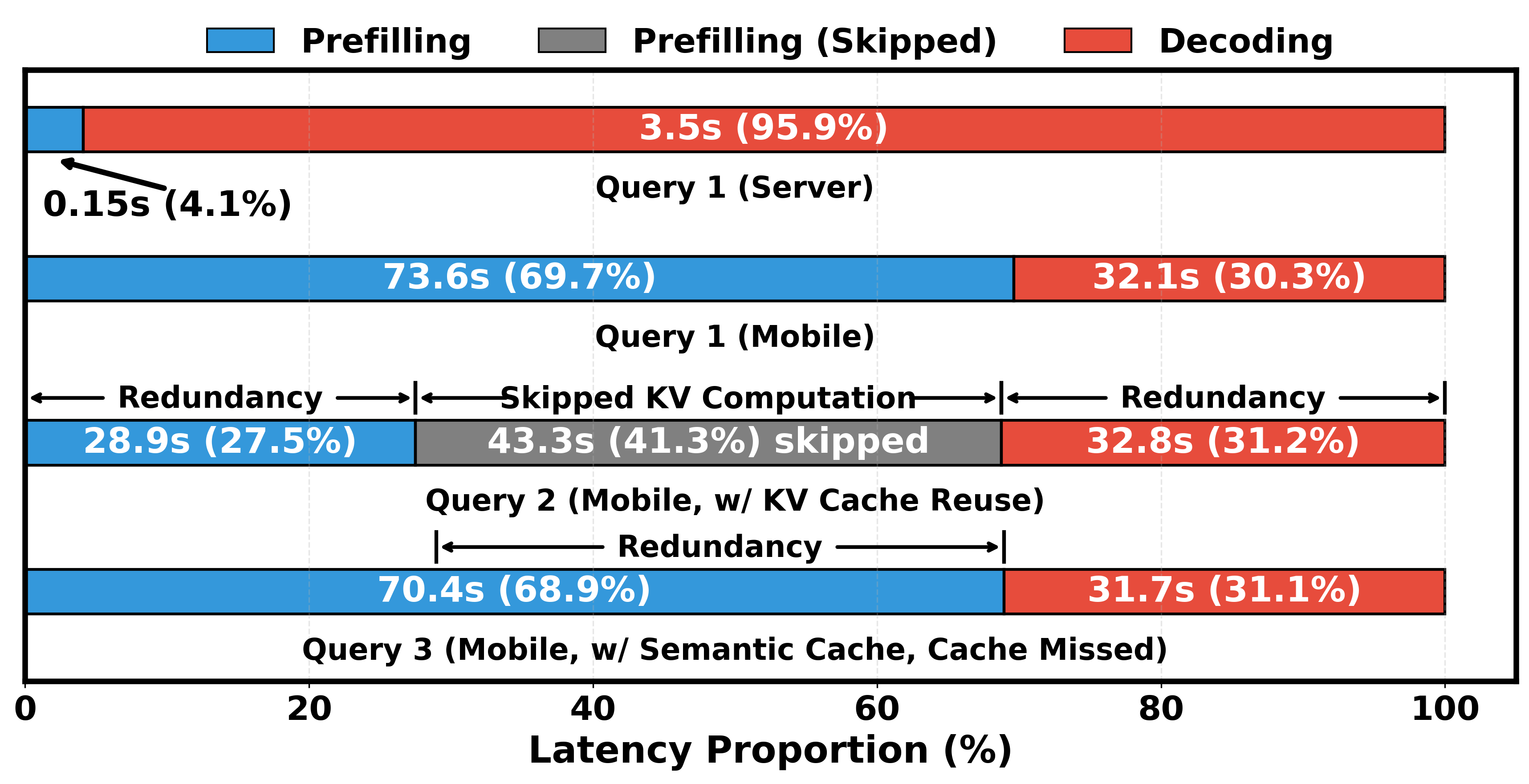}}
\caption{\final{Inference latency breakdown of Llama-3.2-3B on Google Pixel 7 and NVIDIA RTX A6000.}}
\label{fig:inference_time_breakdown}
\end{figure}

Although there exist reusable results at different stages of mobile RAG, \textbf{\textit{existing caching solutions only target individual stages in isolation, failing to fully realize the acceleration potential of each stage.}}
This limitation is demonstrated in Figure~\ref{fig:inference_time_breakdown} through experiments conducted under single-stage caching solutions.
We process three user queries from the Email dataset through the RAG pipeline on a mobile phone (Google Pixel 7) and record the LLM prefilling and decoding latencies for each query. 
Specifically, $\text{Query}_1$ and $\text{Query}_2$ are semantically similar, while $\text{Query}_1$ and $\text{Query}_3$ are dissimilar but retrieve overlapping knowledge chunks.
For $\text{Query}_1$, we additionally record the latencies on a server equipped with an Nvidia RTX A6000 GPU. All experiments use Llama-3.2-3B as the LLM.
The results of $\text{Query}_1$ and $\text{Query}_2$ on the mobile phone indicate the limitation of relying solely on KV cache reuse.
Although these similar queries retrieve identical knowledge chunks and produce highly similar answers, \final{KV reuse reduces prefilling latency by avoiding redundant computation for cached prefixes, but still requires processing uncached tokens and autoregressively generating each answer.}
The results of $\text{Query}_1$ and $\text{Query}_3$ show the limitation of only reusing similar queries. 
Although these two queries retrieve overlapping knowledge chunks, $\text{Query}_3$ fails to hit the cache and undergoes full LLM inference.
The latency caused by repeatedly retrieved chunks remains unreduced and substantial.
These results show that single-stage result reuse is insufficient to reduce the latency caused by computational redundancy.
Comparison between the results of $\text{Query}_1$ on server and mobile device reveals a key difference in latency distribution. 
Unlike server-side inference where decoding dominates the overall latency, mobile inference exhibits significant proportions of both prefilling and decoding latencies. 
This discrepancy primarily stems from the limited parallel computing capabilities of mobile devices.
This unique characteristic of mobile devices necessitates the full exploitation of reusable results from each RAG stage, which is not achieved by existing approaches.

\subsection{Single-User Query Sparsity}
\label{sec:query_sparsity}

\begin{figure}[t]
\centering
\captionsetup{skip=0pt}
\begin{minipage}[t]{0.42\textwidth}
    \centering
    \begin{subfigure}[t]{0.49\columnwidth}
        \centering
        \includegraphics[width=\textwidth]{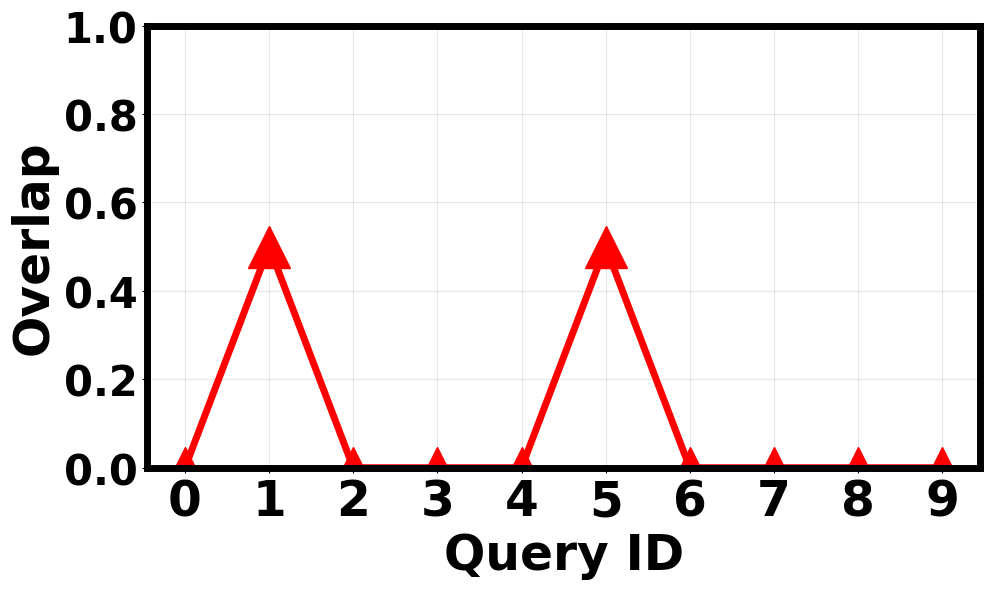}
        \caption{Email Dataset}
        \label{fig:email_overlap}
    \end{subfigure}
    \hfill
    \begin{subfigure}[t]{0.49\columnwidth}
        \centering
        \includegraphics[width=\textwidth]{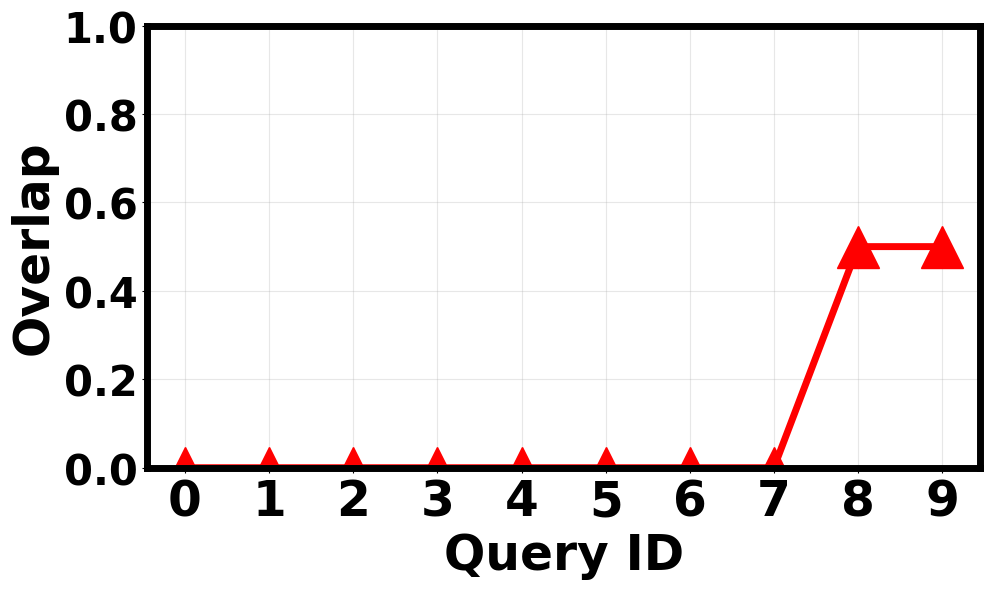}
        \caption{Dialog Dataset}
        \label{fig:overlap_dialog}
    \end{subfigure}
    \caption{Prefix overlap degree of retrieved chunks. \final{Query sparsity results in low prefix overlap ratio for most queries.}}
    \label{fig:overlap}
\end{minipage}
\hfill
\begin{minipage}[t]{0.54\textwidth}
    \centering
    \begin{subfigure}[t]{0.49\columnwidth}
        \centering
        \includegraphics[width=\textwidth]{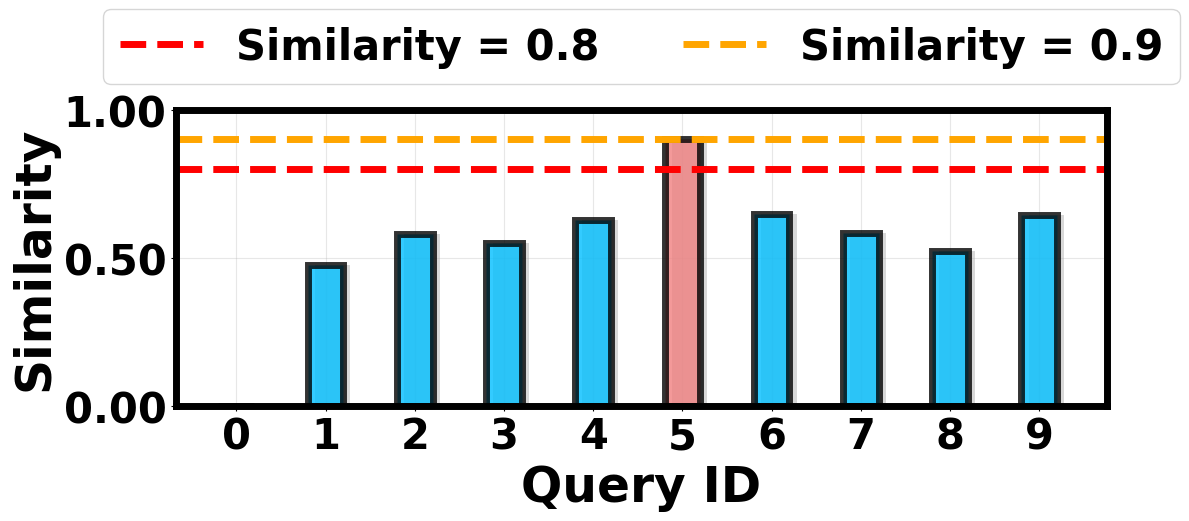}
        \caption{Email Dataset}
        \label{fig:question_similarities_email_in_progress}
    \end{subfigure}
    \hfill
    \begin{subfigure}[t]{0.49\columnwidth}
        \centering
        \includegraphics[width=\textwidth]{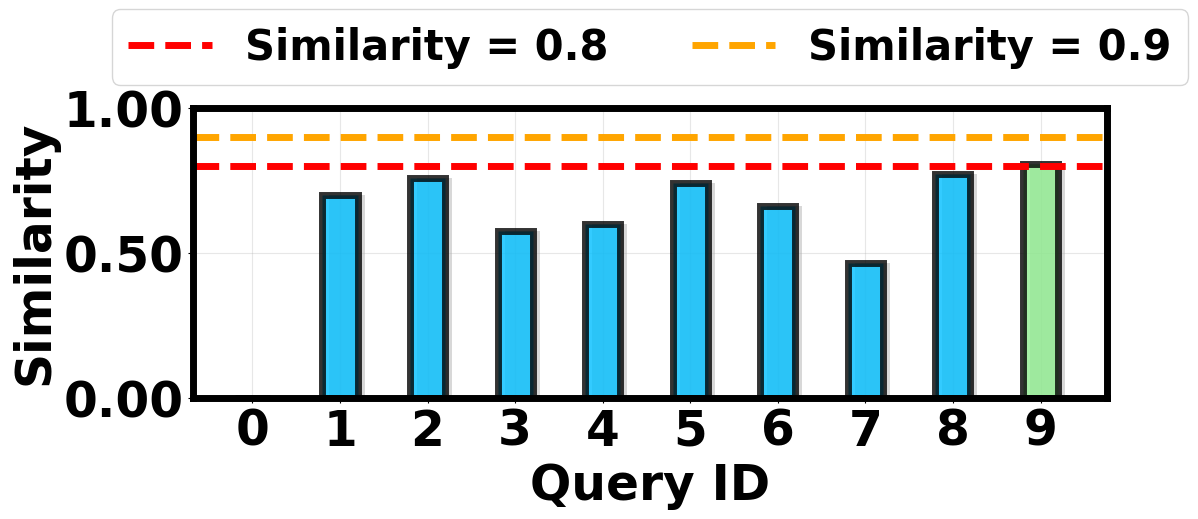}
        \caption{Dialog Dataset}
        \label{fig:question_similarities_dialog_in_progress}
    \end{subfigure}
    \caption{Semantic similarity between each query and its most similar previous query. Query sparsity results in low similarity matches for most queries.}
    \label{fig:question_similarities_in_progress}
\end{minipage}
\vspace{-1em}
\end{figure}

For mobile RAG applications, the knowledge base stores personal data with only a single user accessing it. 
However, individual mobile users typically issue queries at a much lower frequency than multi-user RAG systems deployed on centralized servers, and these queries tend to be semantically varied rather than showing significant repetition (Figure~\ref{fig:question_similarities_in_progress}). \textit{\textbf{This results in an extremely limited number of queries available for cache population, leading to insufficient cache population and consequently low hit rates.}}
For example, a user of a personal meeting assistant might only request meeting summaries every few days.
Existing methods, including KV cache reuse and semantic cache, reactively update the cache based on received queries, populating new cache entries only upon cache misses.
With sparse queries, these approaches struggle to adequately populate the cache in mobile RAG applications, resulting in low cache hit rates.

We illustrate this issue using two cases, corresponding to KV cache reuse and semantic cache, respectively.
We use \modify{data from one user in each of the Email and Dialog datasets}, sequentially processing all queries and reactively updating the cache.
As shown in Figure~\ref{fig:overlap}, for KV cache reuse (e.g., RAGCache~\cite{jin2024ragcache}), 
we retrieve the top-2 relevant knowledge chunks for each query and check their hit rate against cached KV tensors using prefix overlap ratio.
The results show that the prefix overlap ratios are quite low for most queries, with some dropping to zero.
Similarly, for the semantic cache, we store semantic embeddings of previous queries instead of KV tensors. 
Figure~\ref{fig:question_similarities_in_progress} shows that few queries exhibit high similarity to previously cached queries. 
For the user from the Email dataset, only $\text{Query}_5$ \modify{matches a previously cached query with a similarity above 0.9}, while other queries remain below 0.8.
\modify{
Here, we use 0.8 and 0.9 for analysis because 0.8 serves as a practical lower bound below which matched queries often differ substantially in answer content, whereas 0.9 provides a more conservative setting that retains meaningful reuse opportunities.
}
These results indicate that the sparsity of single-user queries leads to low cache hit rates with the reactive cache update strategy used by most existing solutions. 
Although Section~\ref{sec:intermediate_result_reusing} demonstrates that single-user queries show certain similarities and knowledge chunk retrieval overlap, query sparsity prevents converting these opportunities into significant efficiency gains. 
Therefore, improving cache hit rates under sparse query frequencies is essential to unlock the efficiency \modify{benefits of reusing results} in mobile RAG applications.

\section{SYSTEM OVERVIEW}
\begin{figure}[t]
\centering
\captionsetup{skip=0pt}
\resizebox{0.6\columnwidth}{!}{
\includegraphics{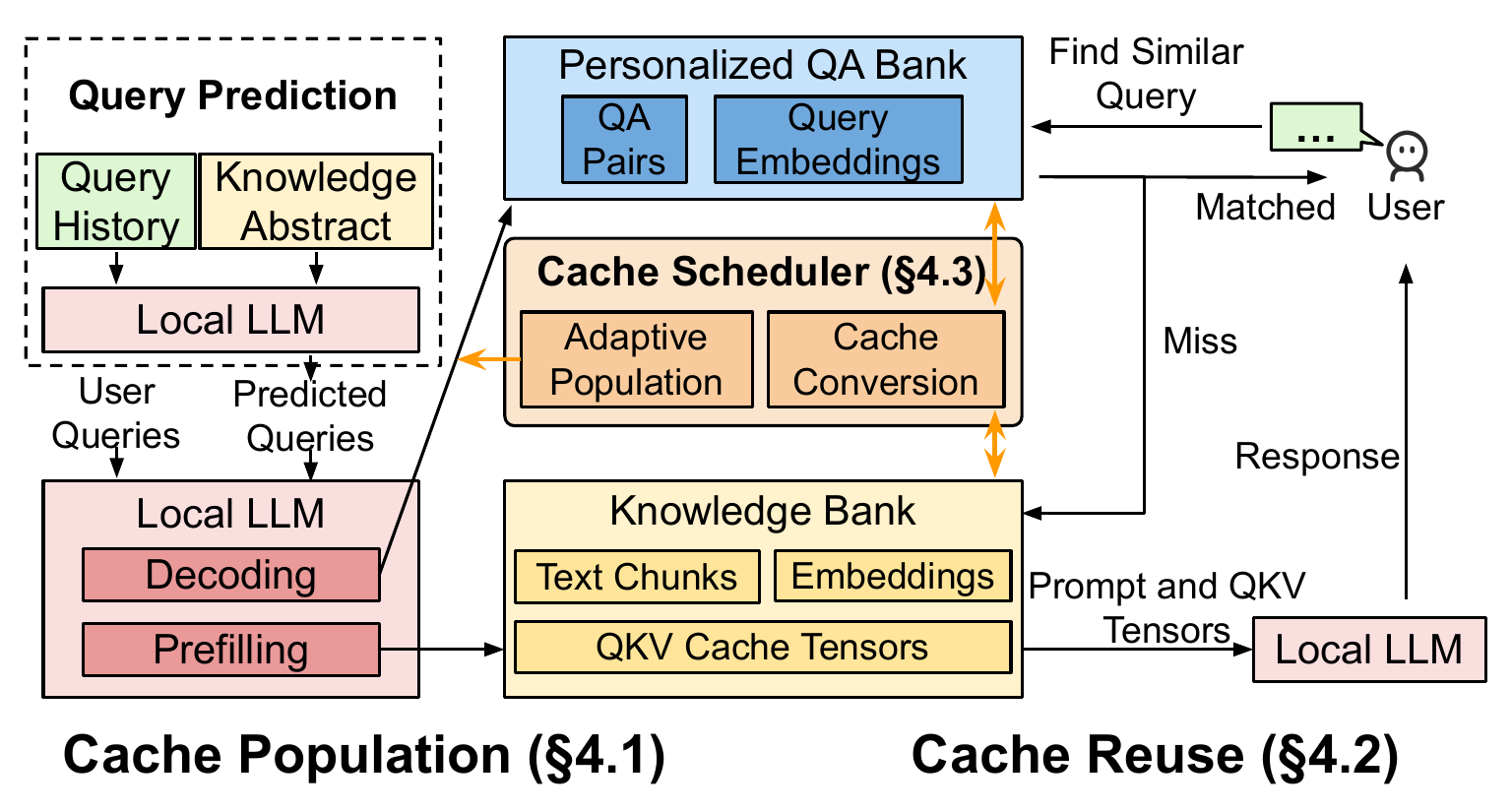}}
\caption{System overview of \workname.}
\label{fig:system_overview}
\end{figure}

Figure~\ref{fig:system_overview} \modify{provides an overview of} the design of our predictive hierarchical cache system, \workname. 
The key design principle is to \modify{precompute} the partial computational load of the mobile RAG workflow in advance to reduce real-time response latency.
\final{\workname~comprises a QA bank, a knowledge bank with QKV cache tensors, a cache scheduler, and a query prediction module.}

The left portion shows the cache population process.
There are two layers in \workname: QA bank and \modify{QKV cache}. 
QA bank stores query-answer pairs and the semantic embeddings of queries, while knowledge bank stores the raw text and embeddings of knowledge chunks and the cached QKV tensors of the chunks retrieved by previous queries. 
QA bank is designed to match similar queries to directly return cached answers, while \modify{QKV cache} is used to skip partial computation during LLM prefilling.
To improve hit rate, \workname~adopts an LLM-based query prediction mechanism that predicts future queries to populate cache during device idle time \modify{from two complementary perspectives}: historical queries and knowledge content.
\workname~also includes a cache scheduler to control cache population strategy and conversion between QA bank and QKV tensors, flexibly adapting to dynamically available computational and storage resources.

The right portion illustrates the cache reuse process. 
Upon receiving a user query, \workname~computes its embedding and compares it with the stored queries in the QA bank. 
If the highest similarity exceeds the predefined threshold, the cached answer is directly returned, bypassing LLM inference.
Otherwise, \workname~ retrieves relevant text chunks from the knowledge bank for this query, and checks whether the QKV tensors of each retrieved knowledge chunk are precomputed and stored in the knowledge bank. 
These retrieved knowledge chunks are then concatenated with the user query to form the LLM prompt. 
During LLM prefilling, the computation of the available QKV tensors is skipped. 
Finally, the LLM output is returned to the user.

\section{SYSTEM DESIGN}
This section details each component of \workname. Section~\ref{sec:cache_construction} describes the hierarchical cache architecture and population process. Section~\ref{sec:cache_reuse} explains the details of cache reuse. 
Section~\ref{sec:cache_scheduler} presents the design of the cache scheduler.

\subsection{Cache Population}
\label{sec:cache_construction}

\begin{figure}[t]
\centering
\captionsetup{skip=0pt}
\begin{minipage}[t]{0.56\textwidth}
    \centering
    \resizebox{\columnwidth}{!}{
    \includegraphics{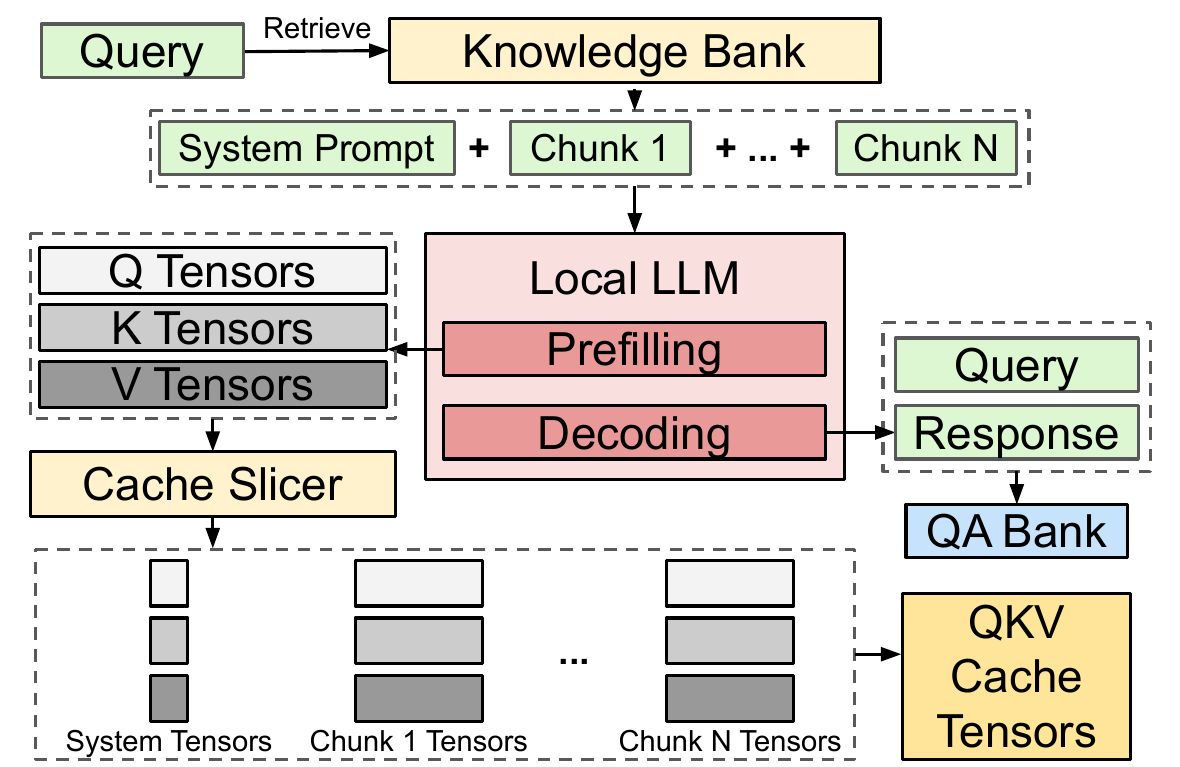}}
    \caption{Overview of cache population.}
    \label{fig:cache_population_details}
\end{minipage}
\hfill
\begin{minipage}[t]{0.40\textwidth}
    \centering
    \resizebox{\columnwidth}{!}{
    \includegraphics{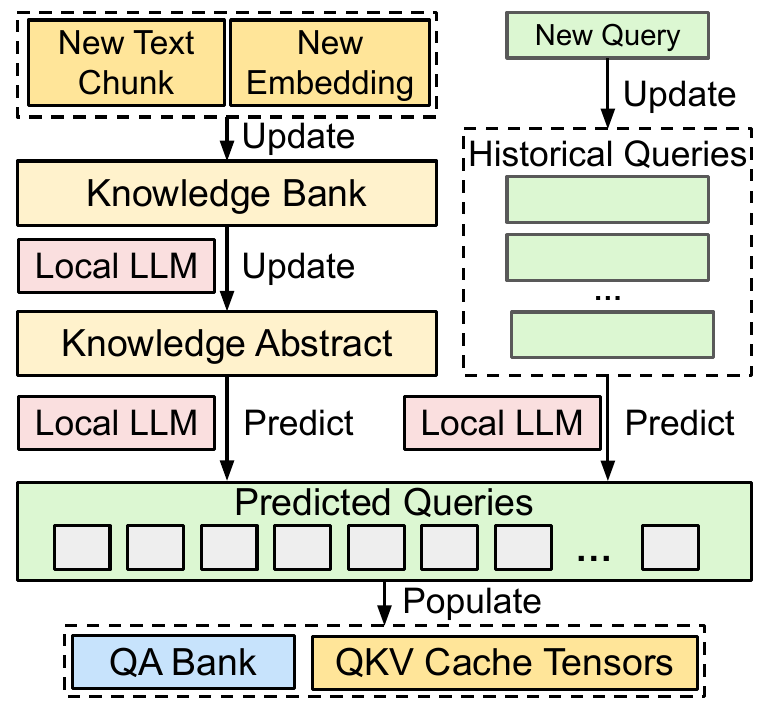}}
    \caption{Query prediction with knowledge-based (left) and history-based (right) methods.}
    \label{fig:query_prediction}
\end{minipage}
\end{figure}

\modify{Reusing results} shows great potential to improve the efficiency of mobile RAG workflows. 
However, prior methods only consider reusing results at a single stage, which becomes particularly problematic on mobile platforms.
They also \modify{overlook query sparsity} in single-user mobile scenarios that causes a low cache hit rate.
For the first issue, we propose a hierarchical cache architecture that caches both query-answer pairs and QKV tensors of the retrieved knowledge chunks to fully exploit reusable results. 
For the second issue, we \modify{use the local LLM to predict future queries and populate the hierarchical cache in advance.}

\subsubsection{Cache Architecture}
\label{sec:cache_architecture}

As shown in Figure~\ref{fig:system_overview}, \modify{\workname stores the knowledge bank and QA bank in separate local-storage directories.}
The knowledge bank stores the user's personal data \modify{segmented into text chunks of a predefined length}, along with the semantic embeddings of these chunks. 
We call each text chunk a \textit{knowledge chunk}.
The knowledge bank also stores the QKV tensors for the previously retrieved chunks. 
\final{In our current prototype, we retain Q tensors alongside K and V, keeping open the option of reconstructing full-sequence QKV and directly reusing the engine's existing execution flow, so that its built-in performance optimizations can be preserved with minimal engineering effort. 
This also leaves room for future optimizations such as offline attention-based KV compression~\cite{zhang2023h2o,li2024snapkv}.}
When these chunks are retrieved again by future queries, the tensors can be directly loaded to skip computation of the cached prefix during prefilling.
The QA bank stores previously processed queries.
It packs each query with its embedding, \modify{answer-availability flag, timestamp} and the answer generated by the LLM as a cache entry.

As shown in Figure~\ref{fig:cache_population_details}, the saving of a new QKV cache is triggered by an input query from either the user or the query prediction module (\S~\ref{sec:speculative_cache_prediction}).
After retrieving relevant knowledge chunks, we concatenate the system prompt, knowledge chunks, and query as input to the LLM.
During LLM prefilling, we collect and save the QKV tensors from the attention module of each transformer block. 
To efficiently utilize the storage space, we adopt the tree structure proposed in~\cite{jin2024ragcache} to organize the cached QKV tensors of knowledge chunks, where each node represents a specific chunk and each path represents a chunk list retrieved by a specific query.
To use the tree structure, we need to save the QKV tensors of each individual chunk in a separate file.
To achieve this goal, we implement a cache slicer to split the QKV tensors of the whole prompt into multiple tensor slices. 
Specifically, the slicer first obtains each chunk's sequence length using the LLM tokenizer, and then calculates start and end positions of \modify{each chunk} in the QKV tensors.
After that, the slicer splits the QKV tensors into tensor slices on the sequence dimension, each of which corresponds to a single chunk. 
Each tensor slice is treated as a tree node and merged into the QKV cache tree.
We regard the Q, K, V tensor slices of the same chunk as a whole and save them in a single file.
\modify{
Each file is named using the node's tree-structured identifier (e.g., \texttt{42}, \texttt{42\_1}, and \texttt{42\_1\_2}), which encodes its parent--child path in the tree and enables efficient prefix lookup and cache-chain reconstruction.
}
In addition to QKV tensors, the QA bank is updated after LLM decoding. Once the LLM output is obtained, the input query, its embedding, and corresponding output are packed into a new cache entry and appended to the QA bank. 

\modify{Since cache matching and loading latency is typically negligible (Table~\ref{tab:system_overhead}), \workname~loads caches on-demand to minimize memory consumption. }
\modify{
The knowledge bank and QA bank are loaded into memory only during retrieval and released before LLM inference.
QKV matching operates on lightweight cache-file identifiers, and only the matched tensors are loaded into the inference engine. 
}
The local storage limit for cache is configurable. 
\workname~allocates a small portion of the total storage (e.g., 50-100 MB) to the QA bank, which is sufficient given the small size of individual entries.
\workname~maintains a \modify{retrieval counter} for each cache entry and employs the Least Frequently Used (LFU) eviction algorithm to ensure the cache storage does not exceed the configured limit.

\subsubsection{Query Prediction}
\label{sec:speculative_cache_prediction}
The efficiency gains achieved by the cache system are highly dependent on the cache hit rate. 
However, existing solutions show suboptimal performance as they only update the cache reactively based on user queries.
Unlike server-side scenarios, user queries in single-user mobile scenarios are sparse, making it difficult to adequately populate cache,
resulting in low cache hit rates.

As shown in Figure~\ref{fig:query_prediction}, to address this issue, \workname~predicts the future queries during mobile device idle periods. 
Idle periods refer to times when users do not interact with their phones for extended durations, such as during overnight or work-hours charging.
\workname~utilizes the on-device LLM to infer potential future user queries during these idle periods.
The predicted queries are then processed sequentially to update the hierarchical cache using the method described in Section~\ref{sec:cache_architecture}.
The number of queries predicted each time is defined as prediction stride, which is a configurable parameter.
To comprehensively capture user query patterns, we perform query prediction from two complementary perspectives: knowledge content and query history. 
For knowledge-based prediction, unlike multi-user RAG systems using diverse public knowledge, mobile RAG applications store domain-specific personal data. 
For example, email assistant applications rely solely on users' personal emails as external knowledge. 
Therefore, mobile user queries are usually highly relevant to knowledge content.
Through this perspective, the LLM analyzes key contents within the knowledge bank and infers likely future queries around them.
For history-based prediction, individual users often show specific linguistic habits and content focus tendencies. 
This perspective enables the LLM to capture users' questioning patterns and recent shifts in content interests.
These two approaches are performed asynchronously.

Specifically, for knowledge-based query prediction, we propose to use knowledge abstracts rather than the raw knowledge chunks for prediction.
Raw knowledge chunks are typically lengthy and detailed, making query prediction computationally expensive. 
In contrast, the knowledge abstract is concise and highly summarized, enabling \workname~to predict queries at a lower computational cost while maintaining a broader view spanning more knowledge chunks. 
The abstract is a collection of key content from all knowledge chunks summarized by the LLM. 
For example, in a meeting memo application, the knowledge bank stores raw audio transcripts of meetings, while the knowledge abstract contains key nouns, important topics, and main participant names.
When new knowledge chunks are added to knowledge bank, \workname~notifies the LLM to extract their key content and \modify{append the extracted key content to the existing abstract collection.}
To avoid frequent model loading and inference, \workname~\modify{batch-processes multiple chunks to generate abstracts} when no new chunk arrives for a period, rather than on every incoming chunk.
During mobile device idle time, \workname~uses the LLM to infer potential future queries related to the abstract.
For history-based query prediction, \workname~maintains a buffer storing recent user queries. 
If \workname~receives new queries from the user and no additional queries are received within a specified time period thereafter, the history-based prediction is triggered.
\workname~first updates the query buffer using these received queries
and then instructs the LLM to predict future queries, mimicking the user's expression style and topical interests.
Unlike knowledge-based prediction, history-based prediction focuses on user query characteristics.
By combining knowledge-based and history-based prediction, \workname~predicts queries more comprehensively from two complementary views.
Since new chunks and queries arrive sparsely and \workname~uses batch processing to \modify{predict queries}, the query prediction module incurs minimal overhead.

\subsubsection{Dynamic Cache Refresh}
\label{sec:cache_refresh}
\modify{
When new knowledge chunks are added to the local knowledge bank, \workname checks whether they rank among the top-\(k\) retrieved chunks for any query in the QA Bank. 
If so, the corresponding QA entry is marked as outdated and refreshed during an idle period.
If a query matches an outdated entry before it is refreshed, \workname uses normal inference instead of returning the cached answer.
}

\subsection{Cache Reuse}
\label{sec:cache_reuse}
In this section, we present two mechanisms in \workname~for reducing LLM response latency: similar query reuse and QKV cache reuse.

\subsubsection{Similar Query Reuse}
\label{sec:similar_query_reuse}
Upon receiving a user query, \workname~attempts to match it with similar queries in the QA bank.
By reusing cached responses from the QA bank, \workname~can skip both prefilling and decoding, enabling near-instantaneous responses for highly similar queries.
Specifically, \workname~employs an on-device text embedding model to convert the query into an embedding and computes the cosine similarity with all previously stored queries.
If the highest similarity exceeds the similarity threshold $\tau_{\mathrm{QA}}$, the response in the corresponding QA pair is directly returned to the user, while the actual response to the new query will be generated during the device’s idle time and stored in the QA bank for future reuse.
The similarity threshold $\tau_{\mathrm{QA}}$ can be adjusted to trade off between latency and response quality. 
A lower threshold increases the QA bank hit rate, allowing more queries to bypass LLM inference. However, setting the threshold too low risks matching queries with low semantic similarity, which may degrade response quality.

\subsubsection{QKV Cache Reuse}
\label{sec:qkv_cache_reuse}
If the user query fails to match any query in the QA bank, \workname~\modify{searches for available QKV tensors.} 
Specifically, it first retrieves relevant chunks from the knowledge bank using the hybrid strategy proposed in~\cite{eversberg2024hybrid}, which combines the BM25 algorithm~\cite{robertson2009probabilistic} with text embeddings.
\modify{After retrieving chunks ranked by relevance score}, \workname~starts to match the available precomputed QKV tensors of these chunks.  
\workname~examines whether there exist QKV tensors whose text matches the first knowledge chunk.
\modify{
This check is implemented by matching the tree-structured filenames of cached QKV files: a child filename extends its parent's identifier (e.g., \texttt{42}, \texttt{42\_1}, and \texttt{42\_1\_2}), allowing PerCache to verify the prefix chain without loading tensor contents.
}
If such tensors are found, \workname~continues to see if there exist such tensors for the second chunk at the position of an immediate child node in the knowledge cache tree. 
This process continues sequentially until a mismatch is encountered.

After finding available QKV tensors, \workname~starts the LLM inference. 
It first loads the matched QKV tensors and then concatenates them in order. 
It also records the total sequence length of the concatenated tensors to calibrate position \modify{embeddings}.
During prefilling, for the input tensor of the attention module within each transformer block, \workname~discards the part corresponding to the loaded QKV tensors and only retains the remaining suffix. 
Q, K, V projection and position embedding computations are performed only on this retained part, significantly reducing computational overhead.
Then, the obtained QKV tensors are concatenated with previously loaded QKV tensors, and \modify{the remainder of the attention module} proceeds normally
\modify{(further details are provided in Appendix~\ref{sec:inference_with_qkv_cache} and~\ref{sec:tokenization_inconsistency})}.
QKV cache reuse enables efficiency improvements during LLM prefilling when no similar queries are matched in the QA bank.

\subsection{Cache Scheduler}
\label{sec:cache_scheduler}

\begin{figure}[t]
\centering
\captionsetup{skip=0pt}
\resizebox{0.6\columnwidth}{!}{
\includegraphics{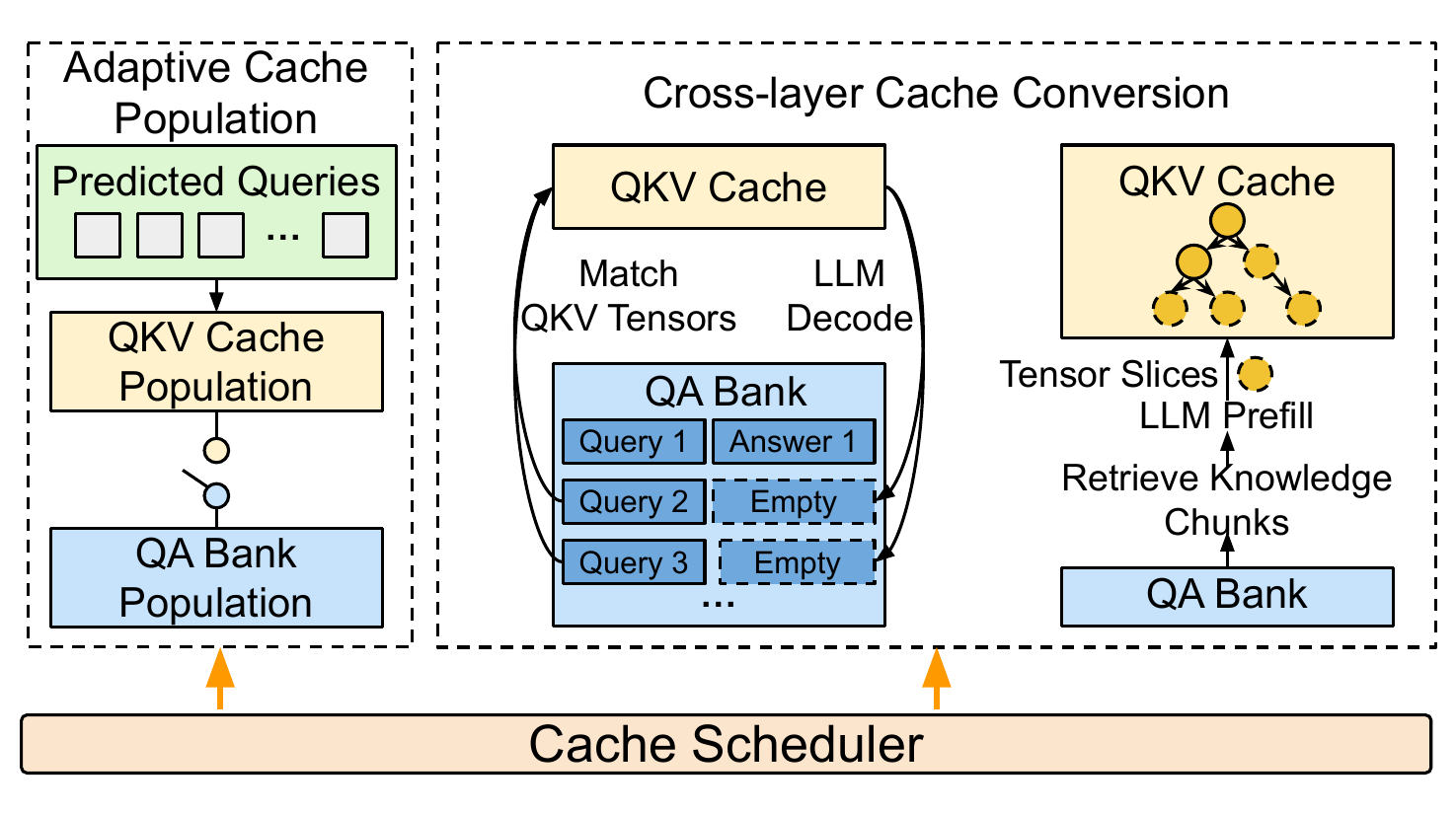}}
\caption{Cache Scheduler.}
\label{fig:cache_scheduler}
\end{figure}

Compared to servers, mobile devices face severe constraints in battery life, computing power, and storage capacity, and their available system resources frequently fluctuate.
Thus, a mechanism is needed to optimize resource consumption while maintaining performance.
To this end, we design a cache scheduler that enables elastic usage of computational and storage resources.

\subsubsection{Trade-off Analysis}
\label{sec:latency_computation_trade_off_analysis}
We identify two primary trade-offs in \workname~as follows.

\noindent \textbf{Latency-Computation Trade-off.} 
As described in Section~\ref{sec:cache_construction}, QKV cache and LLM outputs are stored in the knowledge bank and QA bank, respectively.
For a single query, generating a QA bank entry imposes a higher computational load than only producing a QKV cache entry.
This is because QA bank entries require undergoing both prefilling and decoding for final LLM output, while QKV cache entries only need prefilling.
Higher computational load leads to \modify{greater battery consumption on mobile devices}. 
For this issue, we consider two cache population strategies: 
(1) populate only the QKV cache layer by performing prefilling alone;
(2) populate both QKV cache and QA bank layers by executing both stages.
The first strategy reduces computational overhead but increases latency as fewer cache entries are populated in the QA bank. 
The second strategy lowers latency by populating more QA bank entries but increases computational overhead. 
Thus, a trade-off exists between cache population, computational cost and average response latency.

\noindent \textbf{Latency-Storage Trade-off.} 
\workname~stores both the QA bank and knowledge bank in the local storage of the mobile device. 
For the QA bank, storing more QA pairs increases the likelihood that a new query will match previous ones with semantic similarity exceeding the threshold.
For the knowledge bank, storing more QKV tensors increases the likelihood of finding precomputed tensors for retrieved knowledge chunks. 
Higher hit rates in the two layers respectively increase queries that can skip entire inference and the proportion of computation in the attention module, thereby reducing average latency.
Thus, caching more entries enables faster responses but consumes more storage.

\subsubsection{Adaptive Cache Population}
\label{sec:adaptive_computing_power_allocation}
Considering the trade-offs described above, the cache scheduler dynamically adjusts the cache population strategy to avoid computational overhead that does not yield efficiency gains.
As shown in Figure~\ref{fig:cache_scheduler}, the scheduler switches between the strategies described in Section~\ref{sec:latency_computation_trade_off_analysis}.
It adjusts the population strategy based on the similarity threshold rather than historical hit rates, as query sparsity causes \modify{little historical hit-rate information} with large inter-query gaps that \modify{is weakly correlated} with future hit rates.
When the similarity threshold is high, few queries match previous ones, and performing decoding would waste computational resources. 
In this case, the scheduler selects the first strategy, where \workname~performs only prefilling for predicted queries and populates only the knowledge bank with QKV tensors. 
These queries are also stored in the QA bank, but without responses.
\final{When \workname matches similar queries in the QA bank, these queries without response are filtered out.}
Conversely, when the similarity threshold is low, more queries are likely to match previous ones, and decoding becomes beneficial. 
In this case, the scheduler selects the second strategy, where \workname~performs both prefilling and decoding to populate both banks.
We define a cutoff similarity threshold $\tau_\mathrm{scheduler}$ to guide the scheduler’s decision.
By adjusting cache population strategy, \workname~avoids unnecessary computation while maintaining end-to-end latency.

\subsubsection{Cross-layer Cache Conversion}
To enable elastic utilization of computational and storage resources, we implement a mechanism supporting mutual conversion between the two cache layers, as shown in Figure~\ref{fig:cache_scheduler}.
For the conversion from QKV cache to QA bank, the scheduler checks the QA bank and identifies queries lacking corresponding responses that have not undergone decoding.
It performs decoding for them and adds the newly generated responses to the QA bank.
\final{This process allows \workname~to populate the QA bank with missing responses for previously processed queries, reusing available QKV tensors to reduce prefilling overhead before decoding.}
Such conversion is typically triggered when the similarity threshold $\tau_{\mathrm{QA}}$
becomes low.
For the conversion from QA bank to QKV cache, the scheduler checks if QKV tensors of each QA bank query have been deleted by the cache eviction algorithm and if sufficient storage is available.
If so, it performs prefilling for these queries and restores their QKV tensors in the knowledge bank.
This conversion allows \workname~to maximize available storage space utilization under any storage constraints to reduce the prefilling stage latency of future queries.
Once cache conversion is turned on, it executes automatically \modify{during idle periods}.

\section{EVALUATION}
\subsection{Implementation}
We implement a system prototype using C++ and Python based on mllm~\cite{yi2023mllm}, a mobile LLM inference engine. 
We use Ollama~\cite{ollama2024} to deploy text embedding models and use the BM25~\cite{robertson2009probabilistic} algorithm through the rank\_bm25~\cite{brown2022rank_bm25} Python package. The system runs locally within Termux~\cite{termux2025}.
We evaluate \workname~using two LLMs: Llama-3.2-3B~\cite{meta2024llama32} and Qwen-1.5-1.8B~\cite{qwen1.5}.
We obtain text embeddings using Qwen3-Embedding-0.6B~\cite{zhang2025qwen3}. The mobile devices used in our experiments include Google Pixel 7~\cite{google_pixel7_2024}, Redmi K60 Pro~\cite{gsmarena_redmi_k60_pro}, Samsung Galaxy S22 Ultra~\cite{gsmarena_galaxy_s22_ultra} and OnePlus Ace 6~\cite{gsmarena2025oneplus}.
\modify{
Unless otherwise specified, all on-device experiments are conducted on Google Pixel 7 and the total cache capacity is 12\,GB, with 50\,MB for the QA Bank and the rest for the QKV cache.
The absolute latency values vary with the underlying inference engine implementation and hardware.
Therefore, our evaluation focuses on the relative improvement over baselines rather than the absolute latency.
}
\modify{
\workname targets non-real-time, privacy-sensitive personal-information seeking rather than strictly real-time applications.
}

\subsection{Experiment Setup}
\label{sec:experiment_setup}
\noindent \textbf{Datasets.}  
\modify{We evaluate \workname~on five datasets, covering 608 evaluation units and 59,367 queries in total: 432 meeting dialogs from MISeD, 134 user inboxes from EnronQA, 32 users from PerLTQA, and 10 volunteers from our two self-collected datasets.}
\begin{itemize}[leftmargin=*]
    \item \modify{\textit{MISeD~\cite{golany2024efficient}}. 
    It contains meeting transcripts with corresponding query-answer pairs from three domains: product-design meetings, academic meetings, and public parliamentary committee meetings. 
    We include all 432 meeting-grounded dialog workloads, each comprising a meeting transcript and a set of information-seeking query--response turns about that meeting.}
    \item \modify{\textit{EnronQA~\cite{ryan2025enronqa}}. It consists of email collections organized by user inboxes, each with single-hop query–answer pairs grounded in specific emails. We treat each inbox as one local knowledge base. We exclude 16 users with excessively large data volumes and include all the remaining 134 users.}
    \item \modify{\textit{PerLTQA~\cite{du2024perltqa}}. 
    It contains the personalized long-term memory of multiple users, including semantic memories (world knowledge, profiles, and social relationships) and episodic memories (events and dialogues), together with matching query-answer pairs.
    These memories cover both stable user attributes and past personal interactions.
    We use data of all the 32 users.}
    \item \modify{\textit{Email (Self-collected).\footnote{\label{fn:irb}The data collection and study have been approved by the authors' IRB.}}
    We recruit 6 university students.
   Over three days, whenever they open their inbox and have a real query, they write down the query and its answer.
   We collect their emails along with these query-answer pairs.
   Query-relevant information is scattered across different emails.
    }
    \item \modify{\textit{Daily Dialog Transcript (Self-collected).\footref{fn:irb}}
    We recruit 4 university students to record their daily conversations for one week with portable microphones.
    Whenever they have a real query, they look back at the transcripts and write down the query and answer. 
    Each volunteer's knowledge base is a set of transcript segments of varying length, which are informal and fragmented, shift topics often, and contain speech recognition errors.
    } 
\end{itemize}

\noindent \textbf{Baselines.} We compare \workname~with the following approaches.

\begin{itemize}[leftmargin=*]
    \item \textit{Naive}. The vanilla baseline. Every user query is processed with no caching.
    \item \textit{RAGCache (RC)~\cite{jin2024ragcache}}. This represents a \textbf{KV cache reuse}~\cite{jin2024ragcache, gim2024prompt, yao2025cacheblend} approach that leverages cached KV tensors during prefilling.
    It requires \modify{the order of the reused KV tensors to match that of the retrieved knowledge chunks}. RAGCache reactively updates cached KV tensors based on user queries \modify{and manages them using a prefix-aware Greedy-Dual-Size-Frequency (PGDSF) replacement policy.}
    \item \textit{MeanCache (MC)~\cite{gill2024privacy}}. This represents a \textbf{semantic cache reuse}~\cite{bang2023gptcache, gill2024privacy, li2024scalm} approach that stores historical queries and answers.
    MeanCache also reactively updates the cache based on received queries.
    \item \textit{Sleep-time Compute (SC)~\cite{lin2025sleep}}. This represents a \textbf{query prediction} approach that infers potential user queries. 
    For this baseline, we select and modify the approach from Sleep-time Compute~\cite{lin2025sleep}. 
    The original method predicts queries based on raw context and performs complex reasoning in advance to reduce test-time latency.
    We adjust it to predict queries based on the local knowledge content and generate answers rather than reasoning processes. 
    For fair comparison, we use the same prompts for query prediction as \workname.
    \item \textit{RAGCache  + MeanCache (RC+MC)}. We create a hierarchical cache baseline manually by combining RAGCache and MeanCache. We employ RAGCache to update KV cache tensor tree and MeanCache to update QA Bank.
    \item \textit{RAGCache + Sleep-time Compute (RC+SC)}. We also create a baseline by combining RAGCache and Sleep-time Compute. We use RAGCache to update the KV cache tensor tree and Sleep-time Compute to update the QA bank.
    \item \modify{\textit{RAGCache + Sleep-time Compute + MeanCache + PGDSF (RC+SC+MC+PGDSF)}}. 
    \modify{This baseline combines RAGCache for KV-cache reuse, Sleep-time Compute for query prediction, and MeanCache for QA-cache reuse. It uses RAGCache’s PGDSF policy for KV cache scheduling and eviction under the same storage budget.}
    \item \modify{\textit{RAGCache + Sleep-time Compute + MeanCache + MobiLoRA~\cite{li2025mobilora}-KV-Mgmt (RC+SC+MC+MobiLoRA)}.} 
    \modify{This baseline uses the same cache-reuse and prediction components as RC+SC+MC+PGDSF, but replaces PGDSF with a MobiLoRA-inspired utility-based eviction policy. It assigns each KV chunk a weighted score based on its query-context state, recency of access, and reusable-context length, and evicts the chunk with the lowest score when the cache exceeds its storage budget.}
    \item \modify{\textit{RAGCache + Prompt Reordering~\cite{chung2026agent} + MeanCache + PGDSF (RC+Reorder+MC+PGDSF)}}. 
    \modify{This baseline retains RAGCache, MeanCache, and PGDSF, but replaces query prediction with prompt reordering. It sorts retrieved chunks by their original indices to improve prefix continuity and promote KV-cache reuse.}
    \item \modify{\textit{MobileRAG~\cite{park2025mobilerag}}.}
    \modify{This baseline represents non-cache-based mobile RAG acceleration. It uses efficient vector retrieval and selective context reduction to reduce retrieval and LLM inference costs without KV or QA-cache reuse.}
\end{itemize}

\modify{
RAGCache, MeanCache, Sleep-time Compute, and their combinations are general-purpose LLM/RAG caching methods, which we adapt to our setting. 
MobileRAG, MobiLoRA, and Agent-X are mobile-oriented LLM acceleration frameworks.
}

\subsection{End-to-end Showcases}

\begin{figure}[t]
\centering
\captionsetup{skip=0pt}
\begin{subfigure}[t]{\textwidth}
\centering
\includegraphics[width=\linewidth]{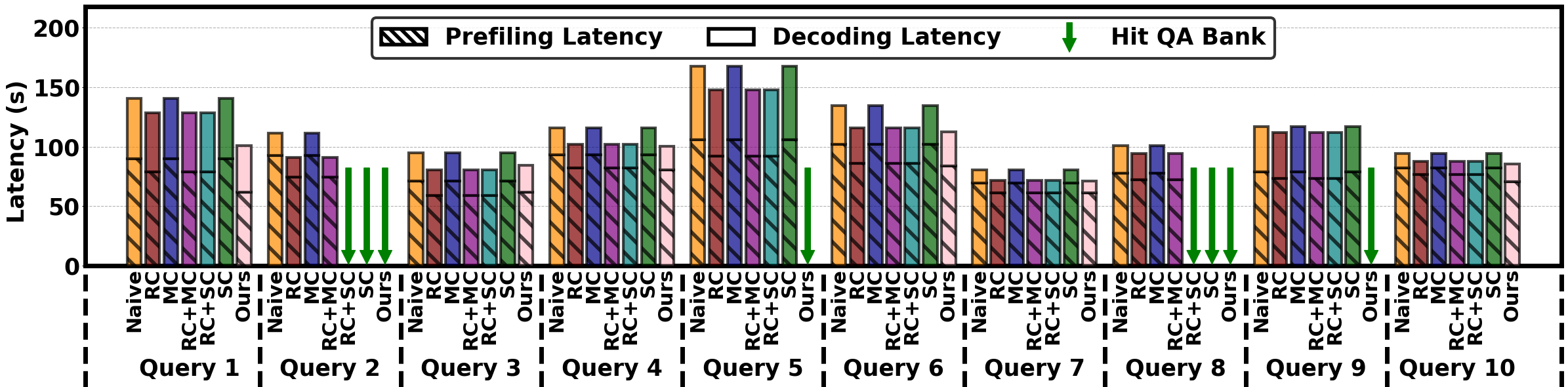}
\caption{$\text{Dialog}_0$ in MISeD}
\label{fig:overall_performance_mised_testset_0}
\end{subfigure}

\vspace{0.3cm}

\begin{subfigure}[t]{\textwidth}
\centering
\includegraphics[width=\linewidth]{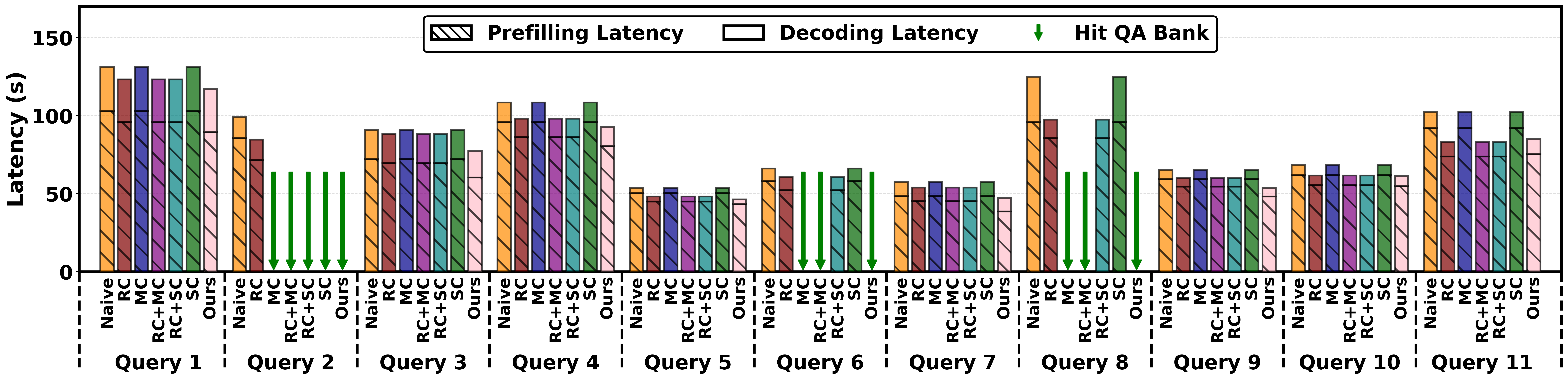}
\caption{$\text{User}_0$ in EnronQA}
\label{fig:showcase_enronqa_train_0}
\end{subfigure}
\caption{
Inference latency breakdown for each query of $\text{Dialog}_0$ in MISeD and $\text{User}_0$ in EnronQA.
}
\label{fig:showcase}
\end{figure}

\begin{figure}[t]
\centering
\captionsetup{skip=0pt}
\resizebox{0.6\columnwidth}{!}{
\includegraphics{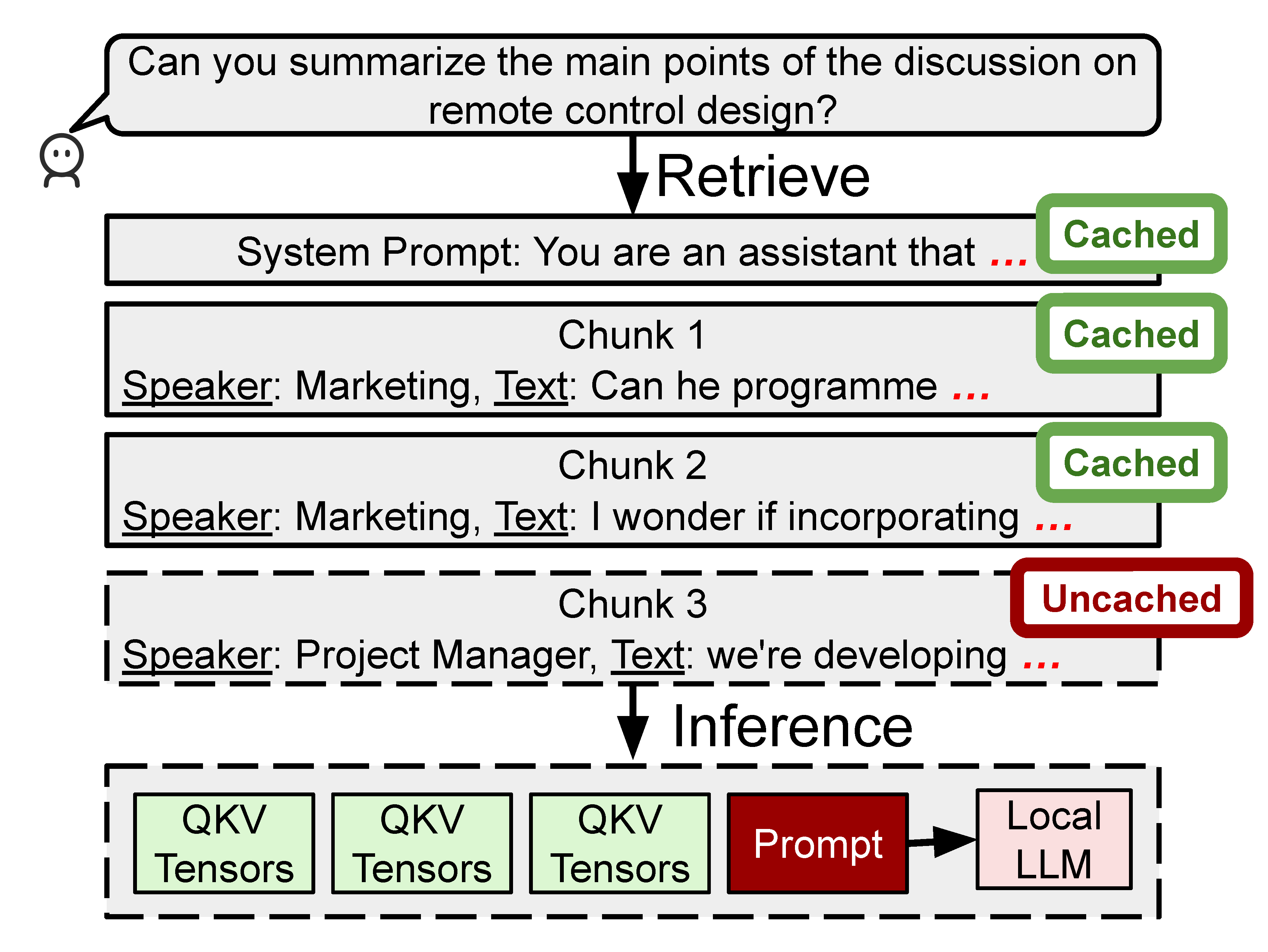}}
\caption{A showcase of QKV cache matching process.}
\label{fig:showcase_details}
\end{figure}

\begin{figure}[t]
\centering
\captionsetup{skip=0pt}
\resizebox{0.6\columnwidth}{!}{
\includegraphics{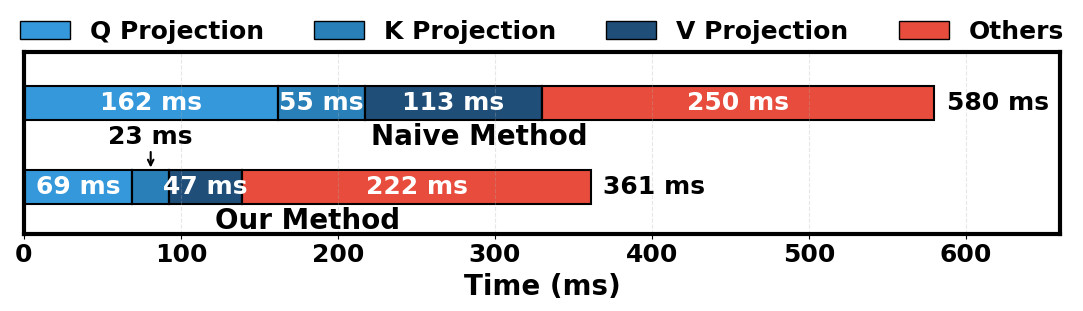}}
\caption{A showcase of latency breakdown of attention module.}
\label{fig:showcase_time_breakdown}
\end{figure}

To demonstrate the processing details of \workname, we run two workloads as end-to-end showcases: one meeting dialog from MISeD with 10 queries, and one EnronQA user with 11 queries. 
In both cases, the local knowledge source is already collected and split into chunks before the queries arrive, and the queries are processed one by one.
Before the first query, \workname runs knowledge-based query prediction twice, generating five queries each time to fill the cache. 
After answering each user query, it adds the query to the history buffer and runs history-based query prediction, again generating five new queries to update the cache.

As shown in Figure~\ref{fig:showcase}, \modify{the latency of each query processed by} \workname is significantly lower than baselines.
On the one hand, more queries hit the QA bank, 4 in MISeD and 3 in EnronQA, allowing more answers to be returned directly without LLM inference.
This improvement stems from the cache population method of \workname, which predicts queries based on both query history and knowledge content, populating the QA bank more comprehensively. 
For a given query, more QA-bank entries are \modify{highly} similar to it, so a hit is more likely.
On the other hand, for queries that miss the QA bank cache (e.g., $\text{Query}_1$ for both datasets), \workname~still achieves much lower latency.
\final{Query prediction populates the cache in advance, increasing the availability of reusable tensors for retrieved knowledge chunks. This enables prefilling reuse even when a query misses the QA bank.}
These results indicate that hierarchical cache architecture allows \workname to match caches level by level and reuse as many intermediate results as possible, and query prediction raises the hit rate of both the QA bank and the QKV cache.

We take a closer look at $\text{Query}_1$ of $\text{Dialog}_0$ in MISeD, \modify{shown} in Figure~\ref{fig:showcase_details}.
Specifically, \workname retrieves three relevant knowledge chunks for this query and concatenates these chunks, placing a system prompt before the first chunk to clarify the task objective.
\workname then checks the knowledge bank and finds that the QKV tensors of the system prompt and the first two chunks have already been precomputed.
These tensors, together with the raw text of the third chunk, are packed for LLM inference.
As shown in Figure~\ref{fig:showcase_time_breakdown}, \workname, with the QKV cache populated by the query prediction module, significantly reduces Q, K, and V projection latencies compared to the Naive method.
For this query, the latencies of projection decrease by 57.4\% (162 → 69 ms), 58.2\% (55 → 23 ms), and 58.4\% (113 → 47 ms), respectively. 
This is because the naive method must perform Q, K, and V \final{computation} for the entire input sequence, whereas \workname performs \final{computation only for the uncached suffix.}

\subsection{Overall Performance}

\begin{figure}[t]
    \centering

    \subfloat[End-to-end latency.\label{fig:llama-latency}]{
        \includegraphics[width=\linewidth]{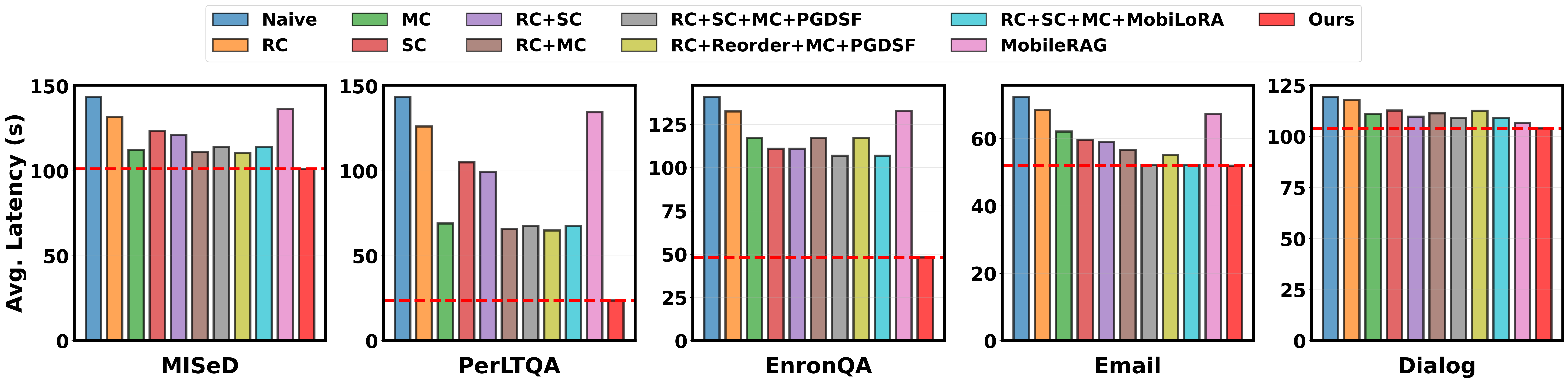}
    }

    \vspace{0.5em}

    \subfloat[LLM-judge recall and precision.\label{fig:llama-llm-judge}]{
        \includegraphics[width=\linewidth]{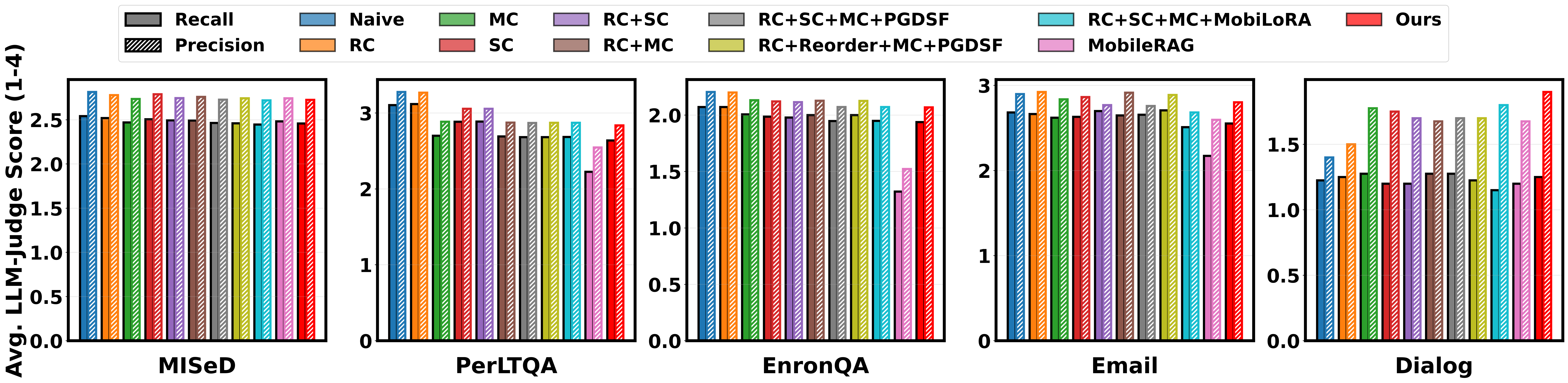}
    }

    \vspace{0.5em}

    \subfloat[ROUGE-L.\label{fig:llama-rouge-l}]{
        \includegraphics[width=\linewidth]{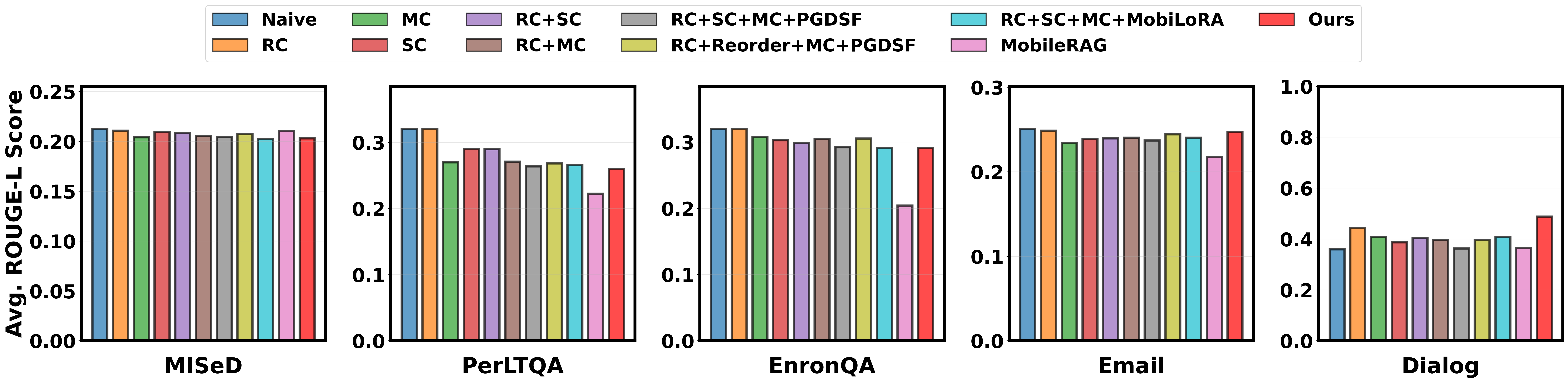}
    }

    \caption{\modify{Overall results on the five datasets using Llama-3.2-3B. (a) Average end-to-end latency, (b) LLM-judge answer quality measured by recall and precision, and (c) ROUGE-L answer quality. \workname achieves low latency while maintaining answer quality comparable to the Naive baseline.}}
    \label{fig:overall-results-llama}
\end{figure}

\modify{
As shown in Figure~\ref{fig:overall-results-llama}, we evaluate \workname~ across five datasets: MISeD, PerLTQA, EnronQA, Email and Dialog.
We measure the end-to-end latency of every query and report the average over all users (or dialogs for MISeD).
To measure response quality, we use two metrics: ROUGE-L~\cite{lin2004rouge} and LLM-judge.
For ROUGE-L, we compare each method's response with the gold answer.
For LLM-judge, we use DeepSeek-V4-Flash~\cite{deepseek2026v4flash}.
Following~\cite{golany2024efficient}, the judge is provided with the query, the method-generated response, and the gold answer, and evaluates two dimensions: \emph{recall}, how much of the key information in the gold answer is covered by the response, and \emph{precision}, how much of the response is consistent with the gold answer. 
Both are rated on a 1--4 scale, from not covered to fully covered.
The full prompt is in Appendix~\ref{sec:prompt_details}.
We first average the scores over the queries of each user, then average over all users.

Figure~\ref{fig:llama-latency} shows that \workname achieves the lowest average end-to-end latency across all datasets, indicating that its efficiency benefit is robust across workloads with different memory sources and query characteristics.
The largest gain is observed on PerLTQA: \workname reduces the average latency from 143.43\,s to 23.73\,s, an 83.5\% reduction (6.04$\times$ speedup) over the Naive baseline. Compared with the strongest baseline, RC+Reorder+MC+PGDSF, \workname reduces latency from 64.94\,s to 23.73\,s, achieving a 63.5\% reduction (2.74$\times$ speedup).
The reason is that each user's memory in PerLTQA is relatively small and well organized, which makes query prediction more accurate and makes the same knowledge chunks more likely to be reused across queries.
The gain on Dialog is smaller.
This is because its speech transcripts are large and diverse while each user has few queries.
Since each query retrieves only a small and dispersed portion of the knowledge chunks, cache reuse is more challenging.
Even so, \workname~still gives the lowest average latency.
Figures~\ref{fig:llama-rouge-l} and~\ref{fig:llama-llm-judge} further show that \workname reaches answer quality close to Naive baseline.
These results indicate that \workname substantially reduces end-to-end latency without much answer quality degradation.
}

\begin{figure}[t]
    \centering
    \begin{subfigure}[t]{0.32\columnwidth}
        \centering
        \includegraphics[width=\textwidth,height=8cm,keepaspectratio]{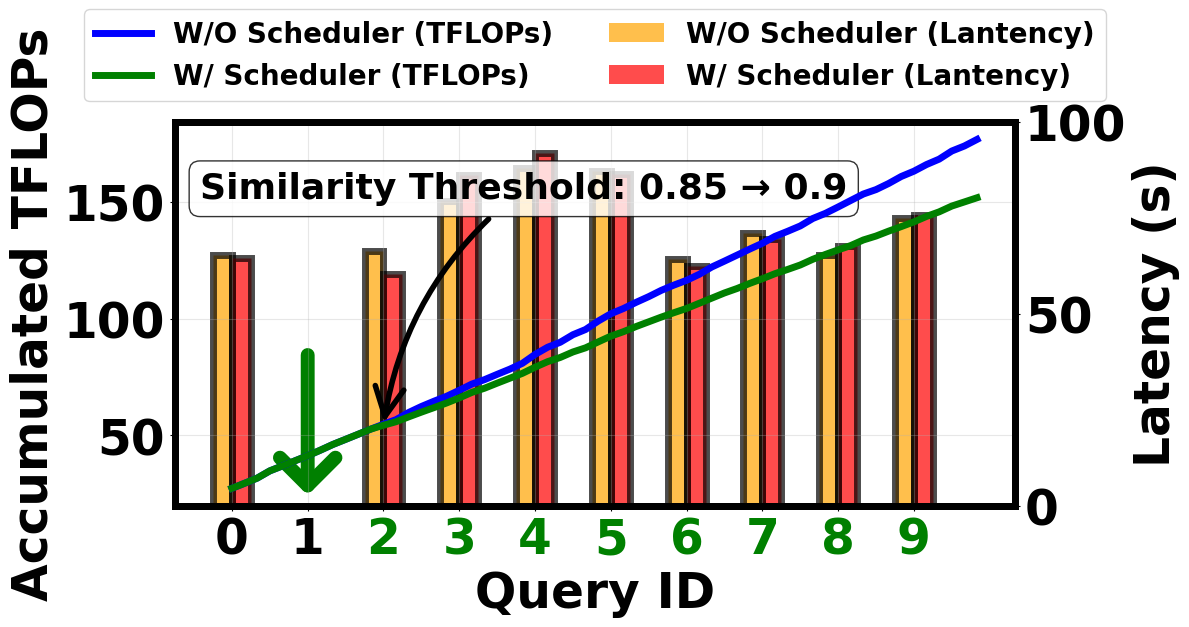}
        \caption{Turn off QA bank population.}
        \label{fig:cache_scheduling_turn_off_decode}
    \end{subfigure}%
    \hfill%
    \begin{subfigure}[t]{0.32\columnwidth}
        \centering
        \includegraphics[width=\textwidth,height=8cm,keepaspectratio]{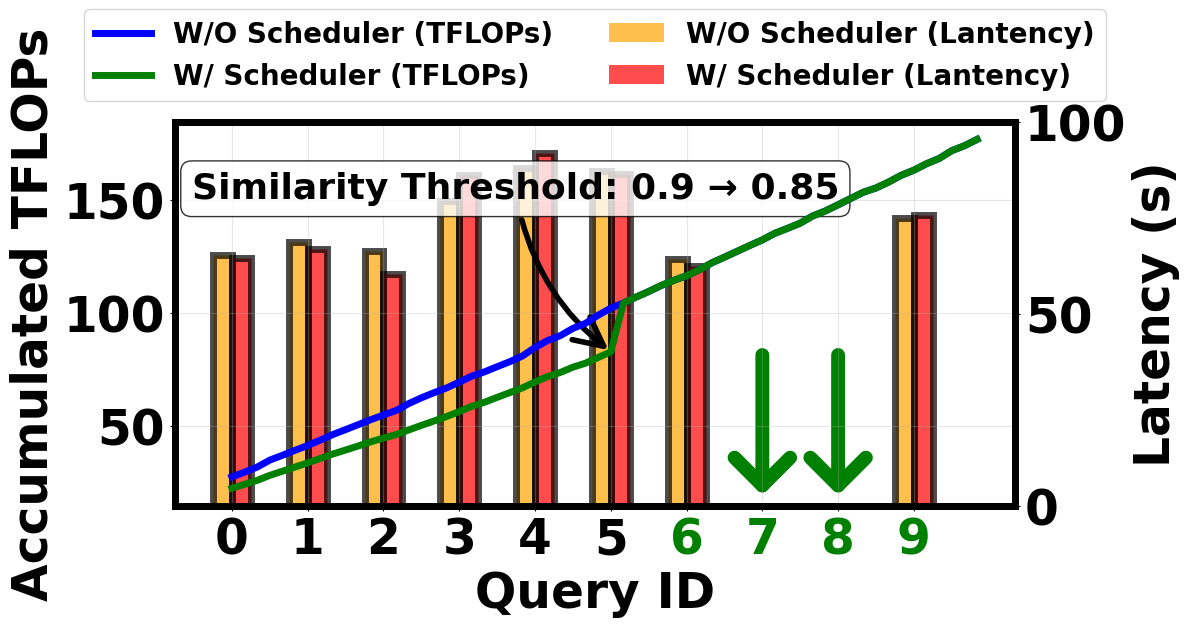}
        \caption{Turn on QA bank population.}
        \label{fig:cache_scheduling_turn_on_decode}
    \end{subfigure}%
    \hfill%
    \begin{subfigure}[t]{0.35\columnwidth}
        \centering
        \includegraphics[width=\textwidth,height=8cm,keepaspectratio]{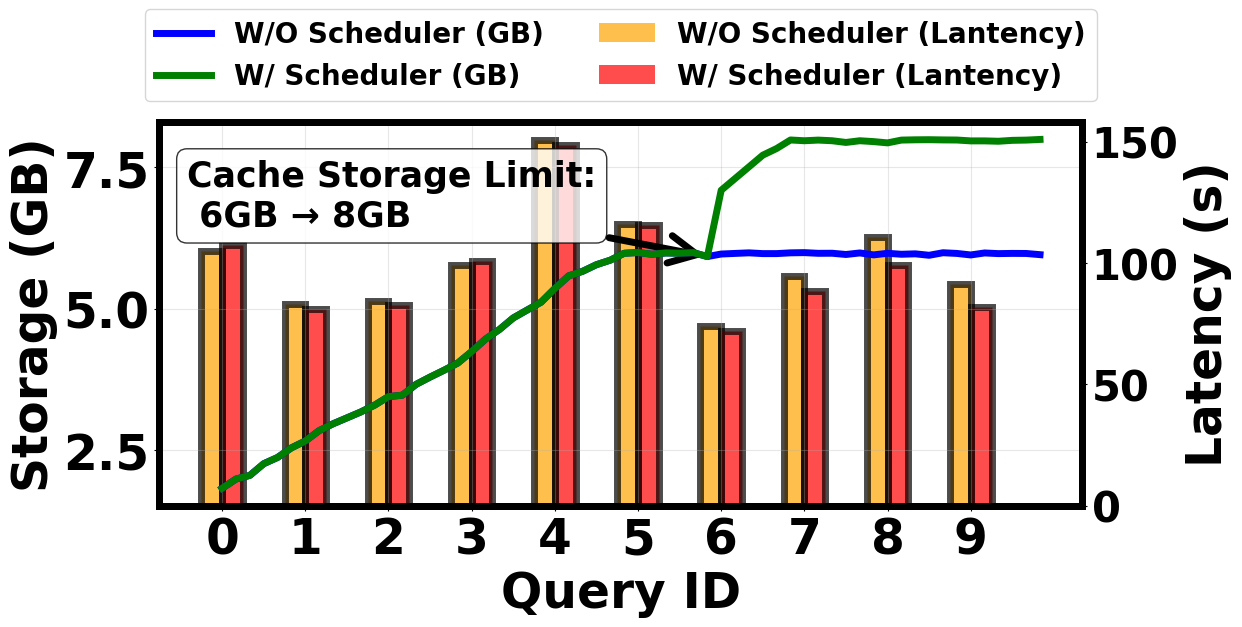}
        \caption{Restore previous QKV cache.}
        \label{fig:cache_scheduling_QKV_cache_restore}
    \end{subfigure}
    \caption{Three cases of cache scheduling. (a) After query 2, the cache scheduler temporarily disables population of the QA bank as the threshold surpasses the predefined cutoff. (b) After query 5, the cache scheduler re-enables population of the QA bank layer as the threshold \modify{falls} below the cutoff. (c) After query 6, the cache scheduler triggers a conversion from QA bank to QKV cache due to an increase in local storage space.}
    \label{fig:cache_scheduling}
\end{figure}

\subsection{Dynamic Cache Scheduling}
To evaluate the effectiveness of the cache scheduler, we construct three micro-benchmarks under varying dynamic conditions, using the subset of $\text{Dialog}_0$ from the MISeD dataset. \modify{We report computational cost in tera floating-point operations (TFLOPs).}

\subsubsection{Adaptive Cache Population}
Figure~\ref{fig:cache_scheduling_turn_off_decode} illustrates a case where the similarity threshold is raised from 0.85 to 0.90 after $\text{Query}_2$ is processed. 
Starting from $\text{Query}_3$, the scheduler detects that the threshold surpasses the predefined cutoff, indicating that subsequent queries are unlikely to hit the QA bank.
To avoid unnecessary computation on the QA bank population, the scheduler disables LLM decoding for predicted queries and performs only prefilling to update the QKV cache.
The green and blue lines in Figure~\ref{fig:cache_scheduling_turn_off_decode} show the accumulated TFLOPs with and without the scheduler, respectively.
They indicate that the cache scheduler effectively reduces computational overhead, and the computational savings grow progressively as the cache population proceeds.
After processing $\text{Query}_9$, the computational savings reach 14.12\% compared to running without the scheduler.
Meanwhile, the action of disabling decoding by the scheduler has minimal impact on latency since few user queries hit the QA bank under high similarity thresholds.

\subsubsection{Conversion from QKV Cache to QA Bank.}
In this experiment, we examine a case where the similarity threshold decreases during query processing and analyze the behavior of the scheduler.
As shown in Figure~\ref{fig:cache_scheduling_turn_on_decode}, the threshold drops from 0.90 to 0.85 after processing $\text{Query}_5$.
The scheduler detects that the threshold has fallen below the cutoff, meaning the QA bank is more likely to be hit by future queries.
To improve the QA bank hit rate, the scheduler then identifies the previous queries that were not decoded and performs decoding for them to populate the QA bank after $\text{Query}_5$.
Through this adjustment, the cache scheduler achieves latency comparable to the baseline without a scheduler, which always performs both prefilling and decoding for every query.

\subsubsection{Conversion from QA Bank to QKV Cache.}
During runtime, storage limits are configured for both the QA bank and the QKV cache to control cache storage consumption.
Since the QA bank requires significantly less space than the QKV cache, it can store a larger number of items than the corresponding QKV tensors. To prevent the cache system from exceeding the \modify{limit}, the QKV tensors can be deleted by the cache eviction algorithm when necessary.
When the storage limit is increased, \workname~can restore QKV tensors from the QA pairs stored in the QA bank, thereby improving the QKV cache hit rate.
In Figure~\ref{fig:cache_scheduling_QKV_cache_restore}, the storage limit for the QKV cache is relaxed from 6GB to 8GB after processing $\text{Query}_6$, which triggers the conversion from the QA bank to the QKV cache.
The results show that \workname~achieves lower latencies than the baseline without a cache scheduler when processing $\text{Query}_7$, $\text{Query}_8$, and $\text{Query}_9$. Specifically, with the cache scheduler, \workname~\modify{enables} these queries to match QKV tensors of two knowledge chunks, while without it, they only match tensors of one chunk.
Thus, the cache scheduler helps skip more computation during LLM prefilling by maximizing the utilization of available storage space.

\subsection{Sensitivity Analysis}
We evaluate the impact of five system parameters: prediction stride, cache capacity, mobile platform, similarity threshold, and on-device LLM selection.

\subsubsection{Impact of Prediction Stride}
We evaluate the performance of \workname~with the prediction stride ranging from 1 to 5. 
Figure~\ref{fig:impact_of_prediction_stride} shows that the average latency slightly decreases as the prediction stride increases.
A larger prediction stride enables more QA pairs and QKV cache tensors to be stored in the cache system, which also improves the diversity of predicted queries and their corresponding retrieved knowledge chunks. As a result, the hit rates for both cache layers increase, thereby reducing the inference latency.

\subsubsection{Impact of Storage Limitation}
\begin{figure}[t]
    \centering
    \begin{minipage}[t]{0.25\textwidth}
        \centering
        \includegraphics[width=\textwidth]{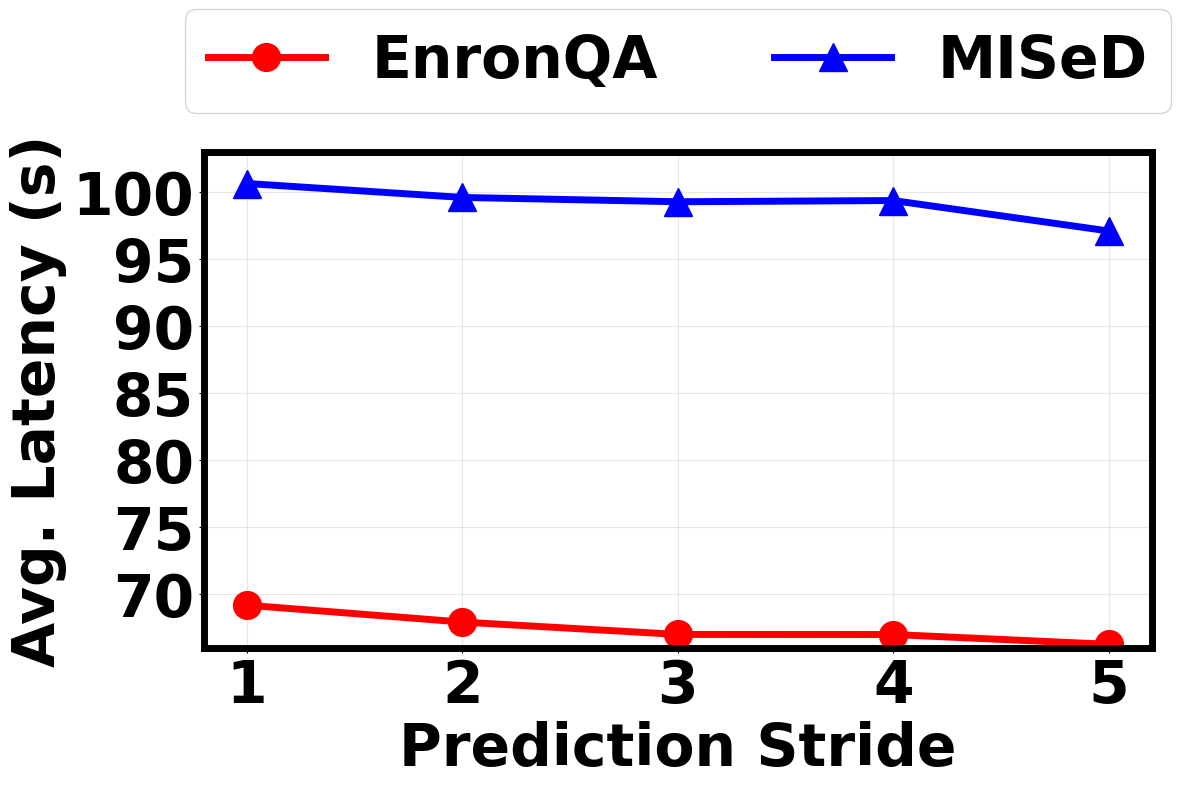}
        \caption{Impact of prediction stride on average latency for MISeD $\text{Dialog}_0$ (10 queries) and EnronQA $\text{User}_0$ (11 queries).}
        \label{fig:impact_of_prediction_stride}
    \end{minipage}%
    \hfill%
    \begin{minipage}[t]{0.25\textwidth}
        \centering
        \includegraphics[width=\textwidth]{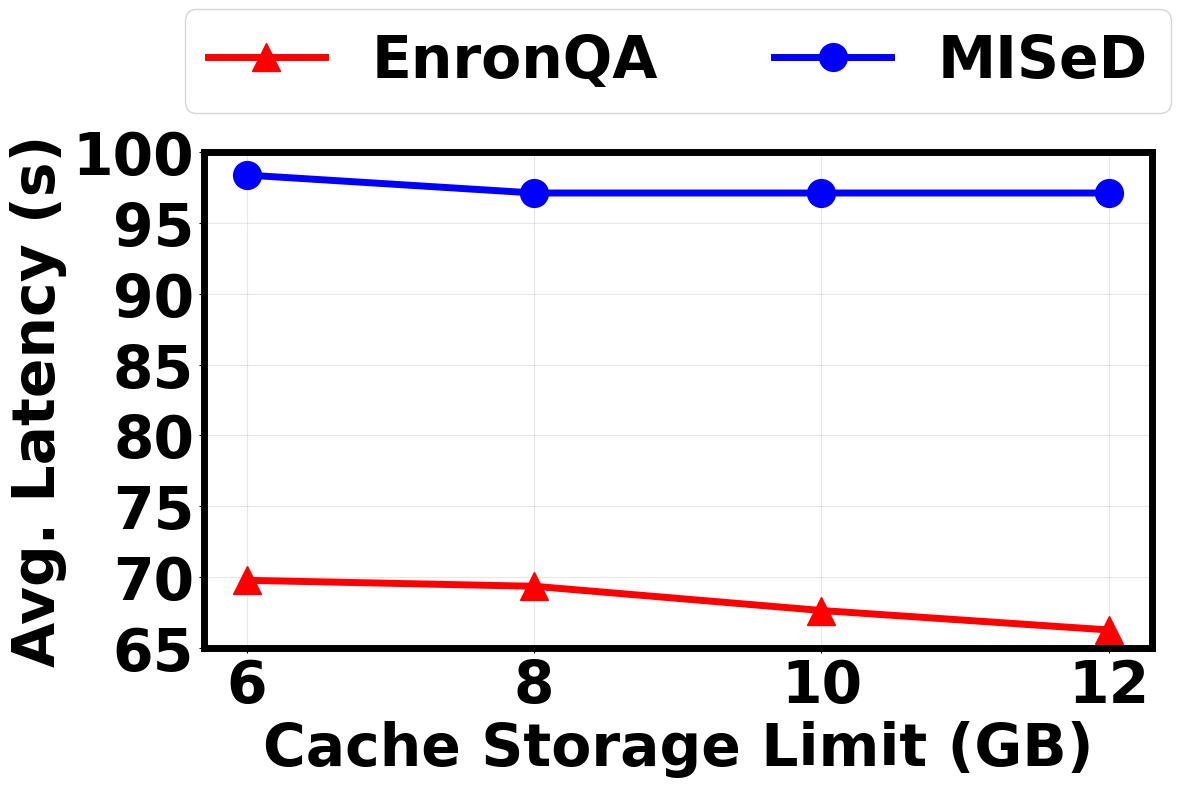}
        \caption{Impact of storage limit on average latency for MISeD $\text{Dialog}_0$ (10 queries) and EnronQA $\text{User}_0$ (11 queries).}
        \label{fig:impact_of_storage_limitation}
    \end{minipage}%
    \hfill%
    \begin{minipage}[t]{0.44\textwidth}
        \centering
        \includegraphics[width=\textwidth]{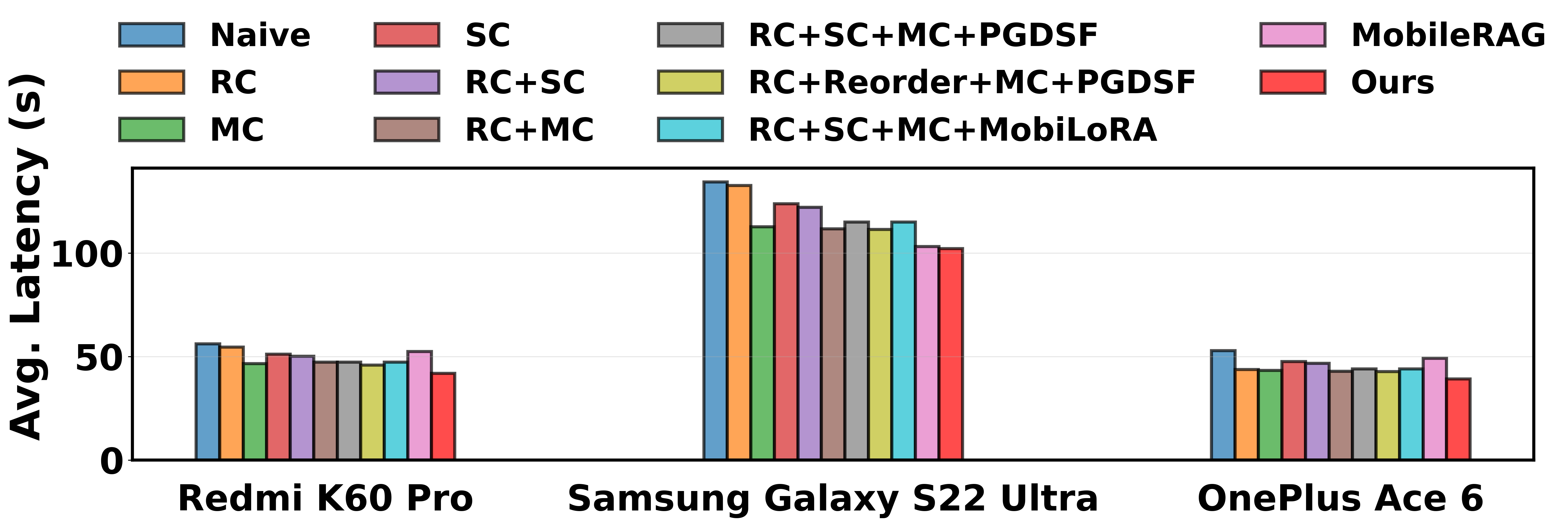}
        \caption{\modify{Overall performance comparison on MISeD dataset across three mobile devices: Redmi K60 Pro, Samsung Galaxy S22 Ultra, and OnePlus Ace 6.}}
        \label{fig:performance_on_different_mobile_platforms}
    \end{minipage}
\end{figure}

\modify{To evaluate the impact of storage limitations, we vary the total cache capacity from 6\,GB to 12\,GB while reserving 50\,MB for the QA Bank and allocating the remaining capacity to the QKV cache.}
As shown in Figure~\ref{fig:impact_of_storage_limitation}, the average latency decreases as the storage limit increases.
This is because, with relaxed storage limitations, fewer QKV tensors are evicted and more tensors are retained in the cache system. 
Consequently, more retrieved knowledge chunks can match their corresponding QKV tensors.

\subsubsection{Impact of Mobile Platforms}

\modify{To evaluate the performance of \workname~across mobile devices with different hardware configurations on the MISeD dataset, we run \workname~and all baseline methods on Redmi K60 Pro, Samsung Galaxy S22 Ultra, and OnePlus Ace 6.
Figure~\ref{fig:performance_on_different_mobile_platforms} shows that while all methods exhibit similar trends in latency variation across devices due to different hardware specifications, \workname~consistently achieves the lowest average latency regardless of the device used.
This demonstrates that our method does not rely on specific hardware features and maintains consistently superior performance across devices of various performance tiers, proving its robustness. Furthermore, its ability to run on consumer-grade devices across different price ranges indicates broad application potential.}

\subsubsection{Impact of Similarity Threshold}

\begin{figure*}[t]
    \centering
    \includegraphics[width=\textwidth]{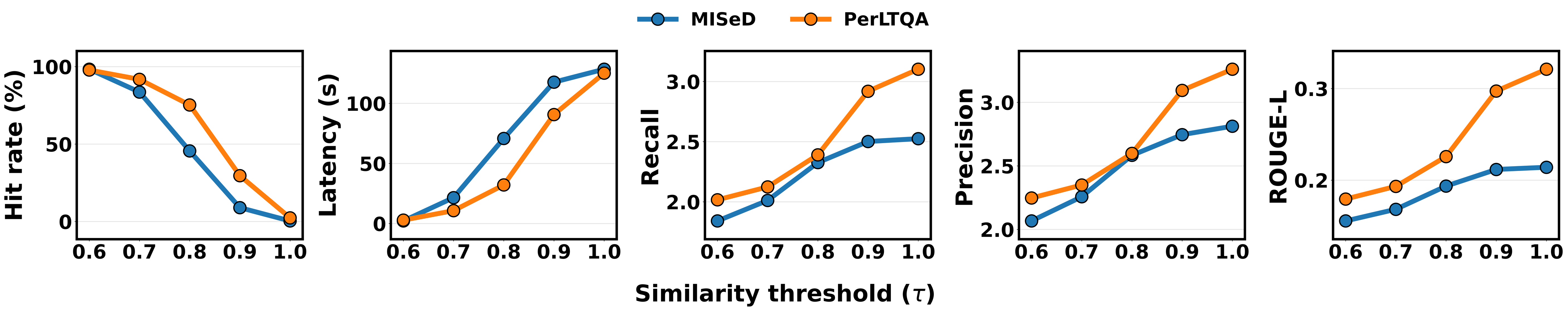}
    \caption{\modify{QA-Bank similarity-threshold sensitivity on MISeD and PerLTQA datasets. 
    From left to right, the panels show QA-Bank hit rate, average end-to-end latency, LLM-judge recall, LLM-judge precision, and ROUGE-L. 
    Each result is averaged over all real queries in the corresponding
    dataset.}}
    \label{fig:qa-bank-threshold-sensitivity}
\end{figure*}

\modify{To study how the similarity threshold affects latency and answer quality, we vary the threshold $\tau_{\mathrm{QA}}$ from 0.60 to 0.95 on MISeD and PerLTQA datasets.
Figure~\ref{fig:qa-bank-threshold-sensitivity} reports the QA-Bank hit rate, average end-to-end latency, LLM-judge recall, LLM-judge precision, and ROUGE-L. 
As $\tau_{\mathrm{QA}}$ increases, QA-Bank reuse becomes more conservative: fewer queries hit the QA Bank, more of them need LLM generation, and average latency goes up. 
In exchange, it is less likely that a poorly matched answer gets reused, which shows up as better answer quality. 
Conversely, lower thresholds increase reuse opportunities and reduce latency, but introduce a larger quality impact. 
These results demonstrate that $\tau_{\mathrm{QA}}$ provides a practical control knob for balancing response quality and end-to-end latency.}

\subsubsection{Impact of On-device LLM Selection}

\begin{figure}[t]
    \centering
    \includegraphics[width=\columnwidth]{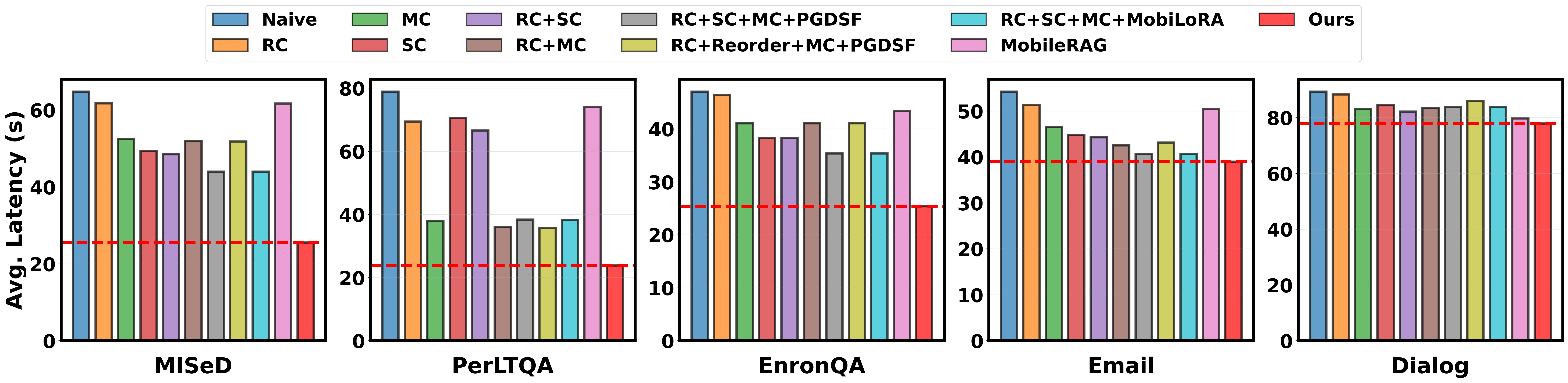}
    \caption{\modify{Average end-to-end latency across the five datasets using Qwen-1.5-1.8B. \workname achieves the lowest average latency among the compared methods.}}
    \label{fig:overall-performance-qwen}
\end{figure}

\modify{
As shown in Figure~\ref{fig:overall-performance-qwen}, we further evaluate \workname using Qwen-1.5-1.8B as the on-device LLM for knowledge-abstract generation, query prediction, and response generation, while keeping all other
settings identical to the Llama-3.2-3B evaluation. 
\workname achieves the lowest average end-to-end latency on all five datasets, demonstrating that its efficiency benefit generalizes across on-device LLM architectures. 
For example, on PerLTQA, \workname reduces latency from 78.88\,s to 23.88\,s relative to Naive, corresponding to a 69.7\% reduction ($3.30\times$ speedup). 
The corresponding Qwen-1.5-1.8B answer-quality results are provided in Section~\ref{sec:answer-quality-qwen} of the appendix.
}

\begin{figure}[t]
    \centering

    \begin{minipage}[t]{0.70\columnwidth}
        \vspace{0pt}
        \centering
        \captionsetup{skip=0pt}

        \begin{subfigure}[t]{0.49\linewidth}
            \vspace{0pt}
            \centering
            \parbox[b][3.5cm][b]{\linewidth}{%
                \centering
                \includegraphics[
                    width=\linewidth,
                    height=3.5cm,
                    keepaspectratio
                ]{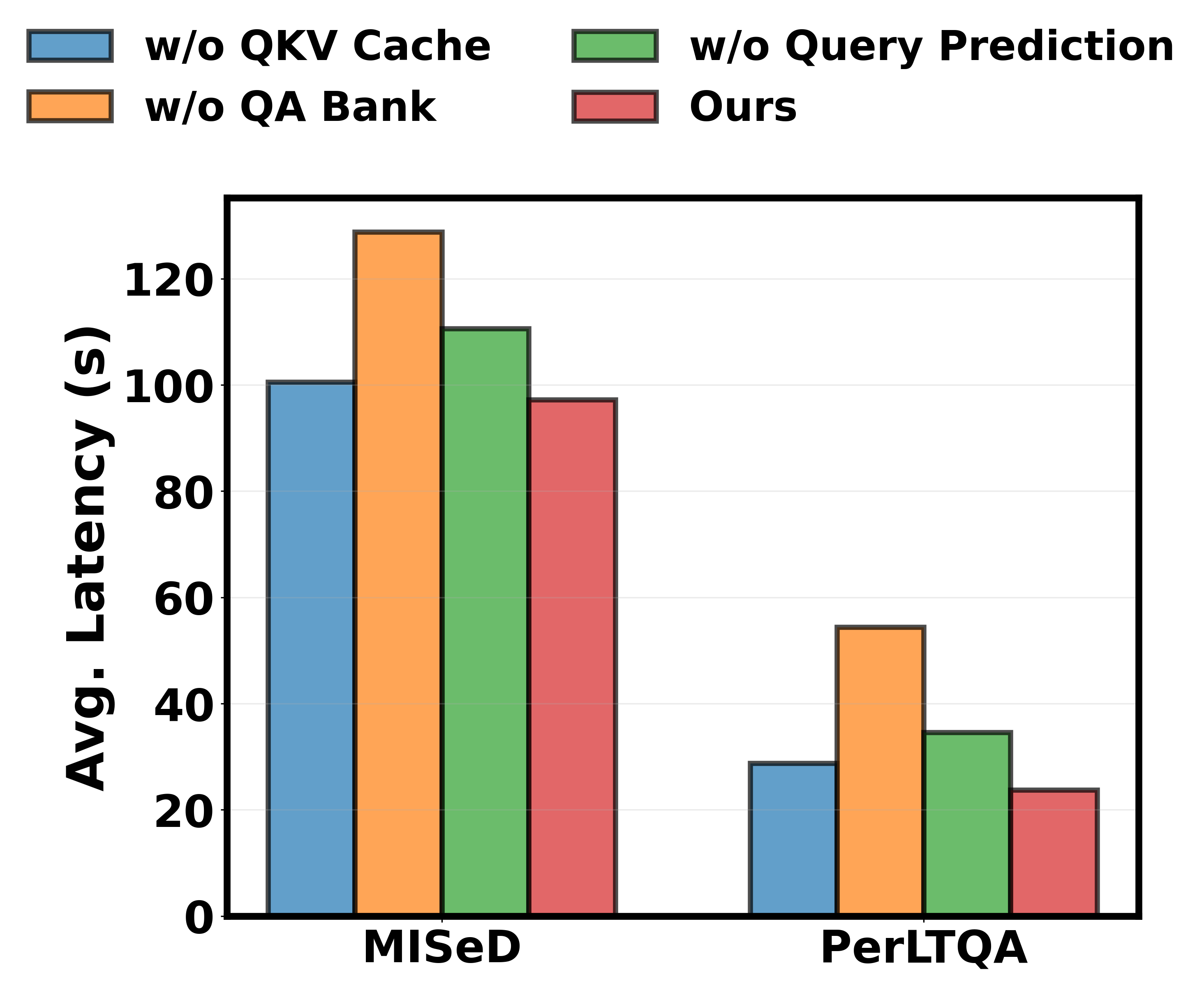}%
            }
            \caption{Average end-to-end latency.}
            \label{fig:ablation-full-latency}
        \end{subfigure}%
        \hfill
        \begin{subfigure}[t]{0.49\linewidth}
            \vspace{0pt}
            \centering
            \parbox[b][3.5cm][b]{\linewidth}{%
                \centering
                \includegraphics[
                    width=\linewidth,
                    height=3.5cm,
                    keepaspectratio
                ]{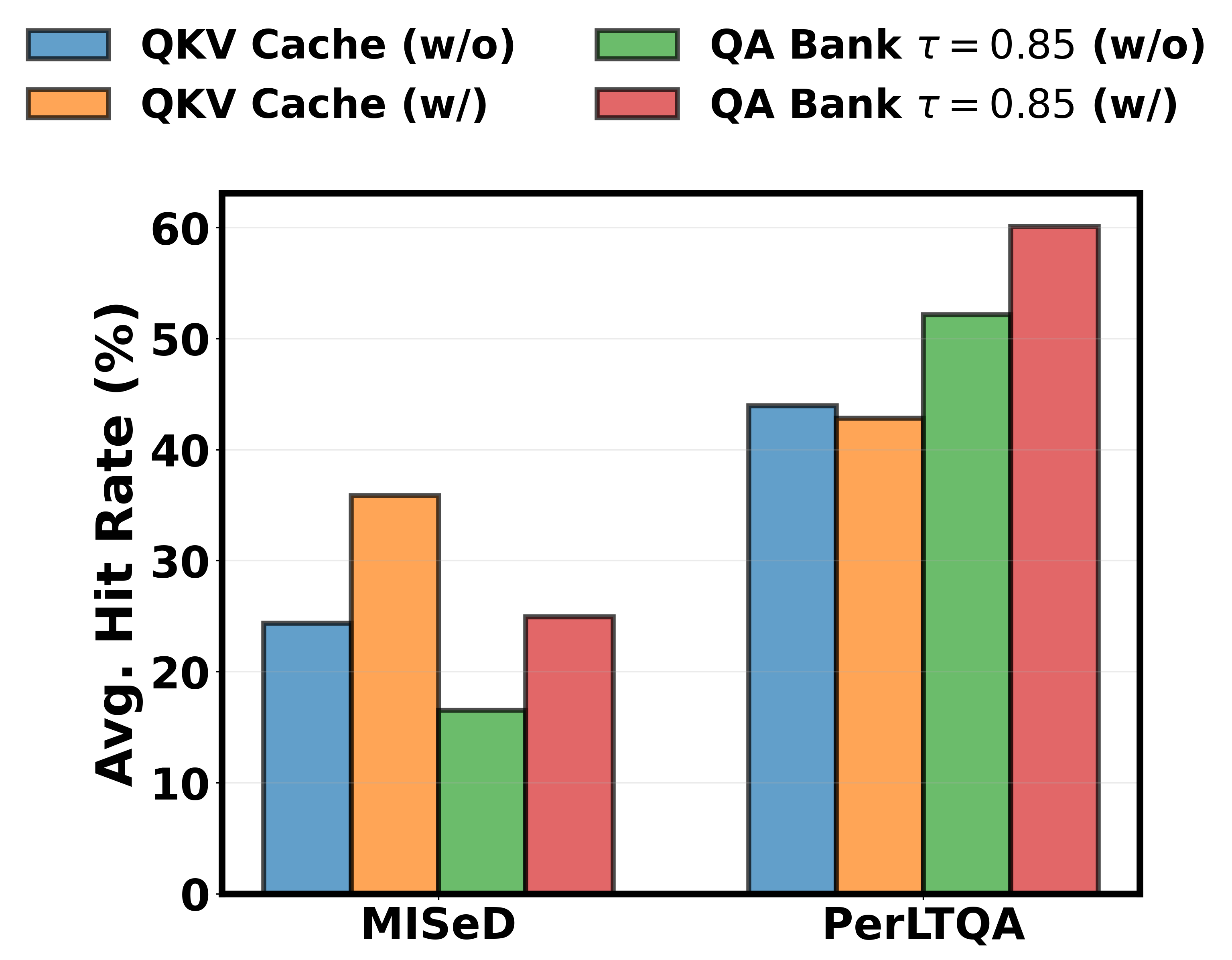}%
            }
            \caption{Average cache hit rate.}
            \label{fig:ablation-full-hit-rate}
        \end{subfigure}
    \end{minipage}%
    \hfill
    \begin{minipage}[t]{0.28\columnwidth}
        \vspace{0pt}
        \centering
        \parbox[b][3.5cm][b]{\linewidth}{%
            \centering
            \includegraphics[
                width=\linewidth,
                height=3.2cm,
                keepaspectratio
            ]{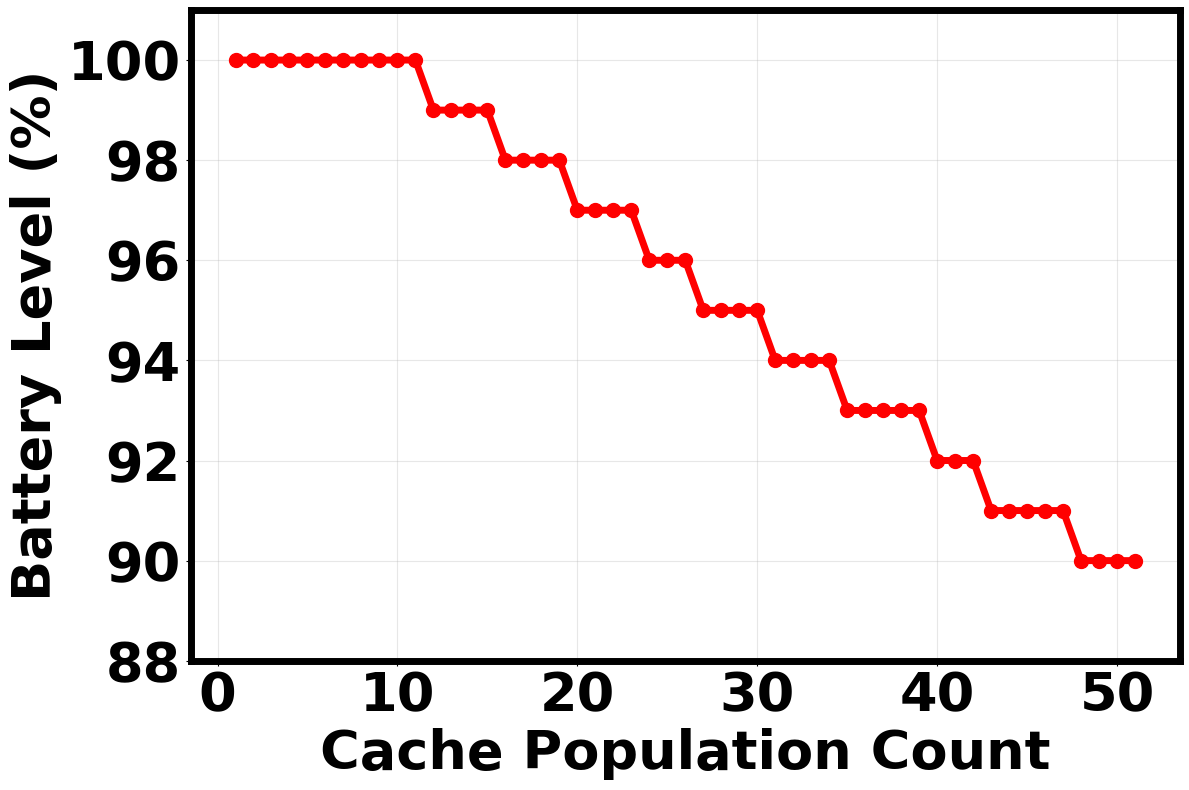}%
        }
    \end{minipage}

    \par\medskip

    \begin{minipage}[t]{0.66\columnwidth}
        \captionsetup{skip=0pt}
        \caption{Ablation results on the full MISeD
        and PerLTQA datasets using Llama-3.2-3B.
        (a) Average end-to-end latency when removing the QKV
        cache, QA Bank, or query-prediction module.
        (b) Average hit rates of the QKV cache and QA Bank \final{with and without query prediction.}}
        \label{fig:ablation-full-datasets}
    \end{minipage}%
    \hfill
    \begin{minipage}[t]{0.31\columnwidth}
        \captionsetup{skip=0pt}
        \caption{Battery level variation with cache population
        count using Llama-3.2-3B on OnePlus Ace 6.}
        \label{fig:battery}
    \end{minipage}

\end{figure}

\subsection{Ablation Study}

\modify{
To evaluate the contribution of each component, we compare \workname with variants that individually remove the QKV cache, QA Bank, or query-prediction module, using \(\tau_{\mathrm{QA}}=0.85\) and a prediction stride of five.
Figure~\ref{fig:ablation-full-latency} shows that \workname achieves the lowest average latency on both datasets.
On MISeD, removing the QKV cache, QA Bank, and query prediction increases latency from 97.20\,s to 100.55\,s, 128.79\,s, and 110.60\,s, respectively; 
\final{on PerLTQA, latency increases from 23.73 s to 28.77 s, 54.39 s, and 34.61 s, respectively.}
The QA Bank provides the largest benefit, while QKV reuse and query prediction provide complementary additional reductions.
Figure~\ref{fig:ablation-full-hit-rate} shows that query prediction increases MISeD's QKV cache hit rate from 24.39\% to 35.88\% and its QA-Bank hit rate from 16.56\% to 24.97\%.
For the PerLTQA dataset, prediction increases the QA-Bank hit rate from 52.15\% to 60.11\%, while the QKV cache hit rate remains comparable (43.98\% versus 42.85\%).
This limited headroom is likely because PerLTQA users have relatively compact and semantically well-structured knowledge, causing retrieved knowledge chunks to be more concentrated and already frequently reused through reactive caching.
Overall, these results demonstrate that the QA Bank, QKV cache, and query prediction jointly contribute to \workname's latency reduction.
}

\subsection{System Resource Profiling}
\label{sec:system_overhead}

\begin{table}[t]
\centering
\captionsetup{skip=1pt}
\caption{System resource profiling of \workname.}
\label{tab:system_overhead}

\renewcommand{\arraystretch}{1.15}
\resizebox{\columnwidth}{!}{
\begin{tabular}{l r l r l r}
\toprule
\multicolumn{2}{c}{\textbf{Latency}} &
\multicolumn{2}{c}{\textbf{Storage}} &
\multicolumn{2}{c}{\textbf{Runtime Memory}} \\
\cmidrule(lr){1-2} \cmidrule(lr){3-4} \cmidrule(lr){5-6}
\textbf{Operation} & \textbf{Time (s)} &
\textbf{Component} & \textbf{Size/Item} &
\textbf{Component} & \textbf{Peak PSS (GiB)} \\
\midrule
\multicolumn{2}{c}{
\multirow[c]{7}{*}{
\begin{tabular}{@{}l@{\qquad}r@{}}
QA-Bank Matching   & 1.61 \\
Knowledge Retrieval & 3.94 \\
QKV Cache Matching  & 0.015 \\
QKV Cache Loading   & 1.03 \\
LLM Prefilling      & 62.14 \\
LLM Decoding        & 10.95
\end{tabular}
}}
&
\multicolumn{2}{c}{
\multirow[c]{7}{*}{
\begin{tabular}{@{}l@{\qquad}r@{}}
QA Bank (per entry)          & 4KB  \\
QKV Cache (per entry)        & 87MB \\
Knowledge Chunk (per chunk)  & 16KB
\end{tabular}
}}
&
LLM Inference
& 2.726 \\

& & & &
QA-Bank Matching
& 0.106 \\

& & & &
Knowledge Retrieval
& 0.111 \\

& & & &
Inference w/ QKV Tensors of 1 Chunk
& 2.804 \\

& & & &
Inference w/ QKV Tensors of 2 Chunks
& 2.918 \\

& & & &
Inference w/ QKV Tensors of 3 Chunks
& 3.072 \\

& & & &
Full \workname
& 2.897 \\
\bottomrule
\end{tabular}
}
\end{table}

\modify{
To measure battery consumption, we repeatedly run \workname's complete cache-population process on OnePlus Ace~6 using a query from $\text{Dialog}_0$ of MISeD (349 input tokens and 136 output tokens). 
Each per-query population process includes query embedding, knowledge retrieval, similar-query matching, QKV cache matching and loading, LLM prefilling and decoding, and saving new QKV tensors and QA pairs.
Figure~\ref{fig:battery} shows that 51 cache-population processes consume 10\% battery capacity, corresponding to approximately 0.196\% per predicted query.
Since each prediction round typically generates one to five queries, a round costs roughly 0.2--1.0\%.
As knowledge chunks and query histories update infrequently and \workname runs when the device is idle (optionally only while charging), its practical battery cost is typically small.

We also measure the latency, storage, and runtime-memory overhead of each component, with the prediction stride set to five and the knowledge chunk length fixed at 100 words.
Table~\ref{tab:system_overhead} shows that LLM prefilling and decoding dominate latency at 77.9\% and 13.7\%, respectively. 
The QKV cache has the largest persistent-storage cost (87\,MB per entry), but its total size is bounded by the configured storage budget.
For runtime memory, we profile each stage using peak total proportional set size (PSS). 
The LLM inference engine dominates at 2.726\,GiB, while QA-Bank matching and knowledge retrieval need only 0.106\,GiB and 0.111\,GiB and do not overlap with LLM inference stage.
Loading QKV tensors for up to three chunks adds at most 0.346\,GiB. The full \workname pipeline uses 2.897\,GiB, only 0.171\,GiB (6.3\%) above model-only inference.
}

\subsection{\modify{Real-World Evaluation}}
\begin{figure}[t]
        \centering
        \includegraphics[width=\linewidth]{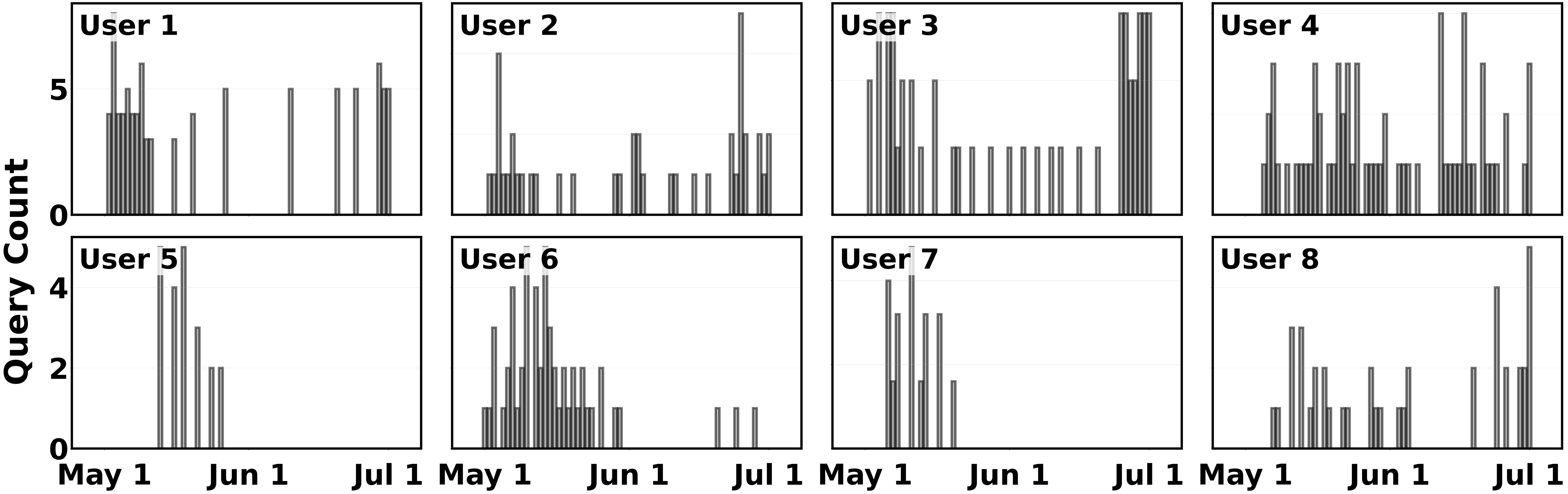}
        \caption{\modify{Chronological daily query counts for all eight users during the two-month in-the-wild deployment.}}
        \label{fig:real-world-chronological-performance}
    \end{figure}

\begin{figure}[t]
    \centering

    \begin{minipage}[t]{0.48\linewidth}
        \centering
        \includegraphics[width=\linewidth]{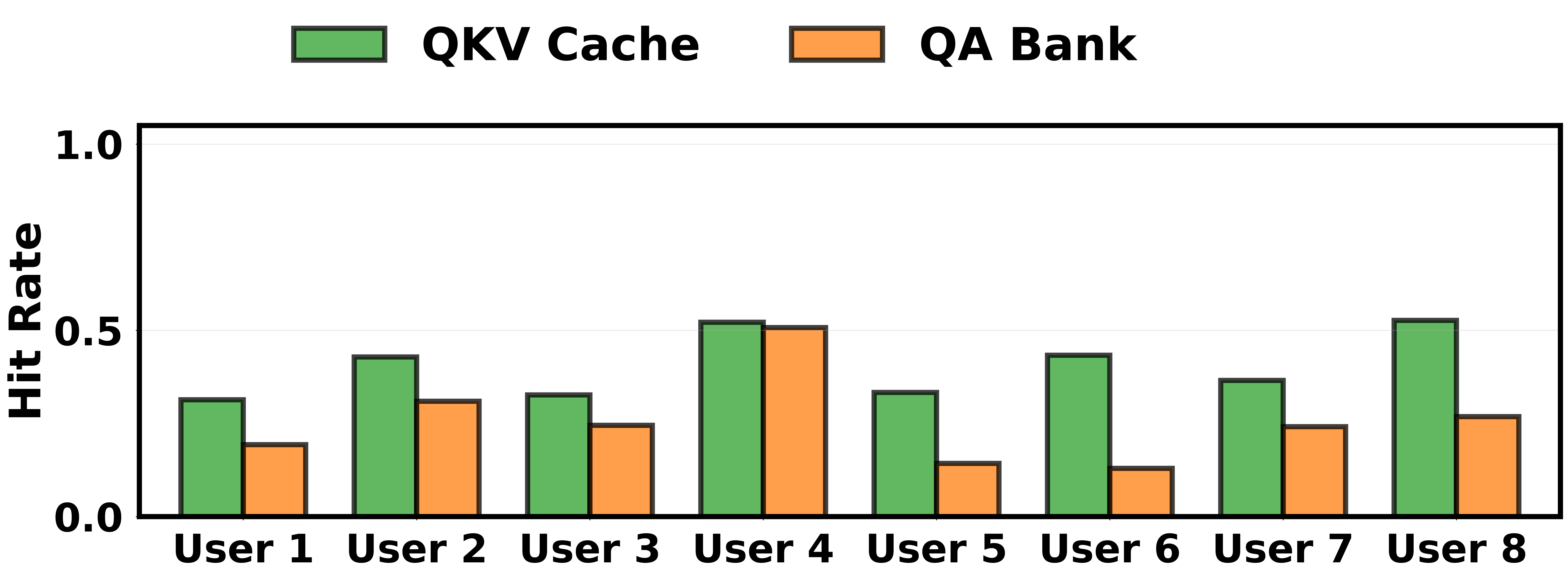}
        \caption{QKV cache and QA Bank hit rates for all eight users during the two-month in-the-wild deployment.}
        \label{fig:real-world-cache-coverage}
    \end{minipage}
    \hfill
    \begin{minipage}[t]{0.48\linewidth}
        \centering
        \includegraphics[width=\linewidth]{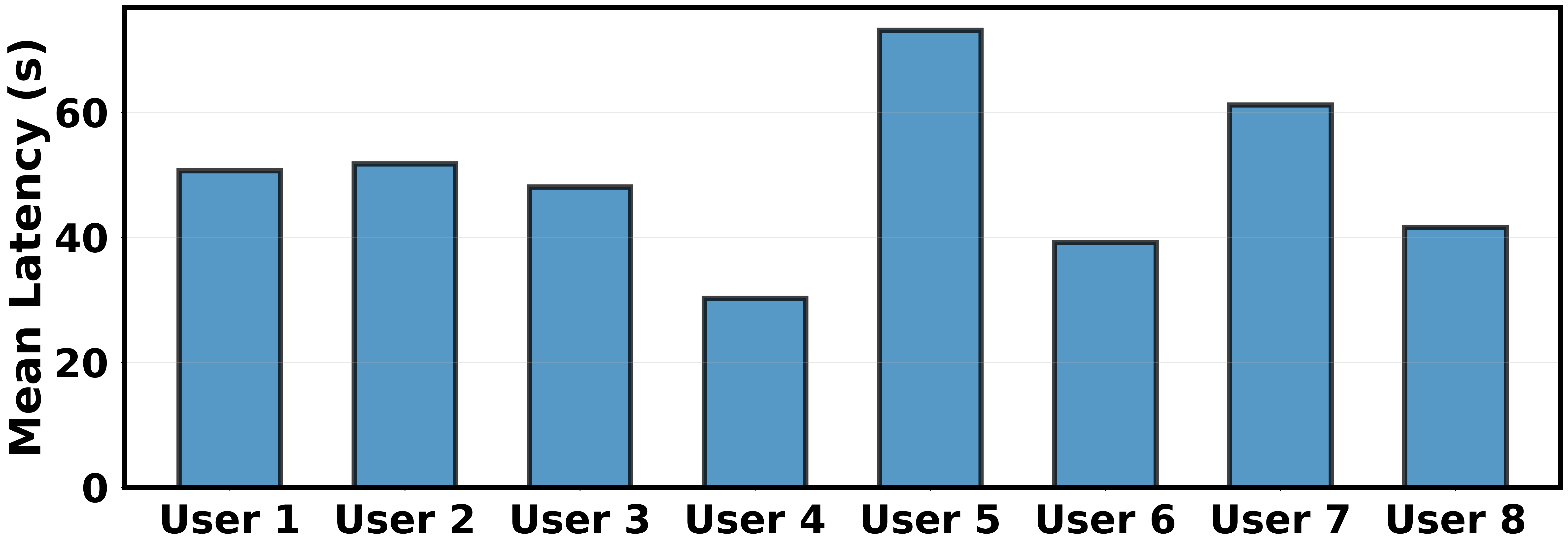}
        \caption{Mean query latency for all eight users during the two-month in-the-wild deployment.}
        \label{fig:real-world-latency}
    \end{minipage}
\end{figure}

\subsubsection{\modify{Workload and Deployment Setup}}
\modify{
To evaluate \workname under realistic workloads, we conducted an additional two-month in-the-wild deployment with eight volunteers, using Llama-3.2-3B as the on-device LLM.
Users~1--3 used their private emails as local knowledge bases. Users~4--8 used non-sensitive private chat logs from their social accounts.
We provided a lightweight terminal-based interaction interface through Termux.
The \workname process started on demand, either when a user issued a query or when cache \final{population} was triggered.
Each user picked a fixed time window in the evening for query prediction and cache population. 
During this window, the system automatically ingested newly arrived knowledge chunks, generated abstracts, performed knowledge-based query prediction, and populated the cache using the predicted queries.
If a user issued real queries during the day, \workname also ran history-based prediction from those recent queries and populated the cache accordingly. 
If there was no new knowledge or real queries, the corresponding prediction step was skipped. 
Users could also manually trigger predictive cache population at a convenient time, such as while charging, at night, or when the phone was idle. 
This deployment examines how \workname performs in real-world use.
}

\begin{figure*}[t]
    \centering
    \includegraphics[width=\textwidth]{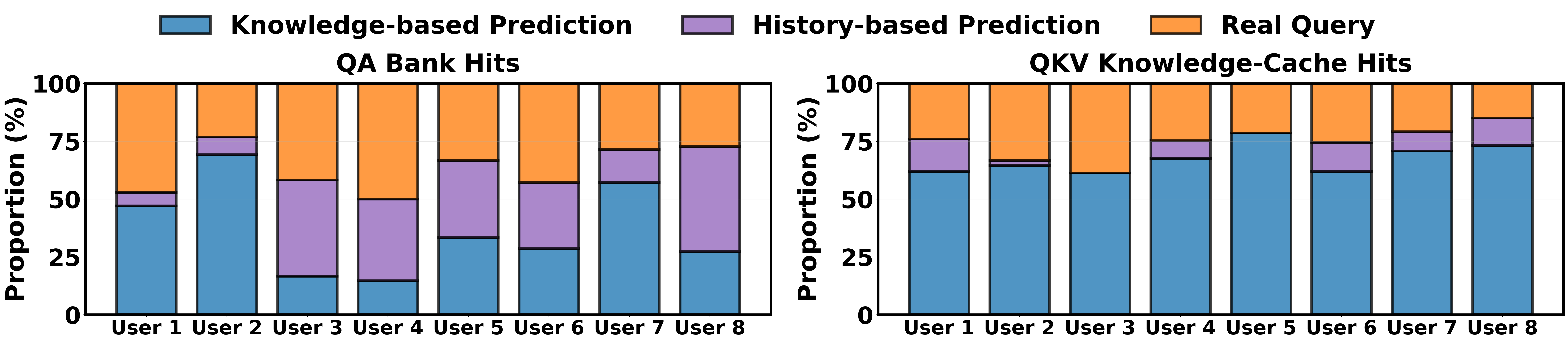}
    \caption{\modify{Sources of QA Bank and QKV cache hits under chronological real-world workloads from eight users.}}
    \label{fig:real_world_cache_hit_origins}
\end{figure*}

\subsubsection{\modify{End-to-End Performance}}
\modify{
Figure~\ref{fig:real-world-chronological-performance} shows how the eight users' queries are distributed over the two-month deployment.
It shows that real-world query activity is sparse and temporally irregular.
Across the two-month deployment, the eight users issued 391 queries, with per-user average daily query rates ranging from 0.34 to 1.42.
Such a workload is hard for purely reactive caching, which fills the cache only when a query arrives: with queries this rare, little useful content accumulates, and most queries are still computed from scratch online.
% hit rate
Despite this sparsity, Figure~\ref{fig:real-world-cache-coverage} shows that \workname achieves cache reuse for all users through both the QA Bank and the QKV cache.
Figure~\ref{fig:real_world_cache_hit_origins} further traces the origin of each cache hit.
It shows that caches populated through knowledge- and history-based prediction contribute approximately 50--77\% of QA Bank hits and 61--85\% of QKV knowledge-cache hits, averaging 63\% and 75\%, respectively.
These results indicate that query prediction contributes the majority of effective cache hits and that query prediction \modify{aligns} with actual usage.
% latency
Figure~\ref{fig:real-world-latency} reports the mean end-to-end latency per user, ranging from 30.22 to 73.10~s.
}

\subsubsection{\modify{Overhead under Real-World Workloads}}

\begin{figure}[t]
        \centering
        \includegraphics[width=\linewidth]{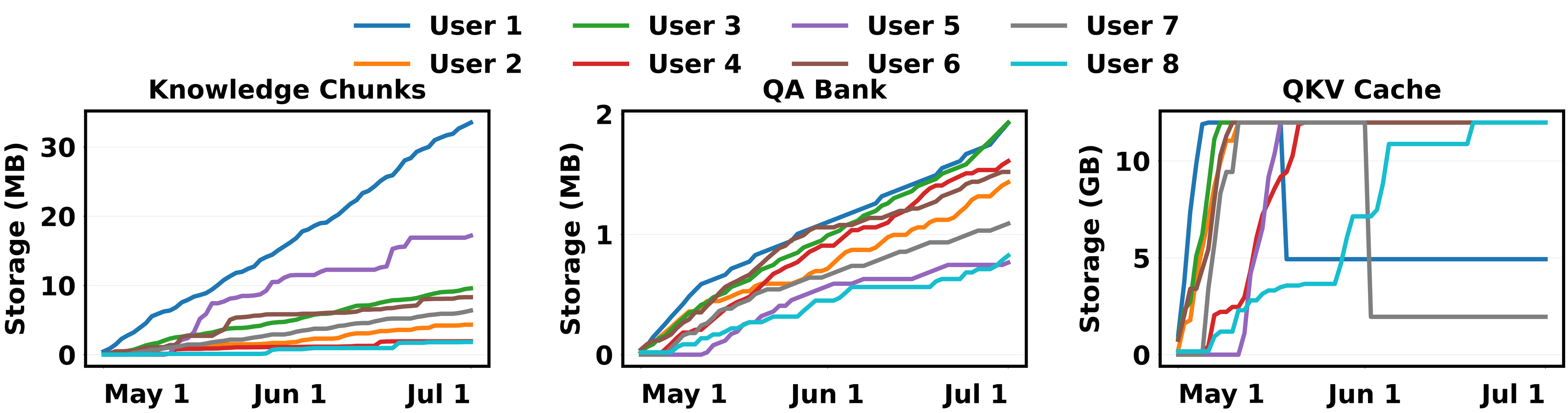}
        \caption{\modify{Storage usage of the knowledge chunks over time, QA Bank, and QKV cache during the two-month real-world deployment across eight users.}}
        \label{fig:real-world-storage-overhead}
    \end{figure}

\begin{figure*}[t]
    \centering
    \begin{minipage}[t]{0.48\textwidth}
        \centering
        \includegraphics[width=\linewidth]{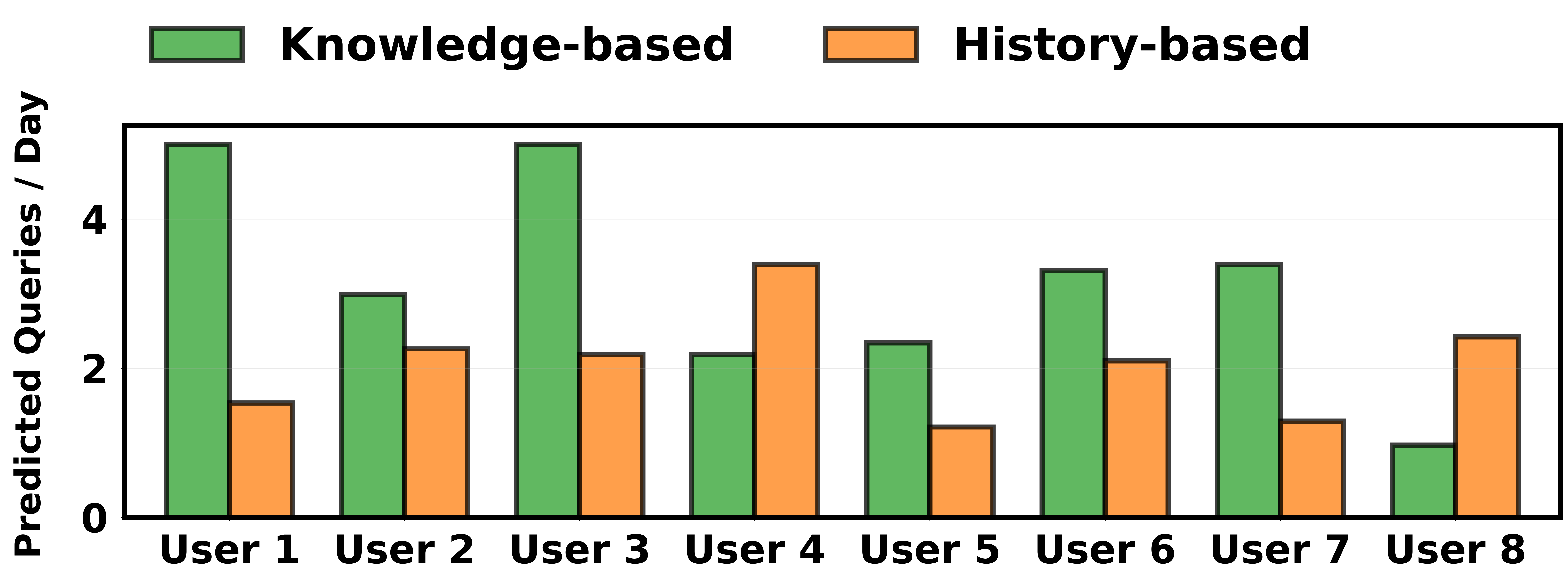}
        \captionof{figure}{\modify{Average number of knowledge-based and history-based predicted queries per day over the two-month deployment.}}
        \label{fig:real-world-prediction-frequency}
    \end{minipage}
    \hfill
    \begin{minipage}[t]{0.48\textwidth}
        \centering
        \includegraphics[width=\linewidth]{figs/real_world_user_study_scores.png}
        \captionof{figure}{\modify{Participants' ratings of \workname across six user-facing dimensions after the two-month real-world deployment.}}
        \label{fig:real-world-user-study}
    \end{minipage}
\end{figure*}

% storage
\modify{
Figures~\ref{fig:real-world-storage-overhead} and~\ref{fig:real-world-prediction-frequency} show that cache population remains bounded under the real-world workload. 
Across users, \workname adds only 1.89--34.60 knowledge chunks and 3.16--7.97 QA items per day on average. 
This variation arises because users' local knowledge chunks grow at different rates as they receive different volumes of emails and chat messages over time.
Storage for the knowledge chunks and QA Bank grows with the accumulated chunks and entries, but both stay small in the end: 1.83--33.52\,MB for the knowledge chunks and 0.77--1.93\,MB for the QA Bank.
The QKV cache instead grows at first and then levels off near its configured budget, showing that the cache-management policy effectively bounds its size.
When Users 1 and 7 lowered their QKV cache budgets mid-deployment, the eviction policy quickly dropped low-priority entries and shrank the cache accordingly, showing that \workname adapts to changing storage constraints.
% energy
Figure~\ref{fig:real-world-prediction-frequency} shows that \workname generates only 0.97--5.00 knowledge-based and 1.21--3.39 history-based predicted queries per user per day, 3.39--7.18 queries in total. 
With a prediction stride of five, this amounts to roughly one to two predictive cache-population rounds per day.
As battery-level indicators have limited resolution and devices may be charging during population, per-round battery changes are usually not directly observable, so we estimate consumption from the calibration in Figure~\ref{fig:battery}.
At about 0.196\% per predicted query, 3.39--7.18 queries per day cost an estimated 0.7--1.4\%.
The actual cost may vary across devices and may be lower when \workname skips decoding during QKV-only population.
Such modest overhead is unlikely to noticeably affect typical daily smartphone use.
}

\subsubsection{\modify{Participant Feedback}}

\modify{
We collected post-deployment feedback from the eight participants, who rated six dimensions on a 1--5 scale: whether response latency was acceptable (\textit{Latency}), whether using \workname interrupted ongoing tasks (\textit{Task Continuity}), whether answers met their information needs (\textit{Answer Quality}), whether background cache population was non-intrusive (\textit{Background Non-interference}), whether battery impact was acceptable (\textit{Battery}), and whether \workname was overall useful (\textit{Usefulness}).
Figure~\ref{fig:real-world-user-study} reports the average ratings in this order: 3.75/5 for latency and task continuity, 3.50/5 for answer quality, 4.25/5 for background non-interference and battery, and 4.00/5 for overall usefulness.
% perceived larency
\final{Users 1, 3, 4, 6, and 8 rated both latency and task continuity positively, giving each dimension scores of 4, 4, 5, 4, and 4 out of 5, respectively.} They reported that \workname often answered directly from QA-Bank hits and that even cache-miss queries stayed acceptable.
Users~2, 5, and~7 gave more moderate ratings for both dimensions (3/5).
They reported fewer direct QA-Bank hits in their workloads, and mentioned that on a few query-intensive days, consecutive cache misses or partial QKV cache hits led to somewhat longer waits, which slightly affected the flow of their information-seeking tasks.
% answer quality
For \textit{Answer Quality}, Users~1, 2, 6, and~7 gave 4/5, while Users~3, 4, 5, and~8 gave 3/5.
We identified three primary answer-quality failure modes.
(i) \textit{Semantic QA-Bank mismatch}: the answer returned from the QA bank covers little of the information the user asked for.
For example, User 2 asked for a professor's talk title but matched a cached answer about the seminar's time and place.
Such cases arise when queries share a nearly identical topic and phrasing but differ in a key attribute, yielding an overly high semantic similarity.
In such cases, \workname gives users the option to continue waiting for an answer generated for the new query.
(ii) \textit{Retrieval miss or ranking error}:  the needed fact exists in the local knowledge but is missing from the retrieved chunks.
For example, User~1 asked about a project mentioned in a GitHub reading digest, and the system replied that the local knowledge lacked the information.
This occurs when a sparse fact is buried in long chunks or a superficially similar chunk outranks the relevant one.
(iii) \textit{Context overload}: the LLM outputs irrelevant, incoherent, or malformed text instead of an answer.
These cases were associated with overly long retrieved prompts, which exceeded the model's effective context capacity.
The latter two modes stem from the retrieval-and-generation pipeline rather than from cache reuse.
% overhead
Finally, participants rated background non-interference and battery acceptability positively, consistent with \workname's lightweight design: it performed query prediction and cache population for only 5--10 minutes during a user-selected evening window each day, resulting in minimal perceived battery impact.
}

\section{RELATED WORK}
\subsection{Caching for LLM Serving}
Numerous works reuse KV cache across different inputs to avoid redundant computation~\cite{jin2024ragcache, gim2024prompt, liu2024cachegen, ye2024chunkattention, yang2025kvlink, chucar, zhao2024blendserve, zhu2024accelerating}.
Prompt Cache~\cite{gim2024prompt} achieves structured reuse of attention states for text segments. 
MobiLoRA~\cite{li2025mobilora} reuses the KV cache of shared contexts for LoRA serving on mobile devices.
Other works use semantic caching~\cite{bang2023gptcache, couturier2025semantic, mohandoss2024context, li2024scalm, gill2024privacy} to store historical queries and skip inference for semantically similar new queries.
GPTCache~\cite{bang2023gptcache} uses centralized storage for query-response pairs across all users. 
MeanCache~\cite{gill2024privacy} employs user-specific caches and uses federated learning for privacy-preserving embedding model training.
To improve cache hit rate, SCALM~\cite{li2024scalm} uses hierarchical semantic clustering to identify significant cache entries.
However, these approaches adopt a reactive strategy to update the cache based on received user queries, resulting in \modify{low cache-hit rates} under sparse query patterns in single-user mobile environments.

\subsection{RAG Performance Optimization}
Significant efforts have focused on the optimization of the performance of the RAG workflow.
RAGCache~\cite{jin2024ragcache} reuses KV cache of external knowledge across multiple requests, computing KV tensors only when absent.
Couturier et al.~\cite{couturier2025semantic} cache compressed document summaries for latency reduction.
However, these works update cache reactively without considering mobile query sparsity. 
We address this through query prediction that predictively populates the cache to improve hit rates.
\modify{Some works also focus on} the acceleration of the knowledge retrieval process. 
Chameleon~\cite{jiang2023chameleon} implements a distributed engine for large-scale vector search.
PipeRAG~\cite{jiang2024piperag} proposes pipelining the retrieval and LLM generation processes to reduce the latency. 
RaLMSpec~\cite{zhang2024accelerating} stores documents in a local cache and speculatively retrieves documents from this cache, verifying them against the retrieval results from the remote knowledge base.
RAGO~\cite{jiang2025rago} abstracts the RAG workload in a unified schema and proposes a systematic framework to optimize scheduling policies for RAG workflow.
\modify{Unlike these approaches}, we focus on single-user mobile scenarios with one-turn QA, where retrieval latency is not a bottleneck since personal knowledge banks are much smaller than server-side knowledge bases and retrieval occurs only once per query.

\subsection{Precomputation for LLM Inference}
Precomputation is a widely adopted strategy for reducing real-time latency in many systems.
Sleep-time Compute~\cite{lin2025sleep} infers potential user questions and conducts LLM reasoning in advance to reduce test-time inference latency. 
PROMTEC~\cite{lee2025promtec} precomputes frequently occurring token sequences offline and accelerates online inference by matching and validating candidate sequences from the precomputed corpus. Pre\textsuperscript{3}~\cite{chen2025pre} precomputes deterministic grammar transition mappings to eliminate runtime parsing overhead for structured LLM outputs. 
Inspired by these works, we predict future user queries and populate the cache based on the predicted results in advance to improve the cache hit rate.

\section{DISCUSSION}

\noindent\textbf{Compatibility.}
\workname~can be extended to other Transformer-based LLMs like Gemma~\cite{team2024gemma}, which share similar attention mechanisms despite different weight distributions. Future work could also apply \workname~to multimodal models such as LLaVA~\cite{liu2023visual} to support diverse modalities of personal data.

\noindent\textbf{More Flexible Query Prediction.}
\modify{The number of queries predicted by} \workname~is currently configured as a system parameter, which may not be the best choice for each query prediction.
Future work will investigate more adaptive approaches that enable the LLM to dynamically determine the appropriate number of queries based on contextual demands.

\noindent\textbf{Fine-grained Cache Organization.}
\final{\workname~currently caches each knowledge chunk as a unit, storing its QKV tensors across all Transformer layers. Future work could explore access-frequency-aware cache organization, retaining tensors from more layers for frequently accessed chunks and fewer layers or selected tensor components (e.g., K or V) for less frequently accessed chunks, to better balance storage usage and online recomputation costs.}

\section{CONCLUSION}

We propose \workname, a hierarchical cache system for mobile RAG applications. 
\workname~integrates semantic cache for user queries and QKV cache reuse for retrieved knowledge chunks, fully exploiting reusable results in mobile RAG workflows.
It utilizes the local LLM to predict future queries to improve cache hit rates under sparse user queries and enables flexible cache scheduling to adapt to dynamic system resources.
\modify{
Across five datasets, \workname achieves the lowest average end-to-end latency while maintaining answer quality comparable to the Naive baseline.
}

\section*{Acknowledgments}

This work is partially supported by the Hong Kong Applied Science and Technology Research Institute under grant UD-1t-TH2318334 and by the Research Grants Council of Hong Kong under grant CUHK C4032-25E.
\section*{Ethics Statement and AI Usage}

This work raises no ethical issues. GPT-5.6 Sol and GPT-5.6 Terra were used to improve writing style, phrasing, and grammar, and to assist with programming and figure preparation. The authors reviewed and verified the resulting text, code, and figures and take full responsibility for the final work.
\vspace{1em}

\bibliographystyle{ACM-Reference-Format}
\bibliography{literature}

@String{Computing = "Computing" }

@String{Computer = "{IEEE} Computer" }

@article{lin2025sleep,
  title={{Sleep-time Compute: Beyond Inference Scaling at Test-time}},
  author={Lin, Kevin and Snell, Charlie and Wang, Yu and Packer, Charles and Wooders, Sarah and Stoica, Ion and Gonzalez, Joseph E},
  journal={arXiv preprint arXiv:2504.13171},
  year={2025}
}

@inproceedings{yao2025cacheblend,
  title={CacheBlend: Fast Large Language Model Serving for RAG with Cached Knowledge Fusion},
  author={Yao, Jiayi and Li, Hanchen and Liu, Yuhan and Ray, Siddhant and Cheng, Yihua and Zhang, Qizheng and Du, Kuntai and Lu, Shan and Jiang, Junchen},
  booktitle={Proceedings of the Twentieth European Conference on Computer Systems},
  pages={94--109},
  year={2025}
}

@article{jin2024ragcache,
  title={{RAGCache: Efficient Knowledge Caching for Retrieval-Augmented Generation}},
  author={Jin, Chao and Zhang, Zili and Jiang, Xuanlin and Liu, Fangyue and Liu, Xin and Liu, Xuanzhe and Jin, Xin},
  journal={arXiv preprint arXiv:2404.12457},
  year={2024}
}

@article{gim2024prompt,
  title={{Prompt Cache: Modular Attention Reuse for Low-Latency Inference}},
  author={Gim, In and Chen, Guojun and Lee, Seung-seob and Sarda, Nikhil and Khandelwal, Anurag and Zhong, Lin},
  journal={Proceedings of Machine Learning and Systems},
  volume={6},
  pages={325--338},
  year={2024}
}

@inproceedings{liu2024cachegen,
  title={{CacheGen: KV Cache Compression and Streaming for Fast Large Language Model Serving}},
  author={Liu, Yuhan and Li, Hanchen and Cheng, Yihua and Ray, Siddhant and Huang, Yuyang and Zhang, Qizheng and Du, Kuntai and Yao, Jiayi and Lu, Shan and Ananthanarayanan, Ganesh and others},
  booktitle={Proceedings of the ACM SIGCOMM 2024 Conference},
  pages={38--56},
  year={2024}
}

@article{ye2024chunkattention,
  title={{ChunkAttention: Efficient Self-Attention with Prefix-Aware KV Cache and Two-Phase Partition}},
  author={Ye, Lu and Tao, Ze and Huang, Yong and Li, Yang},
  journal={arXiv preprint arXiv:2402.15220},
  year={2024}
}

@article{yang2025kvlink,
  title={{KVLink: Accelerating Large Language Models via Efficient KV Cache Reuse}},
  author={Yang, Jingbo and Hou, Bairu and Wei, Wei and Bao, Yujia and Chang, Shiyu},
  journal={arXiv preprint arXiv:2502.16002},
  year={2025}
}

@misc{chucar,
  title={{CaR: An Efficient KV Cache Reuse System for Large Language Model Inference}},
  author={Chu, Kexin and Liu, Tzechinh and Li, Yunding and Yuan, Pengchao and Zhang, Wei},
  url={https://llm-gnn.org/papers/CaR.pdf},
  note={Undated manuscript}
}

@misc{yi2023mllm,
  title = {mllm: fast and lightweight multimodal LLM inference engine for mobile and edge devices},
  author = {Rongjie Yi and Xiang Li and Zhenyan Lu and Hao Zhang and Daliang Xu and Liming Yang and Weikai Xie and Chenghua Wang and Xuanzhe Liu and Mengwei Xu},
  year = {2023},
  publisher = {mllm Team},
  url = {https://github.com/UbiquitousLearning/mllm}
}

@misc{termux2025,
  title = {Termux - a terminal emulator application for Android OS extendible by variety of packages},
  author = {{Termux Contributors}},
  year = {2025},
  howpublished = {\url{https://github.com/termux/termux-app}},
  note = {GitHub repository, accessed July 2025},
  version = {v0.118.3}
}

@inproceedings{bang2023gptcache,
  title={{GPTCache: An Open-Source Semantic Cache for LLM Applications Enabling Faster Answers and Cost Savings}},
  author={Bang, Fu},
  booktitle={Proceedings of the 3rd Workshop for Natural Language Processing Open Source Software (NLP-OSS 2023)},
  pages={212--218},
  year={2023}
}

@misc{ollama2024,
  title = {Ollama: Get up and running with Llama 3.3, DeepSeek-R1, Phi-4, Gemma 3, Mistral Small 3.1 and other large language models},
  author = {{Ollama Team}},
  year = {2024},
  url = {https://github.com/ollama/ollama},
  note = {GitHub repository},
  howpublished = {\url{https://github.com/ollama/ollama}},
  urldate = {2025-07-15}
}

@misc{brown2022rank_bm25,
  title={{rank\_bm25: A Collection of BM25 Algorithms in Python}},
  author={Brown, Dorian},
  year={2022},
  version={0.2.2},
  url={https://github.com/dorianbrown/rank_bm25},
  note={Version 0.2.2}
}

@article{robertson2009probabilistic,
  title={{The Probabilistic Relevance Framework: BM25 and Beyond}},
  author={Robertson, Stephen and Zaragoza, Hugo},
  journal={Foundations and Trends in Information Retrieval},
  volume={3},
  number={4},
  pages={333--389},
  year={2009},
  doi={10.1561/1500000019}
}

@misc{eversberg2024hybrid,
  title={{How to Use Hybrid Search for Better LLM RAG Retrieval}},
  author={Eversberg, Leon},
  year={2024},
  month={August},
  howpublished={Towards Data Science / Medium},
  url={https://medium.com/data-science/how-to-use-hybrid-search-for-better-llm-rag-retrieval-032f66810ebe}
}

@article{park2025mobilerag,
  title={{MobileRAG: A Fast, Memory-Efficient, and Energy-Efficient Method for On-Device RAG}},
  author={Park, Taehwan and Lee, Geonho and Kim, Min-Soo},
  journal={arXiv preprint arXiv:2507.01079},
  year={2025}
}

@misc{meta2024llama32,
  title={Llama 3.2: Revolutionizing edge AI and vision with open, customizable models},
  author={{Meta AI}},
  howpublished={\url{https://ai.meta.com/blog/llama-3-2-connect-2024-vision-edge-mobile-devices/}},
  year={2024},
  month={September},
  day={25},
  note={Accessed: 2025-07-27}
}

@article{zhang2025qwen3,
  title={{Qwen3 Embedding: Advancing Text Embedding and Reranking Through Foundation Models}},
  author={Zhang, Yanzhao and Li, Mingxin and Long, Dingkun and Zhang, Xin and Lin, Huan and Yang, Baosong and Xie, Pengjun and Yang, An and Liu, Dayiheng and Lin, Junyang and others},
  journal={arXiv preprint arXiv:2506.05176},
  year={2025}
}

@misc{google_pixel7_2024,
  title={{Pixel phone hardware tech specs}},
  author={{Google}},
  url={https://support.google.com/pixelphone/answer/7158570?hl=en},
  note={Pixel 7 section}
}

@inproceedings{li2025mobilora,
  title={{MobiLoRA: Accelerating LoRA-based LLM Inference on Mobile Devices via Context-aware KV Cache Optimization}},
  author={Li, Borui and Wang, Yitao and Ma, Haoran and Chen, Ligeng and Xiao, Jun and Wang, Shuai},
  booktitle={Proceedings of the 63rd Annual Meeting of the Association for Computational Linguistics (Volume 1: Long Papers)},
  pages={23400--23410},
  year={2025}
}

@article{zhao2024blendserve,
  title={{BlendServe: Optimizing Offline Inference for Auto-regressive Large Models with Resource-aware Batching}},
  author={Zhao, Yilong and Yang, Shuo and Zhu, Kan and Zheng, Lianmin and Kasikci, Baris and Zhou, Yang and Xing, Jiarong and Stoica, Ion},
  journal={arXiv preprint arXiv:2411.16102},
  year={2024}
}

@inproceedings{lee2025promtec,
  title={PROMTEC: Fast LLM Inference Decoding using Prompt Multi-Lookup with Template Database and Common Sequences},
  author={Lee, Alan Chi-Man and Cheng, Wing-Sun and Chan, Calvin Chun-Kit},
  booktitle={Findings of the Association for Computational Linguistics: ACL 2025},
  pages={6830--6842},
  year={2025}
}

@article{chen2025pre,
  title={{Pre$^3$: Enabling Deterministic Pushdown Automata for Faster Structured LLM Generation}},
  author={Chen, Junyi and Bai, Shihao and Wang, Zaijun and Wu, Siyu and Du, Chuheng and Yang, Hailong and Gong, Ruihao and Liu, Shengzhong and Wu, Fan and Chen, Guihai},
  journal={arXiv preprint arXiv:2506.03887},
  year={2025}
}

@article{li2024personal,
  title={{Personal LLM Agents: Insights and Survey about the Capability, Efficiency and Security}},
  author={Li, Yuanchun and Wen, Hao and Wang, Weijun and Li, Xiangyu and Yuan, Yizhen and Liu, Guohong and Liu, Jiacheng and Xu, Wenxing and Wang, Xiang and Sun, Yi and others},
  journal={arXiv preprint arXiv:2401.05459},
  year={2024}
}

@article{lewis2020retrieval,
  title={{Retrieval-Augmented Generation for Knowledge-Intensive NLP Tasks}},
  author={Lewis, Patrick and Perez, Ethan and Piktus, Aleksandra and Petroni, Fabio and Karpukhin, Vladimir and Goyal, Naman and K{\"u}ttler, Heinrich and Lewis, Mike and Yih, Wen-tau and Rockt{\"a}schel, Tim and others},
  journal={Advances in Neural Information Processing Systems},
  volume={33},
  pages={9459--9474},
  year={2020}
}

@article{christmann2025recursive,
  title={{Recursive Question Understanding for Complex Question Answering over Heterogeneous Personal Data}},
  author={Christmann, Philipp and Weikum, Gerhard},
  journal={arXiv preprint arXiv:2505.11900},
  year={2025}
}

@article{seemakhupt2024edgerag,
  title={{EdgeRAG: Online-Indexed RAG for Edge Devices}},
  author={Seemakhupt, Korakit and Liu, Sihang and Khan, Samira},
  journal={arXiv preprint arXiv:2412.21023},
  year={2024}
}

@article{zhu2024accelerating,
  title={{Accelerating Inference of Retrieval-Augmented Generation via Sparse Context Selection}},
  author={Zhu, Yun and Gu, Jia-Chen and Sikora, Caitlin and Ko, Ho and Liu, Yinxiao and Lin, Chu-Cheng and Shu, Lei and Luo, Liangchen and Meng, Lei and Liu, Bang and others},
  journal={arXiv preprint arXiv:2405.16178},
  year={2024}
}

@article{couturier2025semantic,
  title={{Semantic Caching of Contextual Summaries for Efficient Question-Answering with Language Models}},
  author={Couturier, Camille and Mastorakis, Spyros and Shen, Haiying and Rajmohan, Saravan and R{\"u}hle, Victor},
  journal={arXiv preprint arXiv:2505.11271},
  year={2025}
}

@article{gill2024privacy,
  title={{Privacy-Aware Semantic Cache for Large Language Models}},
  author={Gill, Waris and Elidrisi, Mohamed and Kalapatapu, Pallavi and Anwar, Ali and Gulzar, Muhammad Ali},
  journal={arXiv preprint arXiv:2403.02694},
  year={2024},
  url={https://arxiv.org/abs/2403.02694v1}
}

@inproceedings{li2024scalm,
  title={{SCALM: Towards Semantic Caching for Automated Chat Services with Large Language Models}},
  author={Li, Jiaxing and Xu, Chi and Wang, Feng and von Riedemann, Isaac M and Zhang, Cong and Liu, Jiangchuan},
  booktitle={2024 IEEE/ACM 32nd International Symposium on Quality of Service (IWQoS)},
  pages={1--10},
  year={2024},
  organization={IEEE}
}

@inproceedings{mohandoss2024context,
  title={{Context-based Semantic Caching for LLM Applications}},
  author={Mohandoss, Ramaswami},
  booktitle={2024 IEEE Conference on Artificial Intelligence (CAI)},
  pages={371--376},
  year={2024},
  organization={IEEE}
}

@article{jiang2024piperag,
  title={{PipeRAG: Fast Retrieval-Augmented Generation via Algorithm-System Co-design}},
  author={Jiang, Wenqi and Zhang, Shuai and Han, Boran and Wang, Jie and Wang, Bernie and Kraska, Tim},
  journal={arXiv preprint arXiv:2403.05676},
  year={2024}
}

@article{zhang2024accelerating,
  title={{Accelerating Retrieval-Augmented Language Model Serving with Speculation}},
  author={Zhang, Zhihao and Zhu, Alan and Yang, Lijie and Xu, Yihua and Li, Lanting and Phothilimthana, Phitchaya Mangpo and Jia, Zhihao},
  journal={arXiv preprint arXiv:2401.14021},
  year={2024}
}

@article{jiang2023chameleon,
  title={{Chameleon: a Heterogeneous and Disaggregated Accelerator System for Retrieval-Augmented Language Models}},
  author={Jiang, Wenqi and Zeller, Marco and Waleffe, Roger and Hoefler, Torsten and Alonso, Gustavo},
  journal={arXiv preprint arXiv:2310.09949},
  year={2023}
}

@inproceedings{jiang2025rago,
  title={{RAGO: Systematic Performance Optimization for Retrieval-Augmented Generation Serving}},
  author={Jiang, Wenqi and Subramanian, Suvinay and Graves, Cat and Alonso, Gustavo and Yazdanbakhsh, Amir and Dadu, Vidushi},
  booktitle={Proceedings of the 52nd Annual International Symposium on Computer Architecture},
  pages={974--989},
  year={2025}
}

@article{yi2024phonelm,
  title={{PhoneLM: an Efficient and Capable Small Language Model Family through Principled Pre-training}},
  author={Yi, Rongjie and Li, Xiang and Xie, Weikai and Lu, Zhenyan and Wang, Chenghua and Zhou, Ao and Wang, Shangguang and Zhang, Xiwen and Xu, Mengwei},
  journal={arXiv preprint arXiv:2411.05046},
  year={2024}
}

@article{hu2024minicpm,
  title={{MiniCPM: Unveiling the Potential of Small Language Models with Scalable Training Strategies}},
  author={Hu, Shengding and Tu, Yuge and Han, Xu and He, Chaoqun and Cui, Ganqu and Long, Xiang and Zheng, Zhi and Fang, Yewei and Huang, Yuxiang and Zhao, Weilin and others},
  journal={arXiv preprint arXiv:2404.06395},
  year={2024}
}

@article{yao2024minicpm,
  title={{MiniCPM-V: A GPT-4V Level MLLM on Your Phone}},
  author={Yao, Yuan and Yu, Tianyu and Zhang, Ao and Wang, Chongyi and Cui, Junbo and Zhu, Hongji and Cai, Tianchi and Li, Haoyu and Zhao, Weilin and He, Zhihui and others},
  journal={arXiv preprint arXiv:2408.01800},
  year={2024}
}

@article{team2024gemma,
  title={{Gemma: Open Models Based on Gemini Research and Technology}},
  author={Team, Gemma and Mesnard, Thomas and Hardin, Cassidy and Dadashi, Robert and Bhupatiraju, Surya and Pathak, Shreya and Sifre, Laurent and Rivi{\`e}re, Morgane and Kale, Mihir Sanjay and Love, Juliette and others},
  journal={arXiv preprint arXiv:2403.08295},
  year={2024}
}

@article{bellagente2024stable,
  title={{Stable LM 2 1.6B Technical Report}},
  author={Bellagente, Marco and Tow, Jonathan and Mahan, Dakota and Phung, Duy and Zhuravinskyi, Maksym and Adithyan, Reshinth and Baicoianu, James and Brooks, Ben and Cooper, Nathan and Datta, Ashish and others},
  journal={arXiv preprint arXiv:2402.17834},
  year={2024}
}

@article{abdin2024phi,
  title={{Phi-4 Technical Report}},
  author={Abdin, Marah and Aneja, Jyoti and Behl, Harkirat and Bubeck, S{\'e}bastien and Eldan, Ronen and Gunasekar, Suriya and Harrison, Michael and Hewett, Russell J and Javaheripi, Mojan and Kauffmann, Piero and others},
  journal={arXiv preprint arXiv:2412.08905},
  year={2024}
}

@article{allal2025smollm2,
  title={{SmolLM2: When Smol Goes Big -- Data-Centric Training of a Small Language Model}},
  author={Allal, Loubna Ben and Lozhkov, Anton and Bakouch, Elie and Bl{\'a}zquez, Gabriel Mart{\'\i}n and Penedo, Guilherme and Tunstall, Lewis and Marafioti, Andr{\'e}s and Kydl{\'\i}{\v{c}}ek, Hynek and Lajar{\'\i}n, Agust{\'\i}n Piqueres and Srivastav, Vaibhav and others},
  journal={arXiv preprint arXiv:2502.02737},
  year={2025}
}

@article{christmann2025reqap,
  title={{The ReQAP System for Question Answering over Personal Information}},
  author={Christmann, Philipp and Weikum, Gerhard},
  journal={arXiv preprint arXiv:2508.06880},
  year={2025}
}

@article{yue2025survey,
  title={{A Survey of Large Language Model Agents for Question Answering}},
  author={Yue, Murong},
  journal={arXiv preprint arXiv:2503.19213},
  year={2025}
}

@inproceedings{xu2025fast,
  title={{Fast On-device LLM Inference with NPUs}},
  author={Xu, Daliang and Zhang, Hao and Yang, Liming and Liu, Ruiqi and Huang, Gang and Xu, Mengwei and Liu, Xuanzhe},
  booktitle={Proceedings of the 30th ACM International Conference on Architectural Support for Programming Languages and Operating Systems, Volume 1},
  pages={445--462},
  year={2025}
}

@article{lu2024small,
  title={{Small Language Models: Survey, Measurements, and Insights}},
  author={Lu, Zhenyan and Li, Xiang and Cai, Dongqi and Yi, Rongjie and Liu, Fangming and Zhang, Xiwen and Lane, Nicholas D and Xu, Mengwei},
  journal={arXiv preprint arXiv:2409.15790},
  year={2024}
}

@article{yin2024elms,
  title={{ELMS: Elasticized Large Language Models On Mobile Devices}},
  author={Yin, Wangsong and Yi, Rongjie and Xu, Daliang and Huang, Gang and Xu, Mengwei and Liu, Xuanzhe},
  journal={arXiv preprint arXiv:2409.09071},
  year={2024},
  url={https://arxiv.org/abs/2409.09071v1}
}

@article{yin2024llm,
  title={{LLM as a System Service on Mobile Devices}},
  author={Yin, Wangsong and Xu, Mengwei and Li, Yuanchun and Liu, Xuanzhe},
  journal={arXiv preprint arXiv:2403.11805},
  year={2024}
}

@inproceedings{yuan2024mobile,
  title={Mobile foundation model as firmware},
  author={Yuan, Jinliang and Yang, Chen and Cai, Dongqi and Wang, Shihe and Yuan, Xin and Zhang, Zeling and Li, Xiang and Zhang, Dingge and Mei, Hanzi and Jia, Xianqing and others},
  booktitle={Proceedings of the 30th Annual International Conference on Mobile Computing and Networking},
  pages={279--295},
  year={2024}
}

@article{xue2024powerinfer,
  title={{PowerInfer-2: Fast Large Language Model Inference on a Smartphone}},
  author={Xue, Zhenliang and Song, Yixin and Mi, Zeyu and Zheng, Xinrui and Xia, Yubin and Chen, Haibo},
  journal={arXiv preprint arXiv:2406.06282},
  year={2024}
}

@software{llamacpp2023,
  title = {llama.cpp: {LLM} inference in {C/C++}},
  author = {{ggml-org}},
  year = {2023},
  url = {https://github.com/ggml-org/llama.cpp},
  license = {MIT},
  note = {GitHub repository with 85.3k stars and 12.8k forks}
}

@inproceedings{kwon2023efficient,
  title={{Efficient Memory Management for Large Language Model Serving with PagedAttention}},
  author={Kwon, Woosuk and Li, Zhuohan and Zhuang, Siyuan and Sheng, Ying and Zheng, Lianmin and Yu, Cody Hao and Gonzalez, Joseph and Zhang, Hao and Stoica, Ion},
  booktitle={Proceedings of the 29th Symposium on Operating Systems Principles},
  pages={611--626},
  year={2023}
}

@article{golany2024efficient,
  title={{Efficient Data Generation for Source-grounded Information-seeking Dialogs: A Use Case for Meeting Transcripts}},
  author={Golany, Lotem and Galgani, Filippo and Mamo, Maya and Parasol, Nimrod and Vandsburger, Omer and Bar, Nadav and Dagan, Ido},
  journal={arXiv preprint arXiv:2405.01121},
  year={2024}
}

@article{ryan2025enronqa,
  title={{EnronQA: Towards Personalized RAG over Private Documents}},
  author={Ryan, Michael J and Xu, Danmei and Nivera, Chris and Campos, Daniel},
  journal={arXiv preprint arXiv:2505.00263},
  year={2025}
}

@misc{qwen3-0.6b,
      title={Qwen3-0.6B},
      author={Qwen Team},
      year={2025},
      howpublished={\url{https://huggingface.co/Qwen/Qwen3-0.6B}},
      note={Accessed: 2025-08-26}
}

@article{liu2023visual,
  title={{Visual Instruction Tuning}},
  author={Liu, Haotian and Li, Chunyuan and Wu, Qingyang and Lee, Yong Jae},
  journal={Advances in neural information processing systems},
  volume={36},
  pages={34892--34916},
  year={2023}
}

@article{kozma2024theoretical,
  title={{Theoretical Analysis of Byte-Pair Encoding}},
  author={Kozma, L{\'a}szl{\'o} and Voderholzer, Johannes},
  journal={arXiv preprint arXiv:2411.08671},
  year={2024}
}

@inproceedings{quinn2025drex,
  title={DReX: Accurate and Scalable Dense Retrieval Acceleration via Algorithmic-Hardware Codesign},
  author={Quinn, Derrick and Y{\"u}cel, E Ezgi and Prammer, Martin and Fan, Zhenxing and Skadron, Kevin and Patel, Jignesh M and Mart{\'\i}nez, Jos{\'e} F and Alian, Mohammad},
  booktitle={Proceedings of the 52nd Annual International Symposium on Computer Architecture},
  pages={1108--1124},
  year={2025}
}

@article{chen2025drag,
  title={{DRAG: Distilling RAG for SLMs from LLMs to Transfer Knowledge and Mitigate Hallucination via Evidence and Graph-based Distillation}},
  author={Chen, Jennifer and Myrzakhan, Aidar and Luo, Yaxin and Khan, Hassaan Muhammad and Bsharat, Sondos Mahmoud and Shen, Zhiqiang},
  journal={arXiv preprint arXiv:2506.01954},
  year={2025}
}

@inproceedings{liu2025heterrag,
  title={HeterRAG: Heterogeneous Processing-in-Memory Acceleration for Retrieval-augmented Generation},
  author={Liu, Chaoqiang and Liu, Haifeng and Chen, Dan and Huang, Yu and Zhang, Yi and Xiao, Wenjing and Liao, Xiaofei and Jin, Hai},
  booktitle={Proceedings of the 52nd Annual International Symposium on Computer Architecture},
  pages={884--898},
  year={2025}
}

@article{bezerra2025llmquoter,
  title={{LLMQuoter: Enhancing RAG Capabilities Through Efficient Quote Extraction From Large Contexts}},
  author={Bezerra, Yuri Fa{\c{c}}anha and Weigang, Li},
  journal={arXiv preprint arXiv:2501.05554},
  year={2025}
}

@misc{qwen1.5,
    title = {Introducing Qwen1.5},
    url = {https://qwenlm.github.io/blog/qwen1.5/},
    author = {Qwen Team},
    month = {February},
    year = {2024}
}

@online{gsmarena_redmi_k60_pro,
    title = {Xiaomi Redmi K60 Pro - Full phone specifications},
    author = {{GSMArena}},
    url = {https://www.gsmarena.com/xiaomi_redmi_k60_pro-12046.php},
    year = {2023},
    urldate = {2025-12-04},
    note = {Released 2023, January 01}
}

@online{gsmarena_galaxy_s22_ultra,
    title = {Samsung Galaxy S22 Ultra 5G - Full phone specifications},
    author = {{GSMArena}},
    url = {https://www.gsmarena.com/samsung_galaxy_s22_ultra_5g-11251.php},
    year = {2022},
    urldate = {2025-12-04},
    note = {Released 2022, February 25}
}

@article{su2024roformer,
  title={{RoFormer: Enhanced transformer with Rotary Position Embedding}},
  author={Su, Jianlin and Ahmed, Murtadha and Lu, Yu and Pan, Shengfeng and Bo, Wen and Liu, Yunfeng},
  journal={Neurocomputing},
  volume={568},
  pages={127063},
  year={2024},
  publisher={Elsevier}
}

@misc{gsmarena2025oneplus,
  author = {{GSMArena}},
  title = {OnePlus Ace 6 - Full phone specifications},
  year = {2025},
  month = {October},
  url = {https://www.gsmarena.com/oneplus_ace_6_5g-14259.php},
  note = {Accessed: December 5, 2025}
}

@inproceedings{du2024perltqa,
  title={{PerLTQA: A Personal Long-Term Memory Dataset for Memory Classification, Retrieval, and Fusion in Question Answering}},
  author={Du, Yiming and Wang, Hongru and Zhao, Zhengyi and Liang, Bin and Wang, Baojun and Zhong, Wanjun and Wang, Zezhong and Wong, Kam-Fai},
  booktitle={Proceedings of the 10th SIGHAN Workshop on Chinese Language Processing (SIGHAN-10)},
  pages={152--164},
  year={2024}
}

@inproceedings{chung2026agent,
  title={Agent-X: Full Pipeline Acceleration of On-device AI Agents},
  author={Chung, Jinha and Shin, Byeongjun and Kim, Jiin and Rhu, Minsoo},
  booktitle={Proceedings of the 24th Annual International Conference on Mobile Systems, Applications and Services},
  pages={144--157},
  year={2026}
}

@inproceedings{lin2004rouge,
  title={{ROUGE: A Package for Automatic Evaluation of Summaries}},
  author={Lin, Chin-Yew},
  booktitle={Text Summarization Branches Out},
  pages={74--81},
  year={2004}
}

@misc{deepseek2026v4flash,
  author       = {{DeepSeek}},
  title        = {{DeepSeek V4 Preview Release}},
  year         = {2026},
  month        = apr,
  day          = {24},
  howpublished = {\url{https://api-docs.deepseek.com/news/news260424/}},
  note         = {Accessed: 2026-07-27}
}

@inproceedings{zhang2023h2o,
  title     = {{H2O}: Heavy-Hitter Oracle for Efficient Generative Inference of Large Language Models},
  author    = {Zhang, Zhenyu and Sheng, Ying and Zhou, Tianyi and Chen, Tianlong and Zheng, Lianmin and Cai, Ruisi and Song, Zhao and Tian, Yuandong and R{\'e}, Christopher and Barrett, Clark and Wang, Zhangyang and Chen, Beidi},
  booktitle = {Advances in Neural Information Processing Systems (NeurIPS)},
  volume    = {36},
  pages     = {34661--34710},
  year      = {2023}
}

@inproceedings{li2024snapkv,
  title     = {{SnapKV}: {LLM} Knows What You are Looking for Before Generation},
  author    = {Li, Yuhong and Huang, Yingbing and Yang, Bowen and Venkitesh, Bharat and Locatelli, Acyr and Ye, Hanchen and Cai, Tianle and Lewis, Patrick and Chen, Deming},
  booktitle = {Advances in Neural Information Processing Systems (NeurIPS)},
  volume    = {37},
  pages     = {22947--22970},
  year      = {2024}
}

%TC:ignore
\clearpage
\appendix

\section{Appendix}

\subsection{Inference with QKV Cache}
\label{sec:inference_with_qkv_cache}

\begin{figure}[H]
\centering
\captionsetup{skip=0pt}
\begin{minipage}[t]{0.37\textwidth}
    \centering
    \resizebox{\columnwidth}{!}{
    \includegraphics{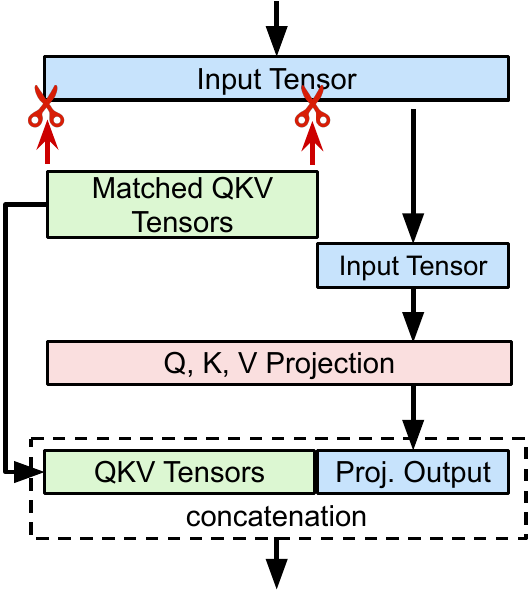}}
    \caption{Inference implementation details.}
    \label{fig:inference_with_cache_details}
\end{minipage}
\hfill
\begin{minipage}[t]{0.59\textwidth}
    \centering
    \resizebox{\columnwidth}{!}{
    \includegraphics{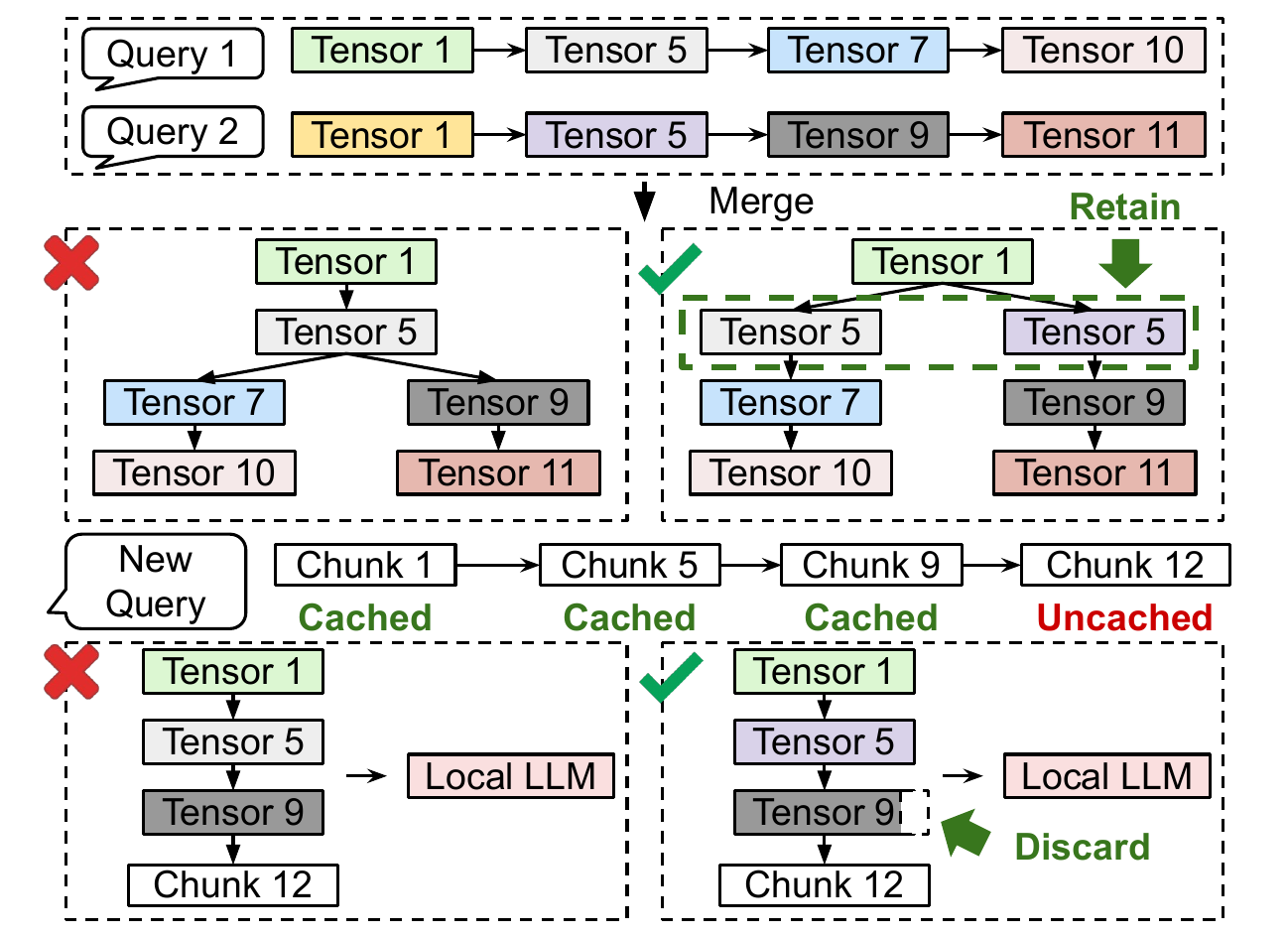}}
    \caption{Tokenization inconsistency and solution.}
    \label{fig:tokenization_issue_and_solution}
\end{minipage}
\end{figure}

Figure~\ref{fig:inference_with_cache_details} shows a simple implementation diagram of how \workname~uses precomputed QKV Cache for inference acceleration during the prefill phase.
The input text is first converted from a string to a tensor by the tokenizer, and then passes through the LLM's embedding layer.
At this \final{stage}, \workname~loads the previously matched QKV tensor from disk and obtains the sequence length value $L_{pre}$ of the QKV tensor. Meanwhile, it also records the sequence length value $L_{total}$ of the embedding layer's output.
Afterwards, this tensor goes through several transformer blocks.
As shown in Figure~\ref{fig:inference_with_cache_details}, within each block, the portion of the tensor with sequence length $L_{pre}$ is first cropped from the beginning of the input tensor of that layer, keeping only the latter part.
Then, the portion with sequence length $L_{total} - L_{pre}$ is used to perform the QKV projection operation.

It is worth noting that since Llama-3.2-3B and Qwen-1.5-1.8B in the mllm repository use Rotary Position Embedding (RoPE)~\cite{su2024roformer} for position encoding, we need to make appropriate adjustments accordingly. Specifically, after obtaining the computation results of the QKV projection on the cropped input tensor, we cannot directly apply RoPE position encoding. Instead, we need to adjust the position index used for retrieving the precomputed sine and cosine values. In the implementation, before applying RoPE to the cropped tensor with length $L_{total} - L_{pre}$, we first offset the position counter by adding $L_{pre}$. This ensures that when accessing the sine and cosine lookup tables, the position encodings correspond to the actual positions in the original full sequence rather than starting from position 0.

\subsection{Tokenization Inconsistency}
\label{sec:tokenization_inconsistency}

Unlike inference engines such as vLLM~\cite{kwon2023efficient} that use fixed-size KV blocks as nodes in the cache prefix tree, each node in our QKV cache tree is an individual knowledge chunk, which facilitates the logical management of the QKV cache tree.
Additionally, to organize the QKV cache according to the semantic content of knowledge chunks, \final{we perform prefix matching on strings encoding sequences of knowledge-chunk identifiers, rather than on token sequences or token hashes.}
However, we notice that some LLM tokenizers like Byte-Pair Encoding (BPE)~\cite{kozma2024theoretical} may cause inconsistencies in subword segmentation.
As shown in Figure~\ref{fig:tokenization_issue_and_solution},  two queries are processed to populate the QKV cache tree.
\workname~merges the common prefix of the QKV tensor chains of their retrieved knowledge chunks and integrates the merged prefix tree into the overall cache tree.
For a new query retrieving chunks 1, 5, 9, and 12, the system loads and concatenates the cached QKV tensors of chunks 1, 5, and 9, then packs the concatenated tensor with chunk 12's raw text for inference.
However, we find that the inference results often differ from naive raw-text-only inference without cache.
This is because: 
(1) The tokenization results near the boundary between tensor 5 (gray) and 7 are different from those between tensor 5 (purple) and 9; 
(2) The tokenization results at the end of tensor 9 are also different from the results when directly inputting the concatenated text of chunks 9 and 12 into the tokenizer.
We take two measures for this issue: 
(1) When merging QKV cache tensor chains, we retain the last node (tensor 5) of the common prefix and only merge up to the second-to-last node of the common prefix; (2) For a matched tensor chain, we discard tensors of the last few tokens in the final node (tensor 9), using raw text inference for this portion instead.
\final{In addition, the cache slicer’s split positions may have minor offsets relative to actual knowledge-chunk boundaries. However, adjacent slices fit together without missing or duplicating tokens, preserving the original tensor sequence when concatenated along the same cached path. Recomputing the boundary suffix maintains inference correctness, provided that the retained cache matches the current input’s token prefix.}

\subsection{Prompt Details}
\label{sec:prompt_details}

In this section, we provide the prompts used in \workname.

\hspace{1em}Figure~\ref{fig:prompt_abstract} shows the prompt template designed to enable the local LLM to extract summaries from one or multiple knowledge chunks. To use this template, \workname~simply replaces the \texttt{[knowledge chunks]} slot with the corresponding knowledge chunk(s) from which to extract summaries.
Once a summary is generated, it will be appended to the overall summary collection.

\hspace{1em}Figure~\ref{fig:knowledge_based_prediction} shows the prompt template used for knowledge-based query prediction. \workname~replaces the \texttt{[knowledge abstract]} slot with the corresponding knowledge abstracts, the \texttt{[personal emails/daily dialog/\\meeting record]} slot with the type of external knowledge, and the \texttt{[prediction stride]} slot with the number of queries to predict.
This template guides the LLM to generate two types of questions: general questions and detail-oriented questions, aiming to cover a comprehensive range of question types.

\hspace{1em}Figure~\ref{fig:history_based_prediction} shows the prompt template used for history-based query prediction. 
When using this template, \workname~replaces the \texttt{[query history]} slot with queries from the query history buffer, the \texttt{[personal emails/daily dialog/meeting record]} slot with the type of external knowledge, and the \texttt{[prediction stride]} slot with the number of queries to predict.
This template guides the local LLM to predict future queries by mimicking the language style, tone, formality level, and interests of real user queries.

\modify{
\hspace{1em}Figure~\ref{fig:llm-judge-prompt} shows the prompt template for LLM-based answer-quality judging. The user query, gold answer, and each method's response fill the \texttt{\{query\}}, \texttt{\{gold\_answer\}}, and \texttt{\{method\_response\}} slots. The judge then rates recall and precision on a 1--4 scale.
}

\subsection{\modify{Answer Quality with Qwen-1.5-1.8B}}
\label{sec:answer-quality-qwen}

\begin{figure}[H]
    \centering
    \subfloat[LLM-judge recall and precision.\label{fig:qwen-llm-judge-quality}]{
        \includegraphics[width=\linewidth]{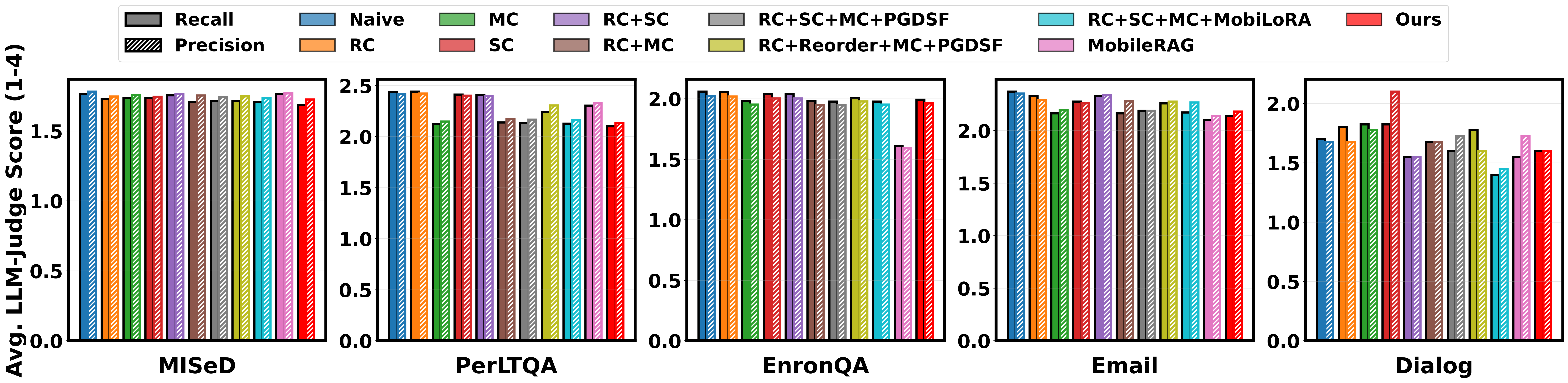}
    }

    \vspace{0.5em}

    \subfloat[ROUGE-L.\label{fig:qwen-rouge-l-quality}]{
        \includegraphics[width=\linewidth]{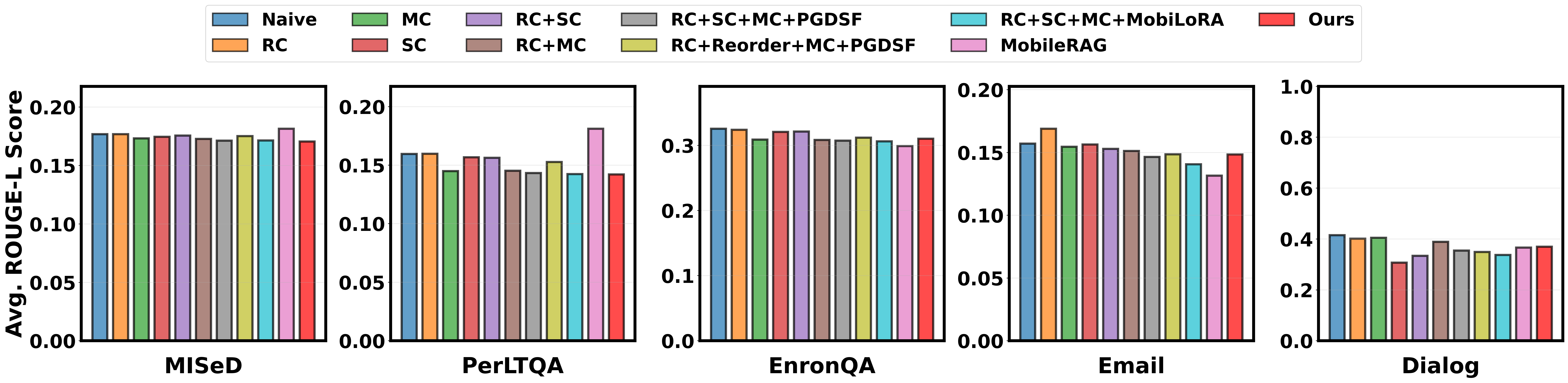}
    }

    \caption{\modify{Answer-quality results of \workname and baselines using Qwen-1.5-1.8B across the five datasets. \workname maintains recall, precision, and ROUGE-L comparable to the Naive baseline.}}
    \label{fig:qwen-answer-quality}
\end{figure}

\modify{Figure~\ref{fig:qwen-answer-quality} reports answer quality with Qwen-1.5-1.8B on the five datasets, using ROUGE-L and LLM-as-a-Judge scores. 
\workname stays close to the Naive baseline on all three scores, with only minor variation across datasets, so its latency gains carry over to Qwen-1.5-1.8B without much loss in answer quality.}

\begin{promptbox}[Knowledge Abstract Generation]
Please summarize the following text into a sentence that captures the key events and key concepts: [{knowledge chunks}]. Focus only on the essential core content and main ideas. Do NOT include any other text or explanations.
\end{promptbox}
\vspace{-1em}
\captionof{figure}{Prompt template for knowledge abstract generation.}
\label{fig:prompt_abstract}

\begin{promptbox}[Knowledge-based Query Prediction]
You are a smartphone AI assistant that helps users by checking their [{personal emails/daily dialog/meeting record}] to answer questions. 
Here is an abstract of the content: [{knowledge abstract}].

Think about what questions users are most likely to ask regarding this content in their daily life.

The user might ask two kinds of questions about their [{personal emails/daily dialog/meeting record}]:

1. General questions, such as "What is the main topic of the [{personal emails/daily dialog/meeting record}] recorded on XXX?"

2. Detailed questions, such as "What activities did the user mention doing on day XXX?"

Now guess [{prediction stride}] questions that users might ask.
Provide them in this format: ``1. ...?; 2. ...?; 3. ...?'' without any other text or explanations.
\end{promptbox}
\vspace{-1em}
\captionof{figure}{Prompt template for knowledge-based query prediction.}
\label{fig:knowledge_based_prediction}

\begin{promptbox}[History-based Query Prediction]
You are a smartphone AI assistant that helps users by checking their [{personal emails/daily dialog/meeting record}] to answer questions. 
Here is their recent query history: [{query history}].

Generate [{prediction stride}] additional questions users might ask.
Carefully analyze and mirror the language style, tone, formality level, and interests shown in the examples.

Please provide them in this format: ``1. ...?; 2. ...?; 3. ...?'' Do not repeat the examples. Do not include any other text or explanations, just the questions.
\end{promptbox}
\vspace{-1em}
\captionof{figure}{Prompt template for history-based query prediction.}
\label{fig:history_based_prediction}

\begin{promptbox}[LLM-based Answer-Quality Judging]
\textbf{[System prompt]}

\begin{lstlisting}[
    basicstyle=\ttfamily\scriptsize,
    breaklines=true,
    columns=fullflexible,
    keepspaces=true
]
You are an answer-quality judge. Evaluate key information overlap, not wording or style.

Evaluation rules:
- Score key information overlap, not wording or style.
- Provide two Likert scores from 1 to 4:
  recall_score: how well the predicted answer covers the key information needed to answer the question.
  precision_score: how much of the predicted answer is supported by or consistent with the gold answer.
- The scale is: 1 = Not Covered; 2 = Slightly Covered; 3 = Mostly Covered; 4 = Fully Covered.
- "Covered" means semantic content overlap or task-relevant equivalence between the gold answer and predicted answer.
- Lower precision only when extra information directly contradicts the gold answer, introduces a key factual error, 
or would mislead the user.
- Give a brief reason only when recall_score is 1 or precision_score is 1. Otherwise use an empty reason.

Return valid JSON only. Do not answer the question. Do not include markdown or explanatory text.

Required schema:
{"recall_score": 1-4, "precision_score": 1-4, "reason": ""}
\end{lstlisting}

\textbf{[User prompt]}

\begin{lstlisting}[
    basicstyle=\ttfamily\scriptsize,
    breaklines=true,
    columns=fullflexible,
    keepspaces=true
]
Question:
{query}

Gold Answer:
{gold_answer}

Predicted Answer:
{method_response}
\end{lstlisting}
\end{promptbox}
\vspace{-1em}
\captionof{figure}{\modify{Prompt template for LLM-based answer-quality judging.}}
\label{fig:llm-judge-prompt}

%TC:endignore

\end{document}